\documentclass[traditabstract]{aa}
\usepackage{graphicx}
\usepackage[varg]{txfonts}
\usepackage{longtable}
\usepackage{lscape}
\usepackage{natbib}
\usepackage{url}
\usepackage{color}
\bibpunct{(}{)}{;}{a}{}{,} 
\newcommand{\feh}{\mathrm{[Fe/H]}}
\newcommand\kms{{\rm\,km\,s^{-1}}}

\newcommand\teff{T_{\rm eff}}

\begin{document}

\title{
Chemical evolution of the Galactic bulge as traced by
microlensed dwarf and subgiant stars\thanks{Based on observations 
made with the European Southern Observatory
telescopes (84.B-0837, 85.B-0399, and 86.B-0757).
This paper also includes data gathered with the 6.5 m Magellan 
Telescopes located at the Las Campanas Observatory, Chile,
and data obtained at the W.M. Keck 
Observatory, which is operated as a scientific partnership among 
the California Institute of Technology, the University of California 
and the National Aeronautics and Space Administration.}
\fnmsep
\thanks{Table 4 is also available in electronic form at the CDS and full Table 5 is only available
in electronic form at the CDS via 
anonymous ftp to {\tt cdsarc.u-strasbg.fr (130.79.128.5)} or via 
{\tt http://cdsweb.u-strasbg.fr/cgi-bin/qcat?J/A+A/XXX/AXX}.}
}
\subtitle{
IV. Two bulge populations
}
\titlerunning{Chemical evolution of the Galactic bulge as traced by
microlensed dwarf and subgiant stars. IV.}

\author{
T.~Bensby\inst{1,2}
\and
D.~Ad\'en\inst{1}
\and
J.~Mel\'endez\inst{3}
\and
A.~Gould\inst{4}
\and
S.~Feltzing\inst{1}
\and
M.~Asplund\inst{5}
\and
J.A.~Johnson\inst{4}\thanks{J.A. Johnson is a guest professor at Lund University}
\and
S.~Lucatello\inst{6}
\and\\
J.C.~Yee\inst{4}
\and
I. Ram\'irez\inst{7}
\and
J.G.~Cohen\inst{8}
\and
I.~Thompson\inst{7}
\and
I.A.~Bond\inst{9}
\and
A.~Gal-Yam\inst{10}
\and
C.~Han\inst{11}
\and
T.~Sumi\inst{12}
\and\\
D. Suzuki\inst{12}
\and
K. Wada\inst{12}
\and
N. Miyake\inst{13}
\and
K. Furusawa\inst{13}
\and
K. Ohmori\inst{13}
\and
To. Saito\inst{14}
\and
P. Tristram\inst{15}
\and
D. Bennett\inst{16}
 }

\institute{Lund Observatory, Department of Astronomy and Theoretical Physics, 
Box 43, SE-221\,00 Lund, Sweden
\and
European Southern Observatory, Alonso de Cordova 3107, 
Vitacura, Casilla 19001, Santiago 19, Chile
\and
Departamento de Astronomia do IAG/USP, Universidade de S\~ao Paulo,
Rua do Mat\~ao 1226, S\~ao Paulo, 05508-900, SP, Brasil
\and
Department of Astronomy, Ohio State University, 140 W. 18th Avenue, 
Columbus, OH 43210, USA
\and
Max Planck Institute for Astrophysics, Postfach 1317, 85741 Garching, Germany
\and
INAF-Astronomical Observatory of Padova, Vicolo dell'Osservatorio 5, 
35122 Padova, Italy
\and
Carnegie Observatories, 813 Santa Barbara Street, Pasadena, CA 91101, USA
\and
Palomar Observatory, Mail Stop 249-17, California Institute of Technology, 
Pasadena, CA 91125, USA
\and
Institute of Information and Mathematical Sciences, Massey University,
Albany Campus, Private Bag 102-904, North Shore Mail Centre,
Auckland, New Zealand
\and
Benoziyo Center for Astrophysics, Weizmann Institute of Science, 
76100 Rehovot, Israel
\and
Department of Physics, Chungbuk National University, Cheongju 361-763, Republic of Korea
\and
Department of Earth and Space Science, Osaka University, Osaka 560-0043, Japan
\and
Solar-Terrestrial Enivironment Laboratory, Nagoya University, Furo-cho, Chikusa-ku, Nagoya, 464-8601, Japan
\and
Tokyo Metropolitan College of Industrial Technology, Tokyo 116-8523, Japan
\and 
Department of Physics, University of Auckland, Private Bag 92019, Auckland 1142, New Zealand
\and
Department of Physics, University of Notre Dame, Notre Dame, IN 46556, USA
}


\date{Received 11 April 2011 / Accepted 26 July 2011}
\offprints{T. Bensby, \email{tbensby@astro.lu.se}}
 \abstract{
 Based on high-resolution ($R\approx42\,000$ to $48\,000$) and high 
 signal-to-noise ($S/N\approx50$ to 150) spectra obtained with UVES/VLT, 
 we present detailed elemental abundances 
 (O, Na, Mg, Al, Si, Ca, Ti, Cr, Fe, Ni, Zn, Y, and Ba) and stellar ages 
 for 12 new microlensed dwarf and subgiant stars in the Galactic bulge. Including
 previous microlensing events, the sample of homogeneously analysed bulge
 dwarfs has now grown to 26.  The analysis is based on equivalent width 
 measurements and standard 1-D LTE MARCS model stellar atmospheres.
 We also present NLTE Li abundances based on line synthesis of 
 the $^7$Li line at 670.8\,nm. The results from the 26 
 microlensed dwarf and subgiant stars show that the bulge metallicity distribution 
 (MDF) is double-peaked; one peak at  $\rm [Fe/H]\approx -0.6$ and one at
 $\rm [Fe/H]\approx +0.3$, and with a dearth of stars around solar
 metallicity. This is in contrast to the MDF derived from red giants in Baade's 
 window, which peaks at this exact value. A simple significance test 
 shows that it is extremely unlikely 
 to have such a gap in the microlensed dwarf star MDF if the dwarf stars are 
 drawn from the giant star MDF. To resolve this issue we discuss
 several possibilities, but we can not settle on a conclusive solution
 for the observed differences. We further find that the metal-poor bulge 
 dwarf stars are predominantly old with ages 
 greater than 10\,Gyr, while the metal-rich bulge dwarf stars show a wide 
 range of ages. The metal-poor bulge sample is very similar to the Galactic 
 thick disk in terms of average metallicity, elemental abundance trends, and 
 stellar ages.
 Speculatively, the metal-rich bulge population might be the manifestation
 of the inner thin disk. If so, the two bulge populations could support the recent 
 findings, based on kinematics, that there are no signatures of a classical bulge
 and that the Milky Way is a pure-disk galaxy.
 Also, recent claims of a flat IMF in the bulge based on the MDF of giant stars
 may have to be revised based on the MDF and abundance trends probed by our
 microlensed dwarf stars.
 }
   \keywords{
   Gravitational lensing: micro ---
   Galaxy: bulge ---
   Galaxy: formation ---
   Galaxy: evolution ---
   Stars: abundances
   }
   \maketitle

\section{Introduction}

Almost 60\,\% of all stellar mass in massive
galaxies in the local Universe is contained in bulges and elliptical
galaxies
\citep[e.g.,][]{gadotti2009}. Being a major component of nearby galaxies
and galaxy populations
and a primary feature that classifies galaxies, it is clear 
that understanding the 
origin and evolution of bulges is integral to the theory of galaxy 
formation.  The central part of our own Milky Way harbours a 
central bulge which enables us to study such a stellar system in a 
detail impossible for any other galaxy  (e.g., \citealt{kormendy2004} for a 
review of bulges in general). For instance, the 
next closest bulge stellar system is that of the Andromeda galaxy which 
is more than a hundred times more distant.
In spite of its ``proximity" and the many detailed spectroscopic
and photometric studies during the last few decades,
the origin and evolutionary history of the Galactic bulge is still
poorly understood. Its generally very old stellar population,
metal-rich nature, and over-abundances of $\alpha$-elements
\citep[e.g.,][]{mcwilliam1994,zoccali2006,fulbright2007,melendez2008,bensby2010}
are consistent with a classical bulge formed during
the collapse of the proto-galaxy and subsequent mergers, which
would have resulted in an intense burst of star formation
\citep[e.g.,][]{white1978,matteucci1990,ferreras2003,rahimi2010}.
Alternatively, the boxy/peanut-like shape of the bulge suggests
an origin through dynamical instabilities in an already
established inner disk
\citep[e.g.,][]{maihara1978,combes1990,shen2010}.
Such secular evolution could possibly explain the recent
discovery of chemical similarities between the bulge and the
Galactic thick disk as observed in the solar neighbourhood 
\citep{melendez2008,alvesbrito2010,bensby2010,gonzalez2011}, 
through the action of radial migration of stars
\citep{sellwood2002,schonrich2009b,loebman2011}.

The metallicity distribution function of the bulge, inferred from photometric 
and spectroscopic studies of red giant stars, peaks around the solar value with 
a significant fraction of super-solar metallicity stars and a low-metallicity 
tail extending down to at least [Fe/H]$\approx -1$
\citep[e.g.,][]{sadler1996,zoccali2003,zoccali2008,fulbright2007}.
Indeed, there are ample indications that the Bulge should harbour substantially
more metal-poor stars \citep[e.g.,][]{bensby2010li}. In fact, the very first 
stars, should any have survived to the present day, may well preferentially be 
found in the central regions of Milky Way-type galaxies 
\citep[e.g.,][]{wyse1992,brook2007,tumlinson2010}. Unfortunately, the high 
stellar densities in the bulge and its generally high metallicity makes finding 
such stellar survivors from the earliest cosmic epochs very challenging although 
several large-scale spectroscopic surveys of the Bulge are currently underway, 
which may discover some of these extremely old stars \citep[e.g.,][]{howard2009}.

Studies of the detailed chemistry of the bulge have so far mainly used 
intrinsically bright stars (as in the references above). However, results
based on spectra from giant stars are not trivial to interpret as evolutionary 
processes erase some of the abundance information. Also, their relatively cool 
atmospheres result in spectra rich in lines from molecules, which are difficult 
to analyse \citep[e.g.,][]{fulbright2006}. The spectra of dwarf stars, on the 
other hand, even metal-rich ones, are fairly straightforward to analyse and are 
the best tracers of Galactic chemical evolution \citep[e.g.,][]{edvardsson1993}. 
The main difficulty with observing dwarf stars in the bulge is their faintness 
($V=19-20$, \citealt{feltzing2000b}), impeding spectroscopic 
observations under normal circumstances. To obtain a 
spectrum of high quality ($R\gtrsim40\,000$ and $S/N\gtrsim75$) of a bulge dwarf 
star would require, even if using the largest 8-10 meter telescopes available 
today, an exposure time of more than 50 hours. Microlensing offers
the unique opportunity to observe dwarf stars in the bulge. In the event that
the bulge dwarf star is lensed by a foreground object, its brightness 
can increase by a factor of several hundred, making it possible to obtain,
in 1 to 2 hours, a spectrum of sufficiently high resolution and $S/N$, 
adequate for an accurate detailed elemental abundance analysis.

By observing microlensed dwarf stars in the bulge several recent studies have 
given a new perspective on the chemical properties of the bulge
\citep{johnson2007,johnson2008,cohen2008,
cohen2009,cohen2010puzzle,bensby2009letter,bensby2009,bensby2010li,
bensby2010,epstein2010}. Main findings so far include the first age-metallicity 
relation in the bulge which shows that metal-poor stars are generally old, 
but metal-rich ones have a wide range of ages \citep{bensby2010}. In addition, 
abundance ratios for 14 elements studied in the bulge dwarfs at sub-solar 
metallicities are in excellent agreement with the abundance patterns in local 
thick disk stars \citep[e.g.,][]{melendez2008,bensby2010,alvesbrito2010}. 
Also, \cite{bensby2010li} presented the first clear detection 
of Li in a bulge dwarf star, showing that the bulge follows the Spite 
plateau \citep{spite1982}. The most striking result to date from microlensed 
dwarf stars is that the metallicity distribution function (MDF) for dwarf 
stars and giant stars in the Galactic bulge differs. In \cite{bensby2010} we 
found that the bulge MDF appeared bimodal for the dwarf stars, with a paucity 
at the metallicity where the MDF based on giant stars in Baade's window 
from \cite{zoccali2008} peaks. Understanding this discrepancy is vital when 
studying external galaxies where dwarf stars can not be studied, where we have 
to rely on the integrated light from all stars, which is dominated by giant stars.

These results illustrate that observations of dwarf stars provide unique
information on the evolution of the bulge. For example, the microlensed
bulge dwarf stars will have an important impact on the modelling of the
bulge, in particular regarding recent suggestions that the initial mass
function (IMF) in the bulge needs to be different from that in the
solar neighbourhood in order to explain the MDF based on red giant stars
\citep{cescutti2011}. Additionally, combining the dwarf abundances and kinematics 
with numerical studies, e.g., \cite{rahimi2010}, points to the 
possibility of disentangling different formation scenarios for the
bulge, e.g. secular versus merger origin.

In this paper we report the most recent findings from our ongoing project 
on the chemical evolution of the Galactic bulge as traced by dwarf and 
subgiant stars that have been observed whilst being optically magnified 
during microlensing events. Adding eleven new events, and a re-analysis 
of MOA-2009-BLG-259S, the sample now consists of in total 26 microlensed 
dwarf and subgiant stars in the bulge that have been homogeneously analysed. 

\begin{figure*}
\resizebox{\hsize}{!}{
\includegraphics[bb=52 40 515 540,clip]{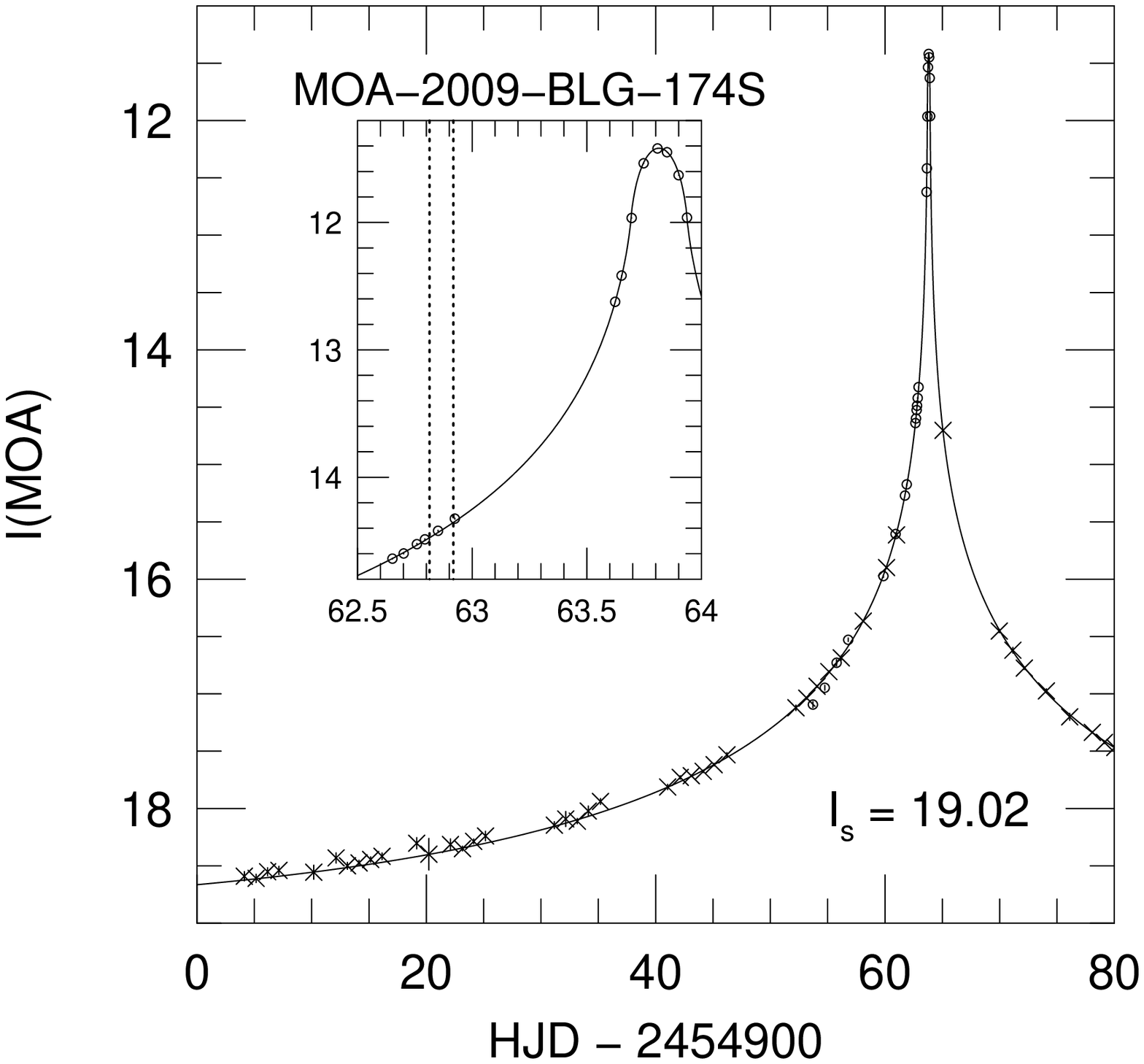}
\includegraphics[bb=52 40 515 540,clip]{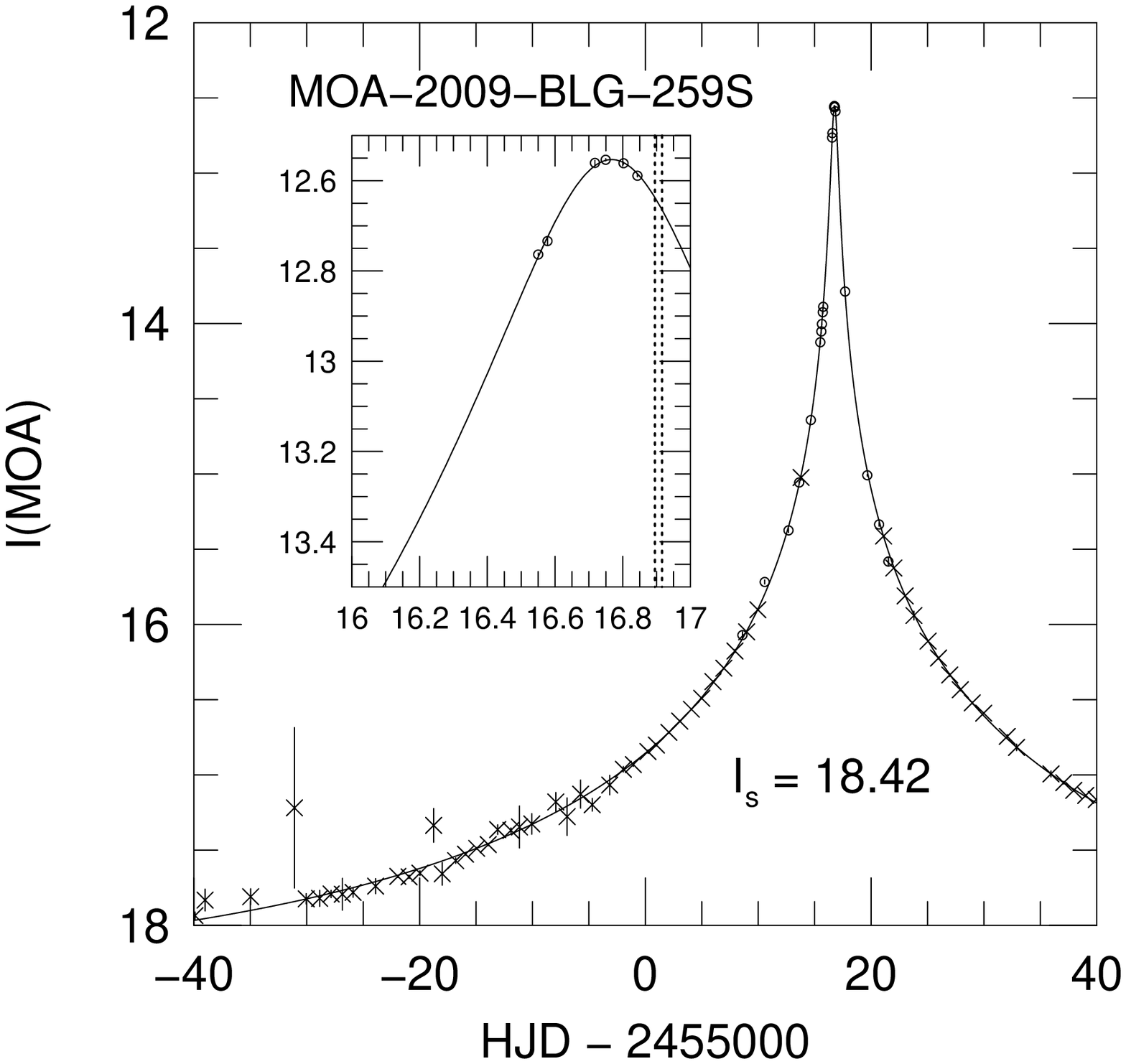}
\includegraphics[bb=52 40 515 540,clip]{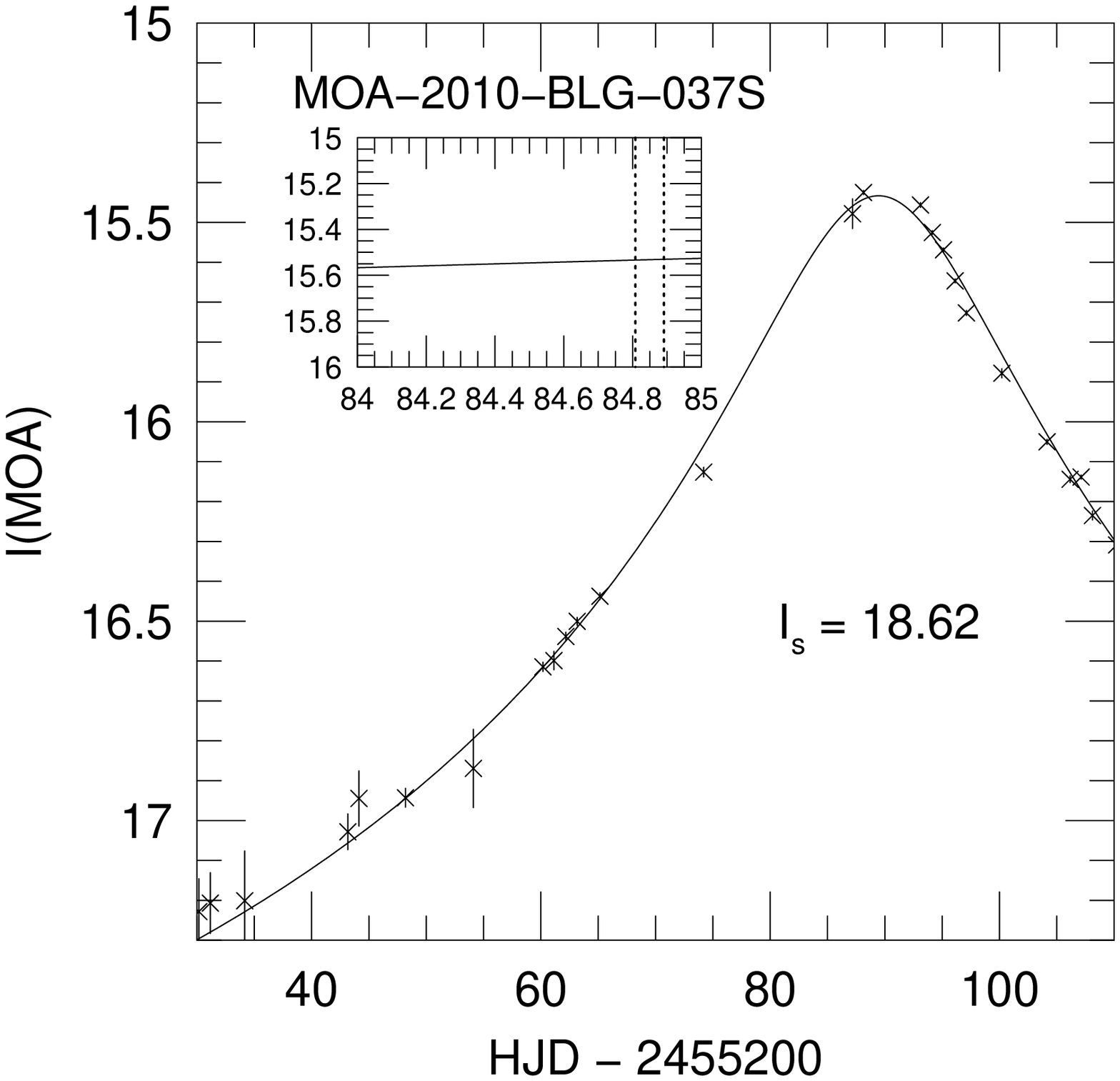}
\includegraphics[bb=52 40 515 540,clip]{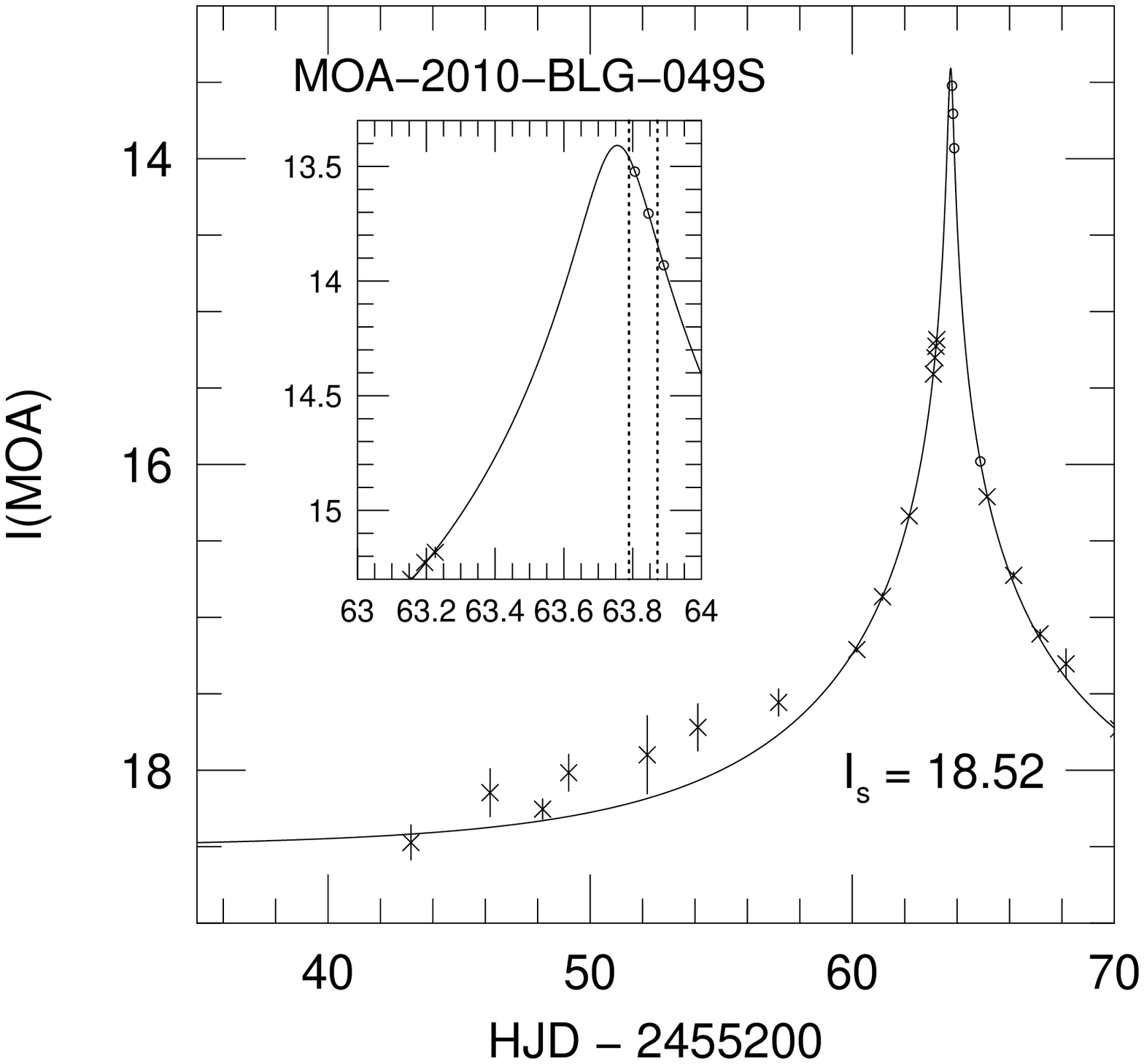}
}
\resizebox{\hsize}{!}{
\includegraphics[bb=52 30 515 540,clip]{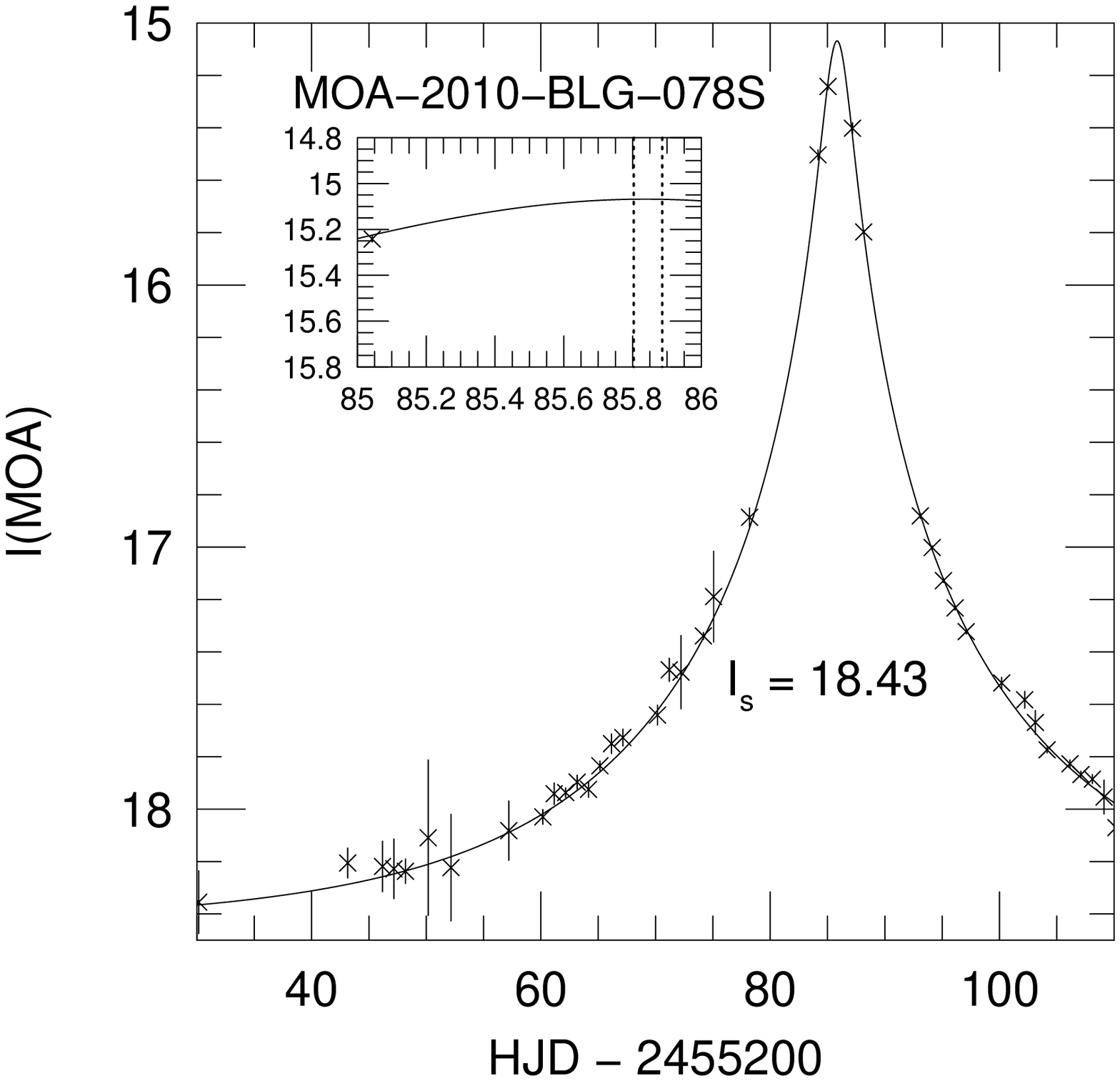}
\includegraphics[bb=52 30 515 540,clip]{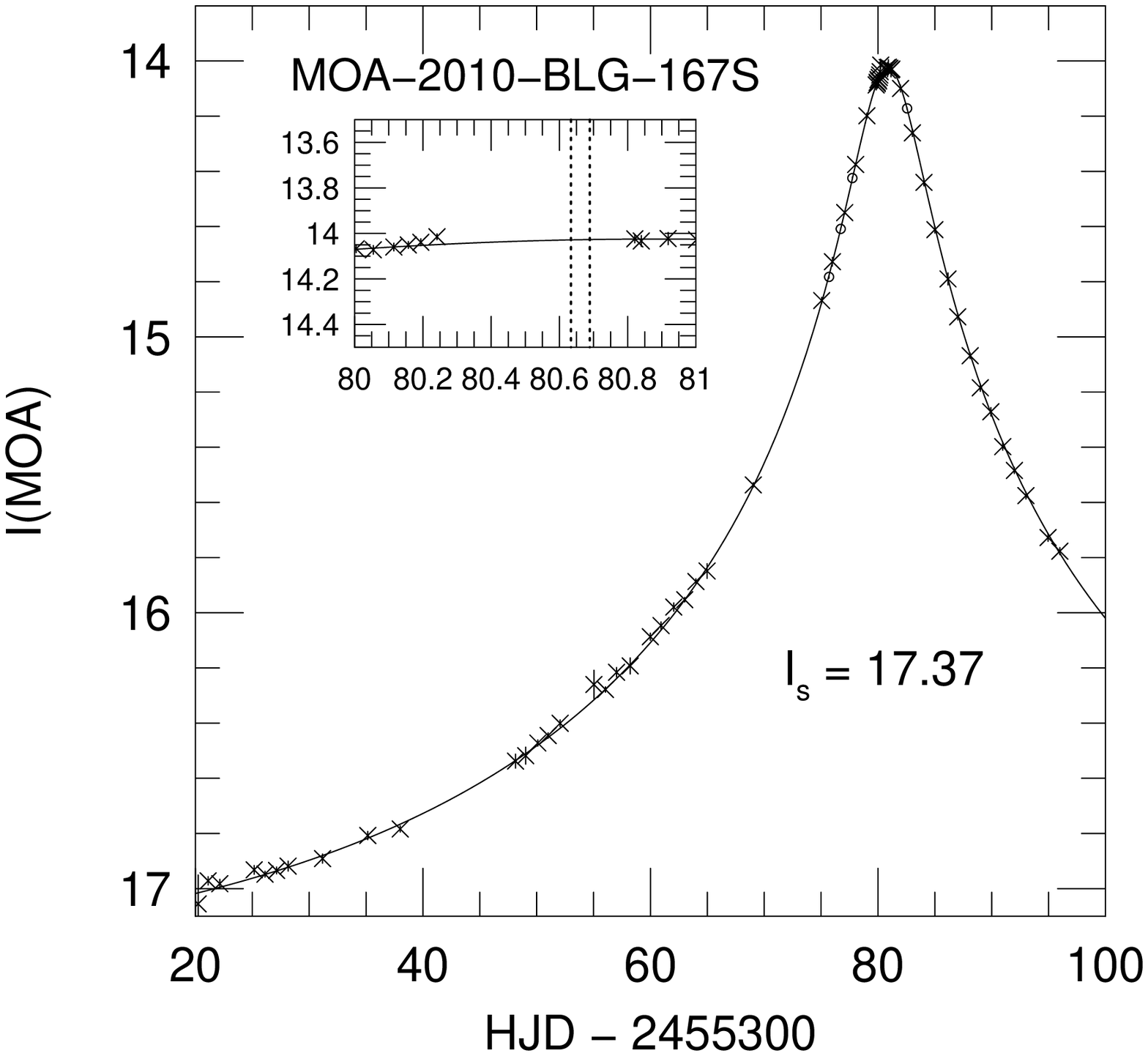}
\includegraphics[bb=52 30 515 540,clip]{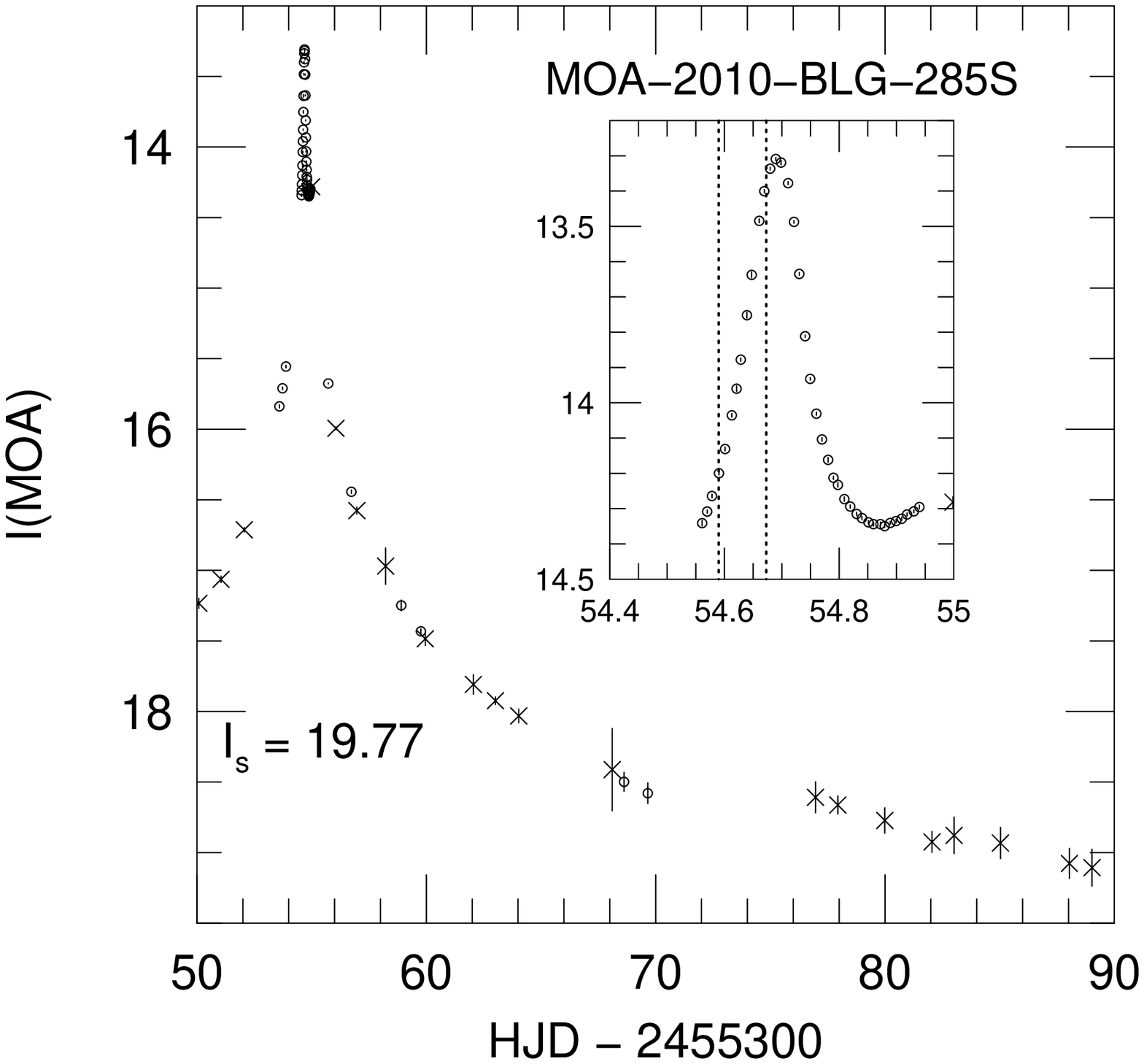}
\includegraphics[bb=52 30 525 540,clip]{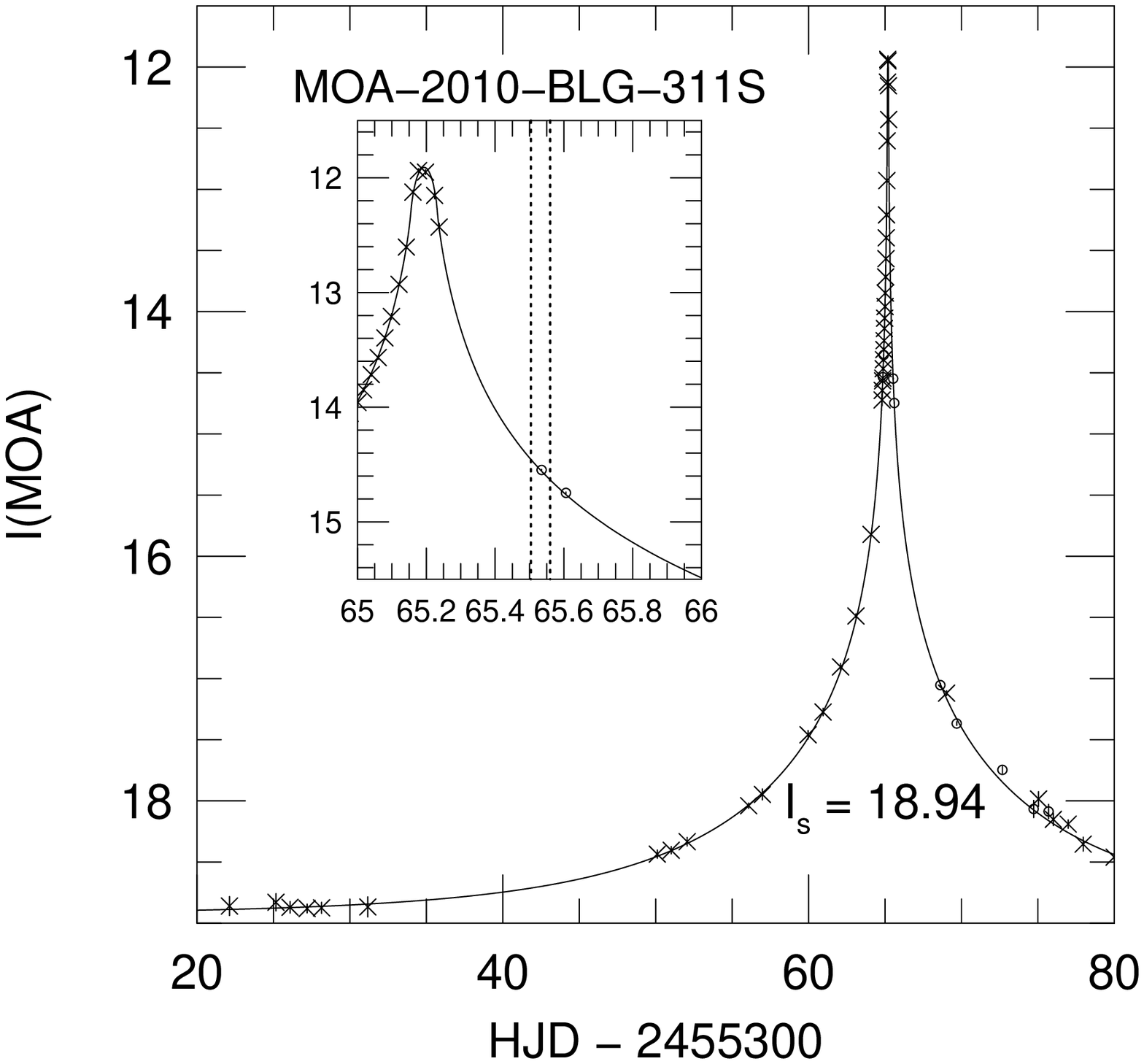}
}
\resizebox{\hsize}{!}{
\includegraphics[bb=52 30 515 540,clip]{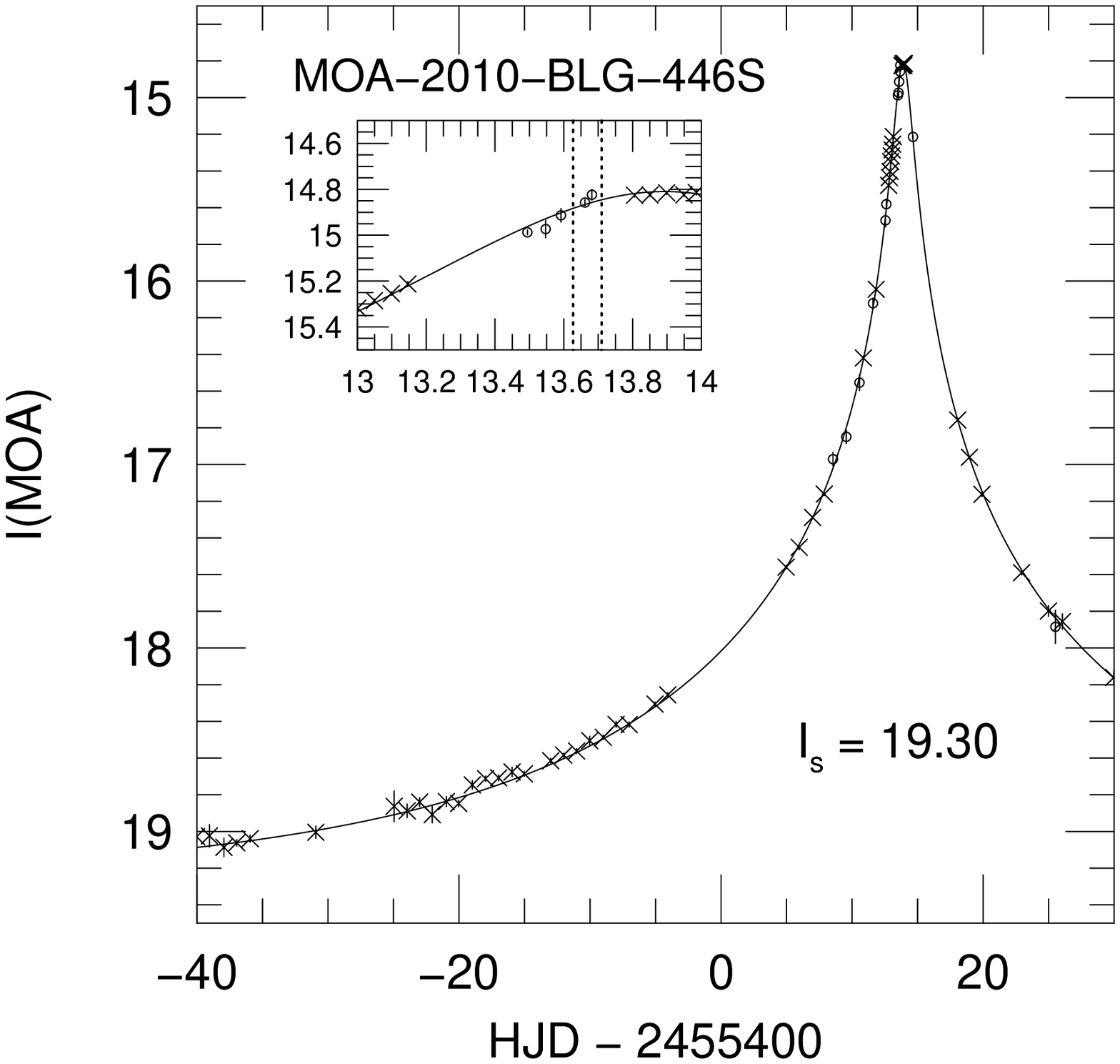}
\includegraphics[bb=52 30 515 540,clip]{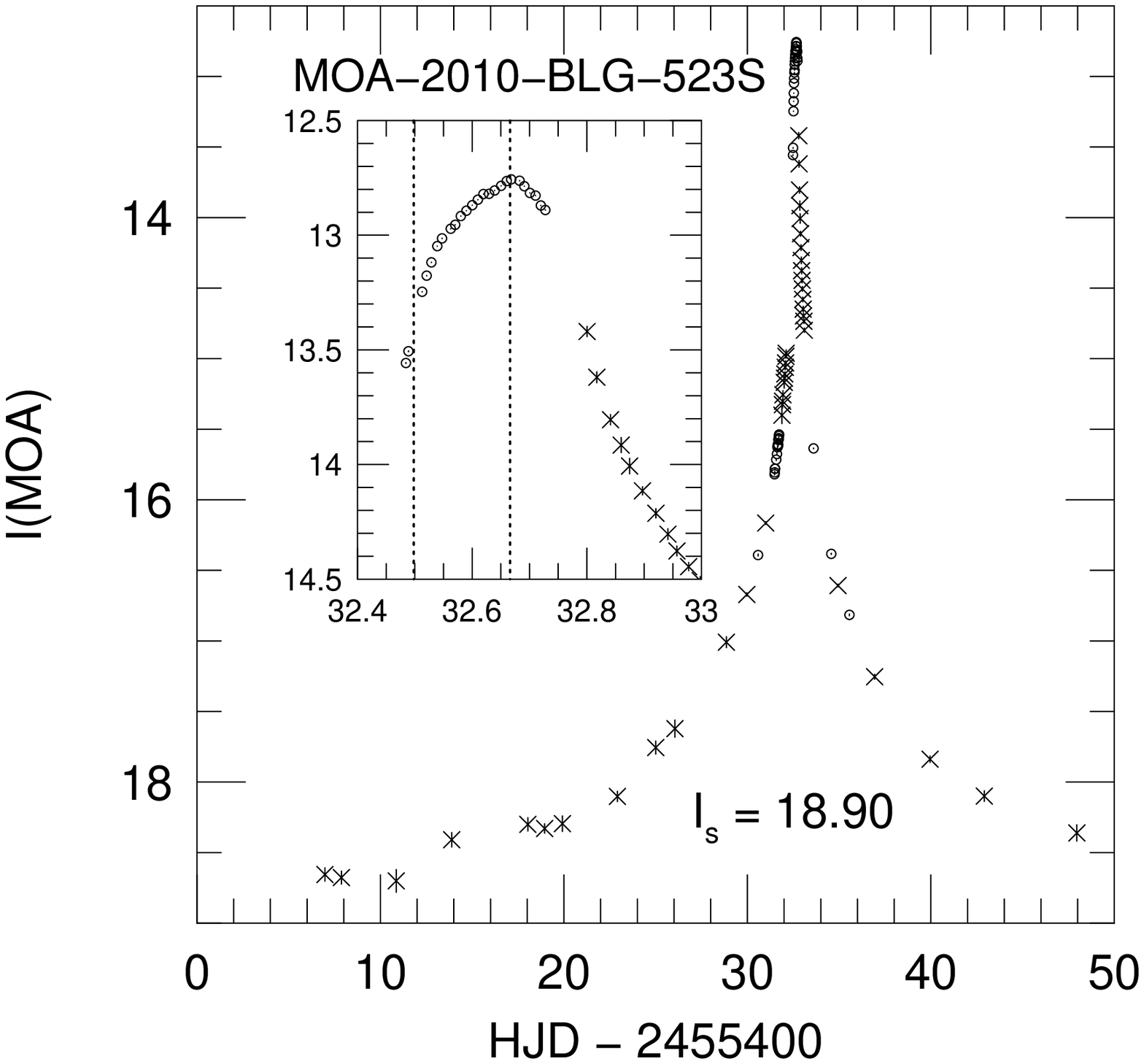}
\includegraphics[bb=52 30 515 540,clip]{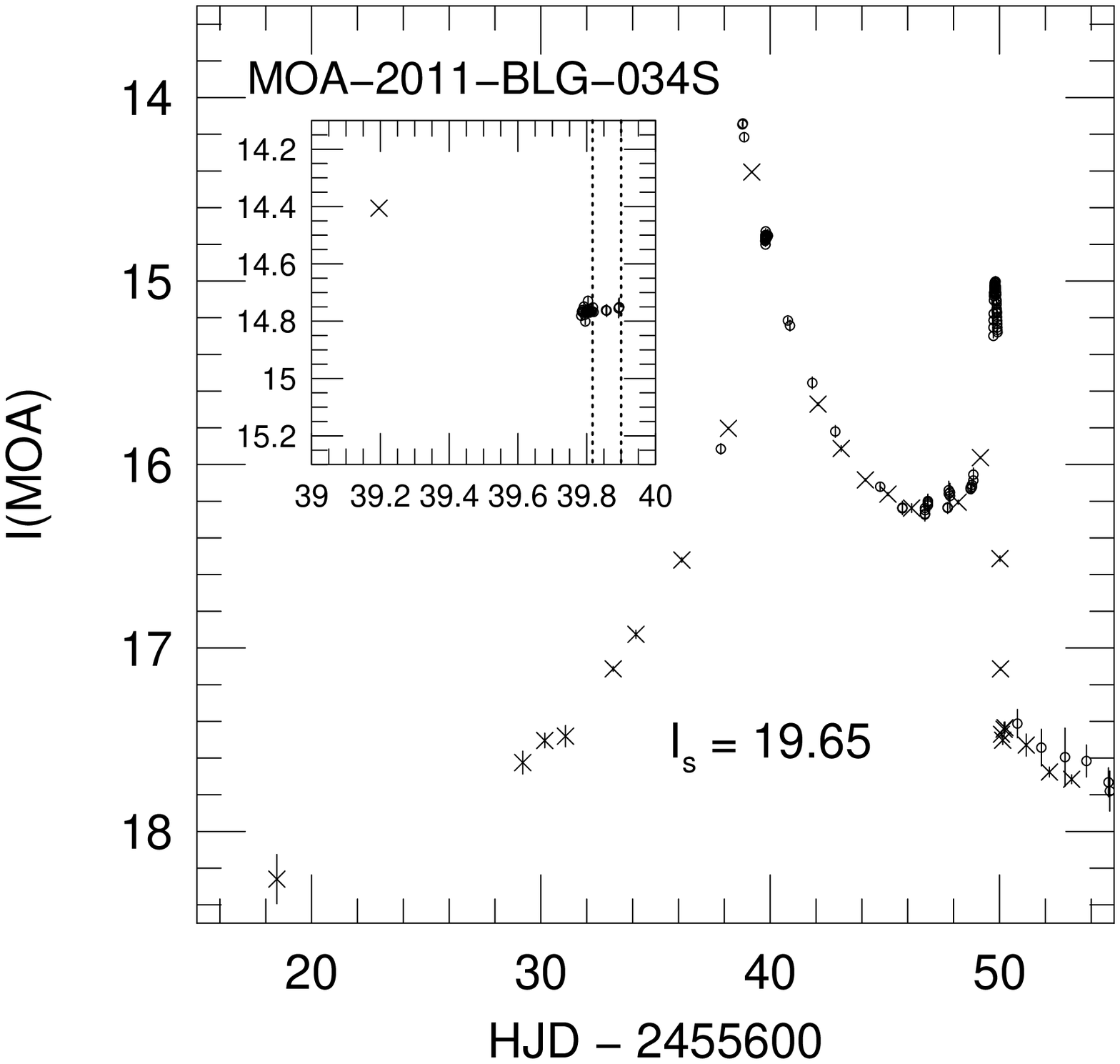}
\includegraphics[bb=52 30 515 540,clip]{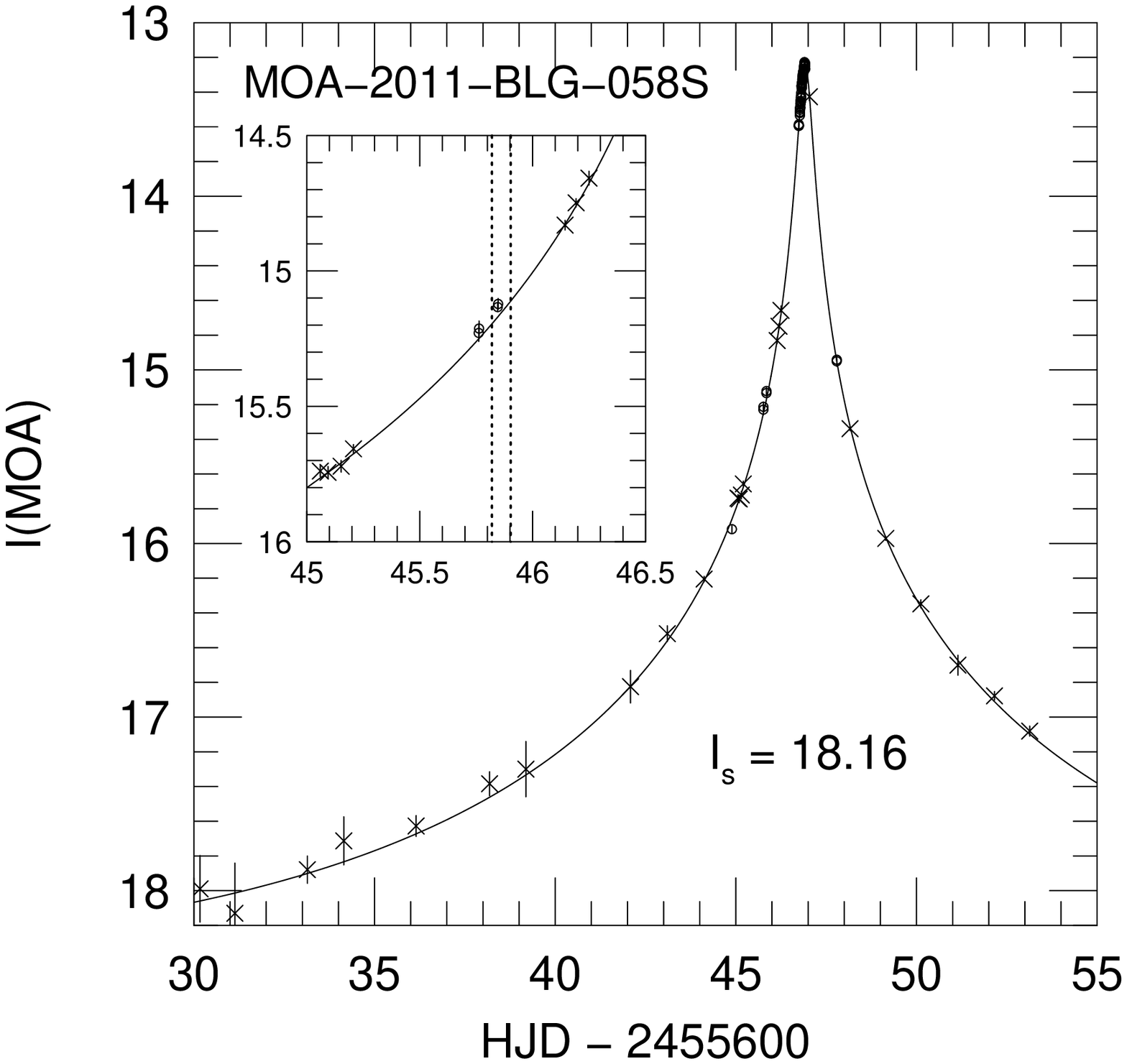}
}
\caption{Light curves for the twelve new microlensing events.
Each plot has a zoom window, showing the time intervals when the source stars
were observed with high-resolution spectrographs. In each plot the
un-lensed magnitude of the source star is also given ($I_{\rm S}$).
\label{fig:events}
}
\end{figure*}

\begin{table*}
\centering
\caption{
Summary of the twelve microlensing events in the Bulge presented in this study$^\dagger$. 
\label{tab:events}
}
\setlength{\tabcolsep}{1.4mm}
\begin{tabular}{rccrccrrrrrll}
\hline\hline
\noalign{\smallskip}
\multicolumn{1}{c}{Object}     &
RAJ2000                        &
DEJ2000                        &
 \multicolumn{1}{c}{$l$}       &
 \multicolumn{1}{c}{$b$}                           &
 \multicolumn{1}{c}{$T_{E}$}   &
 \multicolumn{1}{c}{$T_{max}$}                     &  
 \multicolumn{1}{c}{$A_{max}$}                     &
 \multicolumn{1}{c}{$T_{obs}$} &  
 Exp.                          &
 $S/N$                         &
 Spec.                         &
 \multicolumn{1}{c}{$R$}       \\
                               &
[hh:mm:ss]                     &
[dd:mm:ss]                     &
 [deg]                         &    
 [deg]                         &  
 [days]                        &
 \multicolumn{1}{c}{[HJD]}                         &
                               &   
 \multicolumn{1}{c}{[MJD]}     &  
 \multicolumn{1}{c}{[s]}       &
                               &
                               &
                               \\  
\noalign{\smallskip}
\hline
\noalign{\smallskip}
MOA-2009-BLG-174S  & 18:02:07.70 & $-$31:25:26.40 & $-0.34$ & $-4.34$ & 65  & 4963.8147 & 1100  & 4962.314 & 9000 &  95 & MIKE & 42\,000  \\
MOA-2009-BLG-259S  & 17:57:57.50 & $-$29:11:39.00 &   1.15  & $-2.45$ & 62  & 5016.7674 &  204  & 5016.395 & 1800 & 110 & HIRES& 48\,000  \\
MOA-2010-BLG-037S  & 18:05:17.93 & $-$27:56:13.21 &   3.04  & $-3.23$ & 92  & 5289.4614 &   19  & 5284.308 & 7200 &  35 & UVES & 42\,000  \\
MOA-2010-BLG-049S  & 18:05:07.02 & $-$26:46:12.79 &   4.04  & $-2.63$ & 12  & 5263.7554 &  110  & 5263.289 & 7200 &  90 & UVES & 42\,000  \\
MOA-2010-BLG-078S  & 17:52:01.65 & $-$30:24:25.63 & $-0.64$ & $-1.95$ & 30  & 5285.8370 &   22  & 5285.303 & 7200 &  40 & UVES & 42\,000  \\
MOA-2010-BLG-167S  & 18:11:27.37 & $-$29:41:02.54 &   2.16  & $-5.26$ & 65  & 5380.8834 &   22  & 5380.129 & 4740 & 100 & UVES & 42\,000  \\
MOA-2010-BLG-285S  & 17:56:48.16 & $-$30:00:39.47 &   0.32  & $-2.64$ & 42  & 4354.7421 &  390  & 5354.087 & 7200 & 150 & UVES & 42\,000  \\
MOA-2010-BLG-311S  & 18:08:49.98 & $-$25:57:04.27 &   5.17  & $-2.96$ & 19  & 5365.1937 &  540  & 5365.001 & 4800 &  55 & UVES & 42\,000  \\
MOA-2010-BLG-446S  & 18:07:04.25 & $-$28:03:57.77 &   3.12  & $-3.64$ & 20  & 5413.8937 &   61  & 5413.125 & 7200 &  70 & UVES & 42\,000  \\
MOA-2010-BLG-523S  & 17:57:08.87 & $-$29:44:58.40 &   0.58  & $-2.58$ & 20  & 5432.6283 &  290  & 5431.997 & 7200 & 150 & UVES & 42\,000  \\
MOA-2011-BLG-034S  & 18:10:12.23 & $-$27:32:02.39 &   3.92  & $-3.99$ & --  & 5638.62\phantom{00} &   --  & 5639.317 & 7200 &  45 & UVES & 42\,000  \\
MOA-2011-BLG-058S  & 18:09:00.36 & $-$29:14:12.66 &   2.30  & $-4.57$ & 17  & 5646.92\phantom{00} &   90  & 5645.320 & 7200 &  45 & UVES & 42\,000  \\
\noalign{\smallskip}
\hline
\end{tabular}
\flushleft
{\tiny      
{\bf Notes.} $^\dagger$ Given for each microlensing event is: RA and DE coordinates (J2000) read from the
fits headers of the spectra (the direction where the telescope pointed during observation); 
galactic coordinates ($l$ and $b$); 
duration of the event in days ($T_{E}$); time when maximum magnification occured ($T_{max}$); 
maximum magnification ($A_{max}$); time when the event was observed
with high-resolution spectrograph ($T_{obs}$);
the exposure time (Exp.),
the measured signal-to-noise ration per pixel at $\sim$6400\,{\AA} ($S/N$); the
spectrograph that was used (Spec); the spectral resolution ($R$).
}
\end{table*}

\begin{table*}
\centering
\caption{
Stellar parameters and radial velocities for the 12 new microlensed dwarf stars.$^{\dagger}$
\label{tab:parameters}
}
\setlength{\tabcolsep}{4mm}
\begin{tabular}{rlllccr}
\hline\hline
\noalign{\smallskip}
  \multicolumn{1}{c}{Object}                     &
  \multicolumn{1}{c}{$T_{\rm eff}$}              & 
  \multicolumn{1}{c}{$\log g$}                   &  
  \multicolumn{1}{c}{$\xi_{\rm t}$}              &  
  \multicolumn{1}{c}{[Fe/H]}                     &
  $N_{\ion{Fe}{i},\,\ion{Fe}{ii}}$      &
  \multicolumn{1}{c}{$v_{\rm r}$}                \\
                                                 & 
  \multicolumn{1}{c}{$\rm [K]$}                  &  
  \multicolumn{1}{c}{$\rm [cgs]$}                &   
  \multicolumn{1}{c}{$\rm [\kms]$}               &    
                                                 &
                                                 &
  \multicolumn{1}{c}{[km/s]}                    \\
\noalign{\smallskip}
\hline
\noalign{\smallskip}
  MOA-2010-BLG-285S & $6064\pm129$  & $4.20\pm 0.23$ & $1.85\pm0.38$  &  $-1.23\pm 0.09$ &  \phantom{3}53,           12 &   $ +46.0$  \\
  MOA-2010-BLG-078S & $5231\pm135$  & $3.60\pm 0.31$ & $1.30\pm0.29$  &  $-0.99\pm 0.15$ &  \phantom{3}54, \phantom{3}3 &   $ +52.3$  \\
  MOA-2010-BLG-167S & $5444\pm 58$  & $4.00\pm 0.14$ & $1.17\pm0.11$  &  $-0.59\pm 0.05$ &            118,           17 &   $ -79.4$  \\
  MOA-2010-BLG-049S & $5738\pm 58$  & $4.20\pm 0.12$ & $0.95\pm0.11$  &  $-0.38\pm 0.06$ &  \phantom{3}96,           14 &   $-116.7$  \\
  MOA-2010-BLG-446S & $6327\pm123$  & $4.50\pm 0.19$ & $1.57\pm0.17$  &  $-0.37\pm 0.08$ &  \phantom{3}69, \phantom{3}8 &   $ +56.5$  \\
  MOA-2010-BLG-523S & $5250\pm 98$  & $4.00\pm 0.23$ & $2.10\pm0.24$  &  $+0.10\pm 0.07$ &  \phantom{3}54,           11 &   $ +97.3$  \\
  MOA-2009-BLG-174S & $5620\pm 65$  & $4.40\pm 0.12$ & $1.23\pm0.11$  &  $+0.13\pm 0.05$ &            121,           17 &   $ -21.4$  \\
  MOA-2011-BLG-034S & $5467\pm 67$  & $4.10\pm 0.15$ & $0.87\pm0.14$  &  $+0.13\pm 0.06$ &  \phantom{3}84,           11 &   $+127.0$  \\
  MOA-2009-BLG-259S & $4953\pm 93$  & $3.40\pm 0.24$ & $1.02\pm0.11$  &  $+0.37\pm 0.05$ &  \phantom{3}72,           11 &   $ +81.5$  \\
  MOA-2011-BLG-058S & $5260\pm 75$  & $4.00\pm 0.16$ & $0.67\pm0.14$  &  $+0.39\pm 0.05$ &  \phantom{3}63,           10 &   $-139.7$  \\
  MOA-2010-BLG-311S & $5464\pm108$  & $3.80\pm 0.26$ & $1.29\pm0.19$  &  $+0.49\pm 0.11$ &  \phantom{3}58,           11 &   $ +44.4$  \\
  MOA-2010-BLG-037S & $5745\pm 95$  & $4.00\pm 0.26$ & $1.51\pm0.20$  &  $+0.56\pm 0.10$ &  \phantom{3}56,           13 &   $  -8.4$  \\
\noalign{\smallskip}
\hline
\end{tabular}
\flushleft
{\tiny
{\bf Notes.} 
$^{\dagger}$ 
Given for each star is: effective temperature ($\teff$); surface gravity ($\log g$); microturbulence
parameter ($\xi_{\rm t}$); [Fe/H]; number of \ion{Fe}{i} and \ion{Fe}{ii} lines used in the analysis;
radial velicity ($v_{\rm r}$).
}
\end{table*}

\begin{table*}
\centering
\caption{
Stellar ages and a comparison of colours and effective temperatures as determined
from spectroscopy and microlensing techniques for all 26 microlensed
dwarf stars. The lower part of the table includes the stars
from \cite{bensby2010}.$^\dag$
\label{tab:irfm}
}
\setlength{\tabcolsep}{1.0mm}
\tiny
\begin{tabular}{rccrrrrccccc|cccl}
\hline\hline
\noalign{\smallskip}
  \multicolumn{1}{c}{Object}                     &
  \multicolumn{1}{c}{$T_{\rm eff}$}              & 
  \multicolumn{1}{c}{$\mathcal{M}$}              &
  \multicolumn{1}{c}{$\log L$}              &
  \multicolumn{1}{c}{Age}                        &
  \multicolumn{1}{c}{$-$1$\sigma$}                        &
  \multicolumn{1}{c}{$+$1$\sigma$}                        &
  \multicolumn{1}{c}{$(V$--$I)_{0}$}             &
  \multicolumn{1}{c}{$M_V$}                      &
  \multicolumn{1}{c}{$-$1$\sigma$}                        &
  \multicolumn{1}{c}{$+$1$\sigma$}                        &
  \multicolumn{1}{c|}{$M_I$}                      &
  \multicolumn{1}{c}{$M_I^\mu$}                  &
  \multicolumn{1}{c}{$(V$--$I)_{0}^{\mu}$}       &
  \multicolumn{1}{c}{$\teff^{\mu}$}              &
   Comments                                      \\
                                                 & 
  \multicolumn{1}{c}{$\rm [K]$}                  &
  \multicolumn{1}{c}{$M_{\odot}$}                & 
  \multicolumn{1}{c}{$L_{\odot}$}                & 
  \multicolumn{1}{c}{[Gyr]}                      & 
  \multicolumn{1}{c}{[Gyr]}                      &
  \multicolumn{1}{c}{[Gyr]}                      &
  \multicolumn{1}{c}{[mag]}                      &
  \multicolumn{1}{c}{[mag]}                      &
  \multicolumn{1}{c}{[mag]}                      &
  \multicolumn{1}{c}{[mag]}                      &
  \multicolumn{1}{c|}{[mag]}                     &
  \multicolumn{1}{c}{[mag]}                      &
  \multicolumn{1}{c}{[mag]}                      &
  \multicolumn{1}{c}{[K]}                        &
                                                 \\
\noalign{\smallskip}
\hline
\noalign{\smallskip}
  MOA-2010-BLG-285S & 6064 & 0.81 &    0.04 & 11.6 &  7.4 & 13.3 &  0.63 & 4.80 & 3.78 & 5.06 &  4.18 & 4.84 & 0.58 & 6283  &                          \\
  MOA-2010-BLG-078S & 5231 & 0.84 &    0.75 & 12.7 &  5.0 & 15.5 &  0.86 & 3.04 & 2.23 & 3.45 &  2.18 & 1.86 & 0.85 & 5259  & very high DR                         \\
  MOA-2010-BLG-167S & 5444 & 0.87 &    0.34 & 14.0 &  8.6 & 15.1 &  0.79 & 4.03 & 3.61 & 4.11 &  3.24 & 2.78 & 0.72 & 5678  &                          \\
  MOA-2010-BLG-049S & 5738 & 0.88 &    0.16 & 11.7 &  9.7 & 13.3 &  0.70 & 4.46 & 4.10 & 4.84 &  3.76 & 2.74 & 0.69 & 5785  & some DR                         \\
  MOA-2010-BLG-446S & 6327 & 1.06 &    0.23 &  3.2 &  1.4 &  4.7 &  0.56 & 4.11 & 3.81 & 4.45 &  3.55 &  --  &  --  &  --   & a bit of DR, corrupt CMD              \\
  MOA-2010-BLG-523S & 5250 & 0.99 &    0.26 & 13.1 &  5.5 & 13.2 &  0.86 & 4.29 & 3.60 & 4.67 &  3.43 &  --  & 0.77 & 5510  & a bit of DR                         \\
  MOA-2009-BLG-174S & 5620 & 0.98 & $-$0.05 &  8.7 &  3.8 &  9.7 &  0.73 & 4.99 & 4.62 & 5.12 &  4.26 & 3.57 & 0.71 & 5703  &                          \\
  MOA-2011-BLG-034S & 5467 & 0.96 &    0.17 & 11.6 &  8.0 & 13.2 &  0.77 & 4.46 & 3.85 & 4.86 &  3.69 &      & 0.67 & 5720  &                          \\
  MOA-2009-BLG-259S & 4953 & 1.21 &    0.79 &  3.0 &  1.8 &  7.7 &  1.00 & 3.03 & 2.09 & 3.89 &  2.03 & 2.95 & 0.82 & 5366  &                          \\
  MOA-2011-BLG-058S & 5260 & 0.97 &    0.34 & 12.7 &  6.4 & 13.1 &  0.85 & 4.08 & 3.77 & 4.66 &  3.23 & 3.36 & 0.82 & 5366  &                          \\
  MOA-2010-BLG-311S & 5464 & 1.09 &    0.39 &  4.9 &  3.1 &  9.1 &  0.78 & 3.92 & 2.81 & 4.50 &  3.14 & 3.21 & 0.75 & 5568  & high DR                         \\
  MOA-2010-BLG-037S & 5745 & 1.25 &    0.49 &  4.4 &  2.9 &  5.5 &  0.69 & 3.63 & 2.98 & 4.30 &  2.94 & 3.61 & 0.80 & 5421  &                          \\
\noalign{\smallskip}                 
\hline                               
\noalign{\smallskip}                 
 OGLE-2009-BLG-076S & 5877 & 0.84 & $-$0.04 & 11.6 &  7.0 & 13.1 &  0.67 & 5.00 & 4.26 & 5.17 &  4.33 & 4.24 & 0.70 & 5752  &                          \\ 
  MOA-2009-BLG-493S & 5457 & 0.74 & $-$0.41 & 13.1 &  4.3 & 13.1 &  0.79 & 5.83 & 5.59 & 6.11 &  5.04 & 4.81 & 0.70 & 5752  & some DR                         \\ 
  MOA-2009-BLG-133S & 5597 & 0.81 & $-$0.24 & 12.1 &  5.0 & 15.9 &  0.74 & 5.54 & 4.70 & 5.77 &  4.80 & 4.24 & 0.71 & 5715  & double clump sightline                         \\ 
  MOA-2009-BLG-475S & 5843 & 0.85 & $-$0.08 &  8.7 &  4.5 & 12.3 &  0.68 & 5.10 & 4.12 & 5.33 &  4.42 & 4.30 & 0.62 & 6070  & some DR                         \\ 
MACHO-1999-BLG-022S & 5650 & 0.88 &    0.36 & 13.1 &  7.0 & 13.4 &  0.73 & 3.97 & 3.49 & 4.70 &  3.24 &  --  &  --  & --    &                          \\ 
 OGLE-2008-BLG-209S & 5243 & 0.93 &    0.38 &  8.6 &  5.3 & 12.7 &  0.86 & 3.96 & 3.34 & 4.11 &  3.10 & 2.57 & 0.76 & 5542  &                          \\ 
  MOA-2009-BLG-489S & 5643 & 0.93 & $-$0.09 & 11.6 &  6.1 & 12.8 &  0.73 & 4.97 & 4.33 & 5.22 &  4.24 & 3.44 & 0.88 & 5200  & high DR                         \\ 
  MOA-2009-BLG-456S & 5700 & 1.00 &    0.05 &  8.7 &  4.9 &  9.9 &  0.71 & 4.59 & 4.00 & 4.90 &  4.88 & 2.81 & 0.69 & 5773  &                          \\ 
 OGLE-2007-BLG-514S & 5635 & 1.06 &    0.16 &  6.4 &  4.2 &  9.6 &  0.73 & 4.49 & 3.53 & 4.87 &  3.76 & 4.41 & 0.73 & 5634  & high DR, binary peak              \\ 
  MOA-2008-BLG-311S & 5944 & 1.17 &    0.18 &  2.9 &  1.1 &  3.5 &  0.64 & 4.37 & 4.06 & 4.51 &  3.73 & 3.98 & 0.69 & 5767  & a bit of DR                        \\ 
  MOA-2006-BLG-099S & 5741 & 1.10 &    0.04 &  3.3 &  1.5 &  5.0 &  0.70 & 4.66 & 4.35 & 4.87 &  3.96 & 3.86 & 0.77 & 5508  & very high DR               \\ 
  MOA-2008-BLG-310S & 5704 & 1.09 &    0.07 &  4.9 &  3.0 &  6.3 &  0.71 & 4.69 & 4.04 & 4.76 &  3.98 & 3.51 & 0.72 & 5664  & \cite{janczak2010}       \\ 
 OGLE-2007-BLG-349S & 5229 & 0.95 &    0.15 & 13.1 &  9.6 & 13.9 &  0.87 & 4.58 & 4.34 & 5.30 &  3.71 & 4.21 & 0.81 & 5393  & a bit of DR                   \\
 OGLE-2006-BLG-265S & 5486 & 1.04 &    0.10 &  8.7 &  5.4 &  9.5 &  0.78 & 4.65 & 4.23 & 5.00 &  3.87 & 3.64 & 0.71 & 5696  &                          \\
\noalign{\smallskip}
\hline
\end{tabular}
\flushleft
{\tiny
{\bf Notes.} 
$^{\dagger}$ 
Column 2 gives the spectroscopic temperature; 
col.~3 the stellar mass (inferred from ischrones);
col.~4 the luminosity (inferred from isochrones),
col.~5 the stellar age (inferred from isochrones);
cols.~6 and 7 the 1-sigma lower and upper age limits;
col.~8 the ``spectroscopic"
colours, based on the  colour--[Fe/H]--$\teff$ calibrations
by \cite{casagrande2010};
col.~9 the absolute $V$ magnitude based on the spectroscopic
stellar parameters;
cols.~10 and 11 the 1-sigma lower and upper limits on $M_V$;
col.~12 the absolute $I$ magnitude based on spectroscopic
stellar parameters;
col.~13 the absolute $I$ magnitude based on microlensing techniques;
col.~14 the $(V-I)_0$ colour based on microlensing techniques;
col.~15 the inferred effective
temperatures using the  colour--[Fe/H]--$\teff$ calibrations
by \cite{casagrande2010};
col.~16 Comments: DR stands for differential reddening.
}
\end{table*}

%
\begin{table*}
\centering
\caption{
Elemental abundance ratios, errors in the abundance ratios, and number of lines used,  for the 12 new microlensed dwarf stars$^\dag$.
\label{tab:abundances2}
}
\setlength{\tabcolsep}{1.5mm}
\tiny
\begin{tabular}{rrrrrrrrrrrrrr}
\hline\hline
\noalign{\smallskip}
                    &  [Fe/H] &  [O/Fe]$^\dagger$  &  [Na/Fe] &  [Mg/Fe] &  [Al/Fe] &  [Si/Fe] &  [Ca/Fe] &  [Ti/Fe] &  [Cr/Fe] &  [Ni/Fe] &  [Zn/Fe] &  [Y/Fe] &  [Ba/Fe] \\
\noalign{\smallskip}
\hline
\noalign{\smallskip}
  MOA-2010-BLG-285S  & $  -1.23 $ & $ 0.55 $ & $  -0.06 $ & $   0.42 $ & $    0.30 $ & $    0.30 $ & $  0.35 $ & $  0.38 $ & $ -0.01 $ & $ -0.08 $ & $   0.28 $ & $   0.39 $ & $   0.22 $ \\
  MOA-2010-BLG-078S  & $  -0.99 $ & $ 0.65 $ & $  -0.02 $ & $   0.47 $ & $    0.25 $ & $    0.35 $ & $  0.26 $ & $  0.34 $ & $    -- $ & $  0.00 $ & $     -- $ & $     -- $ & $   0.32 $ \\
  MOA-2010-BLG-167S  & $  -0.59 $ & $ 0.56 $ & $   0.07 $ & $   0.43 $ & $    0.33 $ & $    0.22 $ & $  0.23 $ & $  0.34 $ & $  0.07 $ & $  0.03 $ & $   0.14 $ & $   0.07 $ & $  -0.18 $ \\
  MOA-2010-BLG-049S  & $  -0.38 $ & $ 0.36 $ & $  -0.05 $ & $   0.28 $ & $    0.28 $ & $    0.14 $ & $  0.25 $ & $  0.31 $ & $  0.02 $ & $  0.01 $ & $   0.23 $ & $    0.1 $ & $   0.13 $ \\
  MOA-2010-BLG-446S  & $  -0.37 $ & $ 0.21 $ & $   0.10 $ & $   0.32 $ & $    0.23 $ & $    0.18 $ & $  0.25 $ & $  0.34 $ & $    -- $ & $  0.05 $ & $     -- $ & $     -- $ & $  -0.06 $ \\
  MOA-2010-BLG-523S  & $   0.10 $ & $ 0.25 $ & $   0.19 $ & $   0.14 $ & $    0.33 $ & $   -0.01 $ & $  0.17 $ & $  0.22 $ & $  0.16 $ & $  0.00 $ & $  -0.24 $ & $     -- $ & $  -0.18 $ \\
  MOA-2009-BLG-174S  & $   0.13 $ & $-0.03 $ & $  -0.03 $ & $   0.09 $ & $   -0.04 $ & $    0.03 $ & $  0.02 $ & $  0.08 $ & $  0.04 $ & $ -0.03 $ & $  -0.05 $ & $  -0.03 $ & $  -0.04 $ \\
  MOA-2011-BLG-034S  & $   0.13 $ & $-0.11 $ & $   0.02 $ & $   0.03 $ & $    0.07 $ & $    0.07 $ & $ -0.01 $ & $  0.05 $ & $ -0.02 $ & $  0.03 $ & $   0.03 $ & $     -- $ & $  -0.14 $ \\
  MOA-2009-BLG-259S  & $   0.37 $ & $ 0.03 $ & $   0.42 $ & $   0.36 $ & $    0.35 $ & $    0.10 $ & $  0.17 $ & $  0.14 $ & $  0.11 $ & $  0.18 $ & $   0.41 $ & $   0.38 $ & $  -0.07 $ \\
  MOA-2011-BLG-058S  &     0.39   & $-0.12 $ &     0.20   &     0.11   &      0.05   &      0.08   &    0.00   &    0.00   &    0.10   &    0.13   &     0.28   &       --   & $  -0.12 $ \\
  MOA-2010-BLG-311S  & $   0.49 $ & $-0.24 $ & $   0.26 $ & $   0.24 $ & $    0.18 $ & $    0.08 $ & $  0.13 $ & $  0.07 $ & $  0.03 $ & $  0.14 $ & $   0.14 $ & $     -- $ & $  -0.13 $ \\
  MOA-2010-BLG-037S  & $   0.56 $ & $-0.40 $ & $   0.31 $ & $   0.16 $ & $    0.10 $ & $    0.08 $ & $ -0.04 $ & $  0.14 $ & $  0.06 $ & $  0.13 $ & $  -0.23 $ & $     -- $ & $  -0.10 $ \\
\noalign{\smallskip}
\hline
\noalign{\smallskip}
                      &
$\sigma_{\rm [Fe/H]}$ &
$\sigma_{\rm [O/Fe]}$  &
$\sigma_{\rm [Na/Fe]}$ &
$\sigma_{\rm [Mg/Fe]}$ &
$\sigma_{\rm [Al/Fe]}$ &
$\sigma_{\rm [Si/Fe]}$ &
$\sigma_{\rm [Ca/Fe]}$ &
$\sigma_{\rm [Ti/Fe]}$ &
$\sigma_{\rm [Cr/Fe]}$ &
$\sigma_{\rm [Ni/Fe]}$ &
$\sigma_{\rm [Zn/Fe]}$ &
$\sigma_{\rm [Y/Fe]}$ &
$\sigma_{\rm [Ba/Fe]}$ \\
\noalign{\smallskip}
\hline
\noalign{\smallskip}
\noalign{\smallskip}
  MOA-2010-BLG-285S  & $ 0.095 $ & $   0.26 $ & $    0.05 $ & $    0.07 $ & $    0.07 $ & $    0.07 $ & $    0.06 $ & $   0.12 $ & $     0.19 $ & $    0.11 $ & $    0.10 $ & $   0.24 $ & $   0.21 $ \\
  MOA-2010-BLG-078S  & $ 0.15  $ & $   0.49 $ & $    0.10 $ & $    0.13 $ & $    0.12 $ & $    0.17 $ & $    0.19 $ & $   0.05 $ & $      --  $ & $    0.15 $ & $     --  $ & $    --  $ & $   0.14 $ \\
  MOA-2010-BLG-167S  & $ 0.05  $ & $   0.19 $ & $    0.09 $ & $    0.08 $ & $    0.04 $ & $    0.06 $ & $    0.09 $ & $   0.07 $ & $     0.07 $ & $    0.06 $ & $    0.14 $ & $   0.20 $ & $   0.09 $ \\
  MOA-2010-BLG-049S  & $ 0.06  $ & $   0.18 $ & $    0.15 $ & $    0.06 $ & $    0.03 $ & $    0.05 $ & $    0.10 $ & $   0.09 $ & $     0.09 $ & $    0.06 $ & $    0.10 $ & $   0.21 $ & $   0.08 $ \\
  MOA-2010-BLG-446S  & $ 0.08  $ & $   0.18 $ & $    0.05 $ & $    0.06 $ & $    0.07 $ & $    0.05 $ & $    0.07 $ & $   0.03 $ & $      --  $ & $    0.09 $ & $     --  $ & $    --  $ & $   0.24 $ \\
  MOA-2010-BLG-523S  & $ 0.07  $ & $   0.22 $ & $    0.13 $ & $    0.12 $ & $    0.10 $ & $    0.08 $ & $    0.15 $ & $   0.11 $ & $     0.08 $ & $    0.07 $ & $    0.13 $ & $    --  $ & $   0.11 $ \\
  MOA-2009-BLG-174S  & $ 0.05  $ & $   0.15 $ & $    0.08 $ & $    0.07 $ & $    0.05 $ & $    0.04 $ & $    0.09 $ & $   0.07 $ & $     0.05 $ & $    0.05 $ & $    0.12 $ & $   0.26 $ & $   0.11 $ \\
  MOA-2011-BLG-034S  & $ 0.06  $ & $   0.19 $ & $    0.03 $ & $    0.09 $ & $    0.06 $ & $    0.06 $ & $    0.10 $ & $   0.07 $ & $     0.04 $ & $    0.06 $ & $    0.14 $ & $    --  $ & $   0.10 $ \\
  MOA-2009-BLG-259S  & $ 0.05  $ & $   0.24 $ & $    0.22 $ & $    0.19 $ & $    0.10 $ & $    0.08 $ & $    0.20 $ & $   0.09 $ & $     0.07 $ & $    0.06 $ & $    0.14 $ & $   0.57 $ & $   0.20 $ \\
  MOA-2011-BLG-058S  &   0.04    &     0.18 &        0.09   &      0.09   &      0.10   &      0.06   &      0.10   &     0.10   &       0.08   &      0.05   &      0.09   &      --    &     0.05   \\
  MOA-2010-BLG-311S  & $ 0.11  $ & $   0.27 $ & $    0.12 $ & $    0.19 $ & $    0.14 $ & $    0.09 $ & $    0.16 $ & $   0.09 $ & $     0.08 $ & $    0.09 $ & $    0.17 $ & $    --  $ & $   0.28 $ \\
  MOA-2010-BLG-037S  & $ 0.10  $ & $   0.29 $ & $    0.16 $ & $    0.18 $ & $    0.11 $ & $    0.07 $ & $    0.12 $ & $   0.10 $ & $     0.12 $ & $    0.09 $ & $    0.24 $ & $    --  $ & $   0.28 $ \\
\noalign{\smallskip}
\hline
\noalign{\smallskip}
\noalign{\smallskip}
                    &
$N_{\rm \ion{Fe}{i}}$ &
$N_{\rm O}$    &
$N_{\rm Na}$   &
$N_{\rm Mg}$   &
$N_{\rm Al}$   &
$N_{\rm Si}$   &
$N_{\rm Ca}$   &
$N_{\rm Ti}$   &
$N_{\rm Cr}$   &
$N_{\rm Ni}$   &
$N_{\rm Zn}$   &
$N_{\rm Y}$    &
$N_{\rm Ba}$   \\
\noalign{\smallskip}
\hline
\noalign{\smallskip}
  MOA-2010-BLG-285S  & $     53 $ & $ 3 $ & $      1 $ & $      6 $ & $       3 $ & $      13 $ & $    10 $ & $    15 $ & $     3 $ & $    11 $ & $      2 $ & $      3 $ & $      4 $ \\
  MOA-2010-BLG-078S  & $     54 $ & $ 2 $ & $      1 $ & $      4 $ & $       4 $ & $      16 $ & $    10 $ & $     2 $ & $    -- $ & $     9 $ & $     -- $ & $     -- $ & $      2 $ \\
  MOA-2010-BLG-167S  & $    118 $ & $ 3 $ & $      2 $ & $      6 $ & $       6 $ & $      23 $ & $    15 $ & $    25 $ & $     9 $ & $    33 $ & $      3 $ & $      3 $ & $      4 $ \\
  MOA-2010-BLG-049S  & $     96 $ & $ 3 $ & $      2 $ & $      7 $ & $       6 $ & $      25 $ & $    13 $ & $    13 $ & $     2 $ & $    34 $ & $      3 $ & $      2 $ & $      3 $ \\
  MOA-2010-BLG-446S  & $     69 $ & $ 3 $ & $      2 $ & $      5 $ & $       5 $ & $      19 $ & $    10 $ & $     3 $ & $    -- $ & $    15 $ & $     -- $ & $     -- $ & $      3 $ \\
  MOA-2010-BLG-523S  & $     54 $ & $ 3 $ & $      2 $ & $      4 $ & $       6 $ & $      20 $ & $    11 $ & $     2 $ & $     5 $ & $    25 $ & $      1 $ & $     -- $ & $      3 $ \\
  MOA-2009-BLG-174S  & $    122 $ & $ 3 $ & $      4 $ & $      4 $ & $       7 $ & $      27 $ & $    18 $ & $    16 $ & $    12 $ & $    34 $ & $      1 $ & $      3 $ & $      3 $ \\
  MOA-2011-BLG-034S  &       84   &   3   &        2   &        5   &         6   &        23   &      11   &       6   &       2   &      30   &        1   &     $ --$  &        3   \\
  MOA-2009-BLG-259S  & $     64 $ & $ 3 $ & $      3 $ & $      3 $ & $       3 $ & $      17 $ & $    11 $ & $    13 $ & $    13 $ & $    29 $ & $      1 $ & $      2 $ & $      2 $ \\
  MOA-2011-BLG-058S  &       63   &   2   &        2   &        5   &         6   &        24   &      12   &       6   &       5   &      30   &        1   &       --   &        3   \\
  MOA-2010-BLG-311S  & $     58 $ & $ 3 $ & $      2 $ & $      5 $ & $       6 $ & $      27 $ & $    10 $ & $     6 $ & $     5 $ & $    26 $ & $      1 $ & $     -- $ & $      3 $ \\
  MOA-2010-BLG-037S  & $     56 $ & $ 3 $ & $      2 $ & $      5 $ & $       6 $ & $      25 $ & $    11 $ & $     6 $ & $     3 $ & $    25 $ & $      1 $ & $     -- $ & $      3 $ \\
\noalign{\smallskip}
\hline
\end{tabular}
\flushleft
{\tiny
$^\dagger$ Note that the abundance ratios for oxygen have been corrected for NLTE effects according to the empirical
formula given in \cite{bensby2004}.
}
\end{table*}

\section{Observations and data reduction}

The twelve (11 new and one re-analysis) microlensing events presented 
in this study were all discovered by the MOA microlensing 
alert system\footnote{\tt https://it019909.massey.ac.nz/moa/alert/index.html} 
\citep[e.g.,][]{bond2001}. Most stars were observed on the night when 
the microlensing events peaked, and as close to peak brightness 
($A_{\rm max}$) as possible (the light curves are shown in
Fig.~\ref{fig:events}). Exceptions are MOA-2009-BLG-174S and 
MOA-2011-BLG-058S which were observed one day ahead of peak brightness 
as the full Moon would be very close on the night of the predicted 
$A_{\rm max}$. As can be seen in Fig.~\ref{fig:events}, the brightnesses 
for most events were almost constant, or only slowly varying,  
when the spectroscopic observations were carried out. 

The event that showed most variation during the observations is 
MOA-2010-BLG-285S; from $I\approx 14.3$ at the start of the 2 hour 
exposure to $I\approx 13.3$ at the end, i.e., it more than doubled its 
brightness during the observations. Furthermore, this microlensing
event has a binary lens and the UVES observations were carried out 
during, or very close to, a caustic crossing. In \cite{bensby2010li}, 
where we presented the first results for MOA-2010-BLG-285S and
reported its Li abundance, we showed that even though the 
three UVES exposures not only differed in magnification, but also in 
magnification across the stellar surface, the joint effect is an even 
magnification across the stellar surface. Hence, the final co-added 
spectrum should not be affected. As the model light curve for this event
requires more analysis, which will be published elsewhere, the light
curve in Fig.~\ref{fig:events} currently only contains the observed data 
points and no model fit. Due to a possible planet detection in the lens, 
the model light curve for MOA-2010-BLG-523S also needs more analysis, and therefore 
also only have observed data points in Fig.~\ref{fig:events}. MOA-2011-BLG-034S
is an event with a binary lens and the microlensing model needs more work.
Hence, we have currently no $A_{max}$, $T_E$, or model curve 
(but only the observed data points in Fig.~\ref{fig:events})
for MOA-2011-BLG-2011-034S.

Ten of the twelve events were observed in 2010 and 2011 with the UVES 
spectrograph \citep{dekker2000} on the Very Large Telescope as part of 
our ongoing Target of Opportunity (ToO) program. They were observed with 
a 1.0\,arcsec wide slit and dichroic \#2, giving spectra with a resolution 
of $R \approx42\,000$ and a spectral coverage between 3760-4980\,{\AA}
(blue ccd), 5680-7500\,{\AA} (lower red ccd), and 7660-9460\,{\AA} (upper
red ccd). A featureless spectrum of a rapidly rotating B star 
(either HR\,6141 or HR\,7830) was obtained each time. This spectrum was used 
to divide out telluric lines in the bulge star spectrum. The signal-to-noise 
ratios at 6400\,{\AA} are listed in Table~\ref{tab:events} and range between 
35 and 150. Reductions were carried out with the UVES pipeline 
\citep{ballester2000}, version 4.7.8, using the gasgano interface.

The UVES data were complemented with data for two stars observed in 2009.
MOA-2009-BLG-259S was observed with the HIRES-R spectrograph \citep{vogt1994} 
on the Keck telescopes on Hawaii.  The slit was 
0.86\,arcsec wide, giving a spectral resolution of $R \approx 48\,000$,
a wavelength coverage from 3900 to 8350\,{\AA}, with small gaps between the 
orders beyond 6650\,{\AA}. The signal-to-noise ratio is $S/N \approx 110$ 
per pixel around  6400\,{\AA}. Details of the reduction procedure 
can be found in \cite{cohen2008}. 
Actually, an analysis of MOA-2009-BLG-259S was first presented 
in \cite{bensby2010}. However, that analysis was based on a spectrum 
that was obtained with UVES and when only the UVES blue arm was available due to
maintenance work on the red arm. That spectrum had a very limited 
blue wavelength coverage and as bulge stars are heavily reddened,
it also had poor S/N. As a result only a few \ion{Fe}{i} and 
\ion{Fe}{ii} could be measured and the stellar parameters were poorly
constrained with large uncertainties. Hence, in \cite{bensby2010},
MOA-2009-BLG-259S was excluded when discussing the metallicity distribution
and abundance trends of the microlensed bulge dwarf stars. The first metallicity 
result, based on this HIRES spectrum, was presented in \cite{cohen2010puzzle}.
For consistency, it has here been re-analysed with the same methods
as for all the other events.

MOA-2009-BLG-174S was observed with the MIKE spectrograph \citep{bernstein2003}
on the Magellan Clay telescope at the Las Campanas observatory in Chile. 
A 0.7 arcsec wide slit was used giving a spectrum with $R\approx42\,000$ and 
continuous coverage between 3500 to 9200\,{\AA}.  The signal-to-noise per pixel
at 6400\,{\AA} is around 95. Reductions were made with 
the Carnegie python pipeline\footnote{Available at 
{\tt http://obs.carnegiescience.edu/Code/mike}}.

In total we are now in possession of high quality spectra for 26
microlensed dwarf and subgiant stars in the bulge. Table~\ref{tab:events} 
lists the details for the twelve new events, while the data for the
other 14 events can be found in \cite{bensby2010}.

\begin{figure*}
\resizebox{\hsize}{!}{
\includegraphics[angle=-90,bb=120 30 550 450,clip]{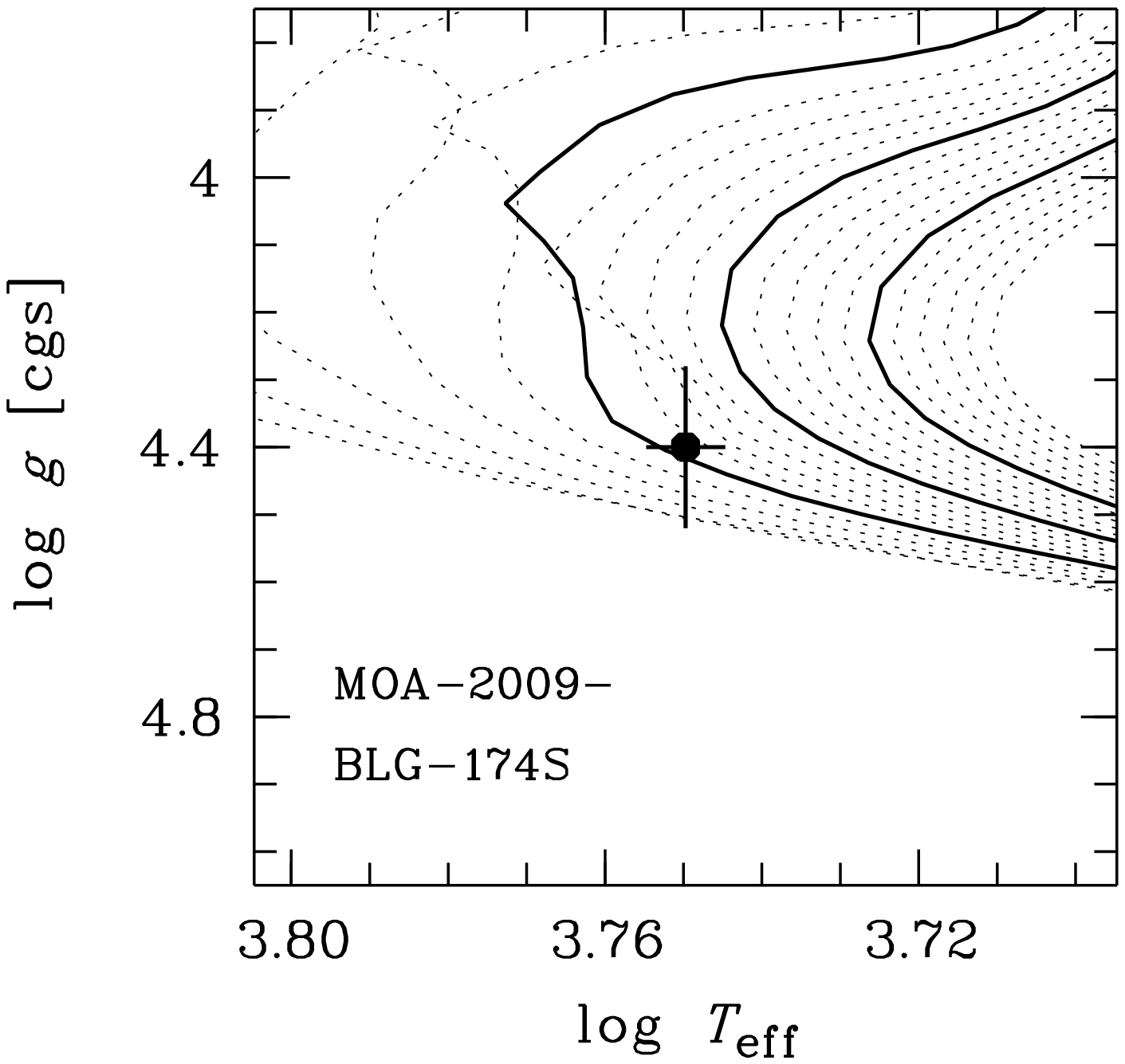}
\includegraphics[angle=-90,bb=120 85 550 450,clip]{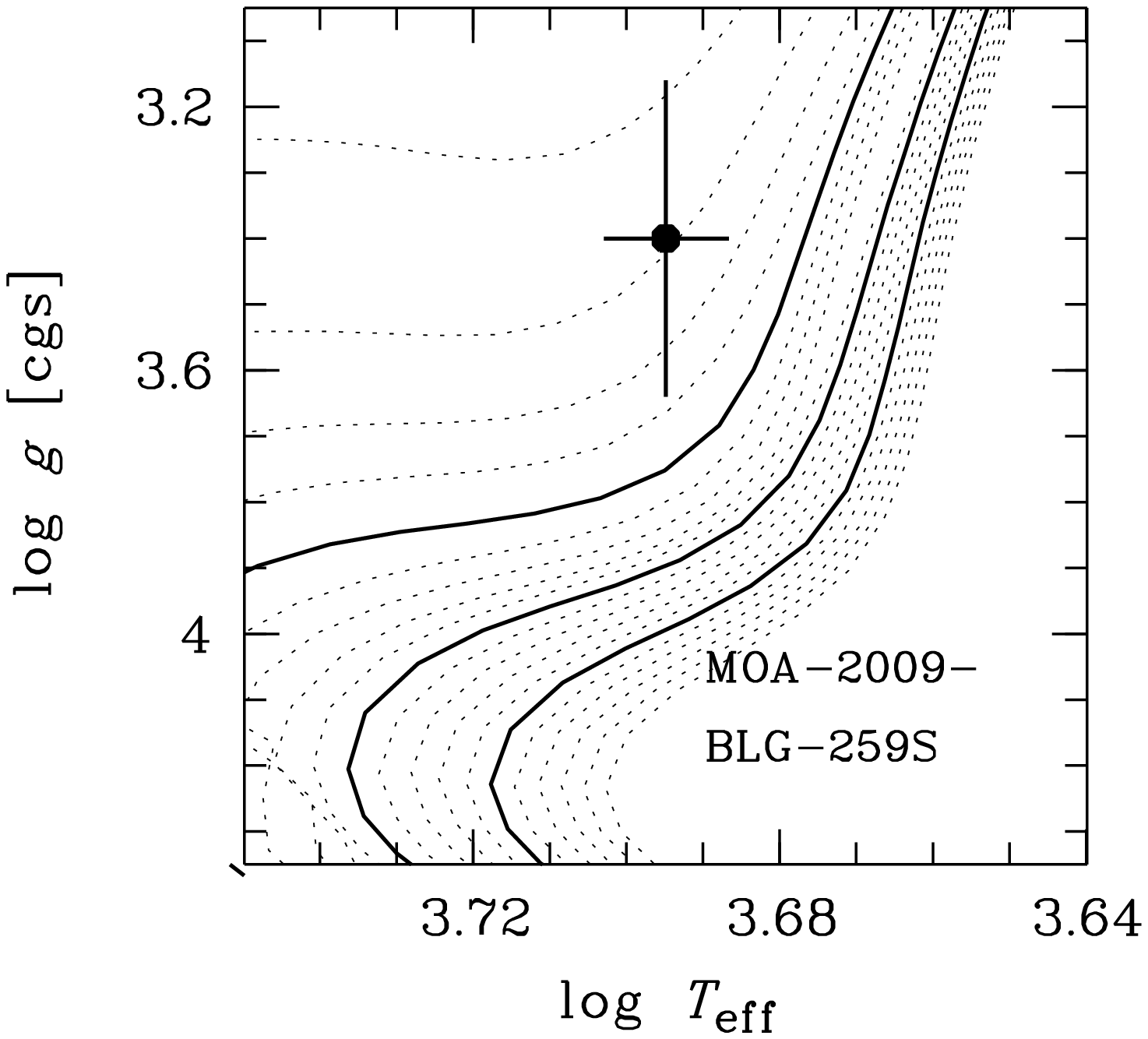}
\includegraphics[angle=-90,bb=120 85 550 450,clip]{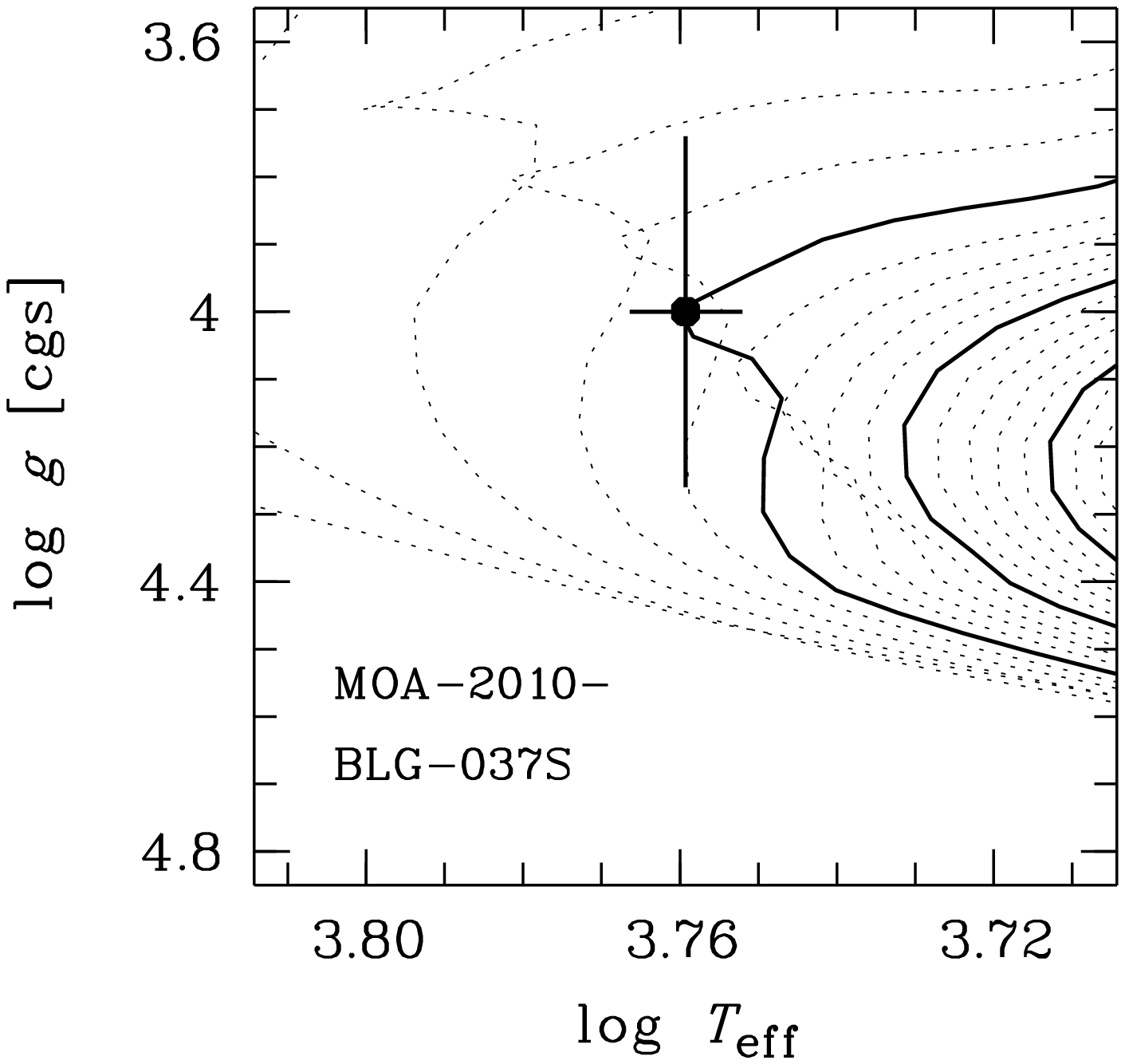}
\includegraphics[angle=-90,bb=120 85 550 485,clip]{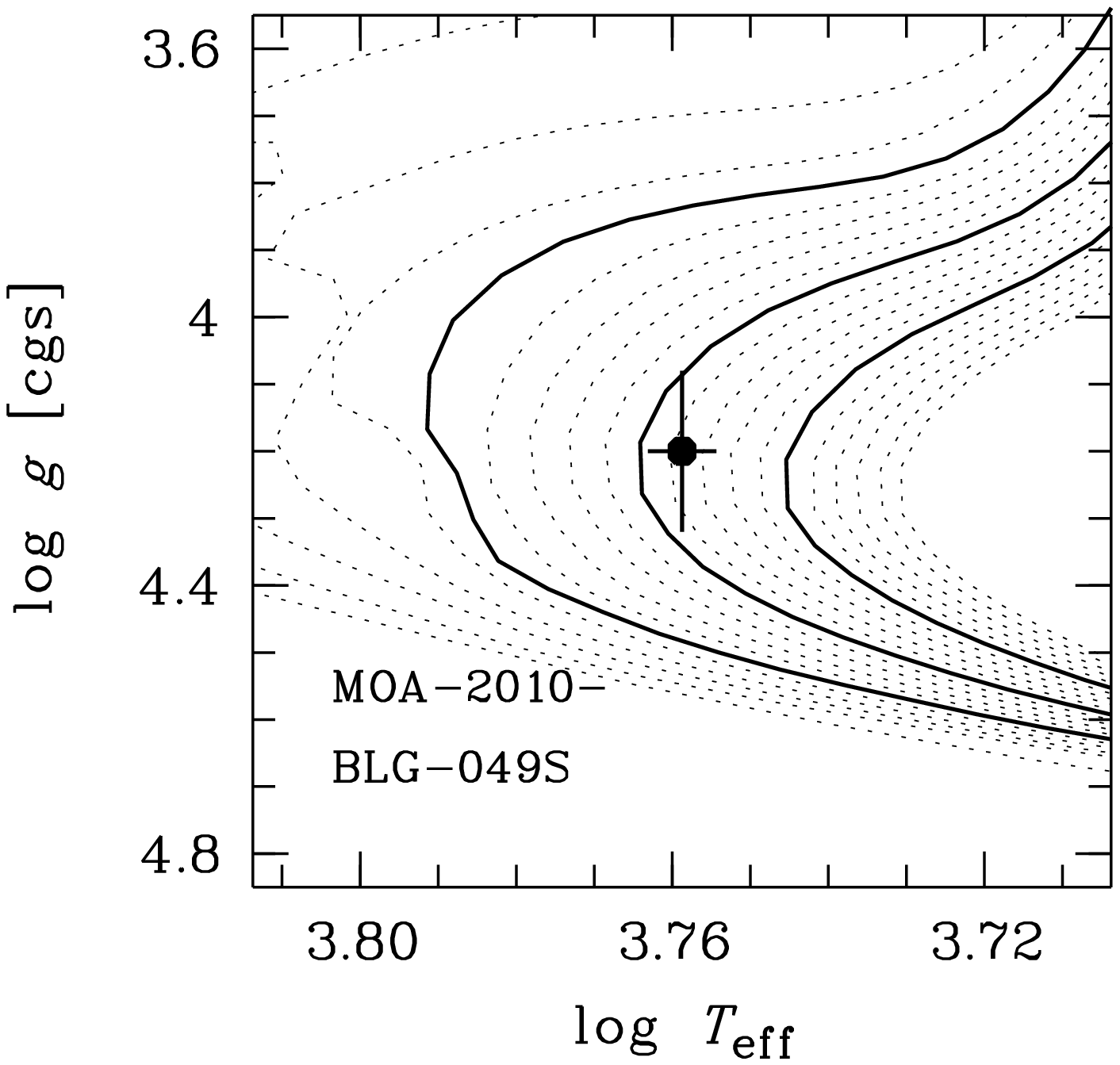}
}
\resizebox{\hsize}{!}{
\includegraphics[angle=-90,bb=140 30 550 450,clip]{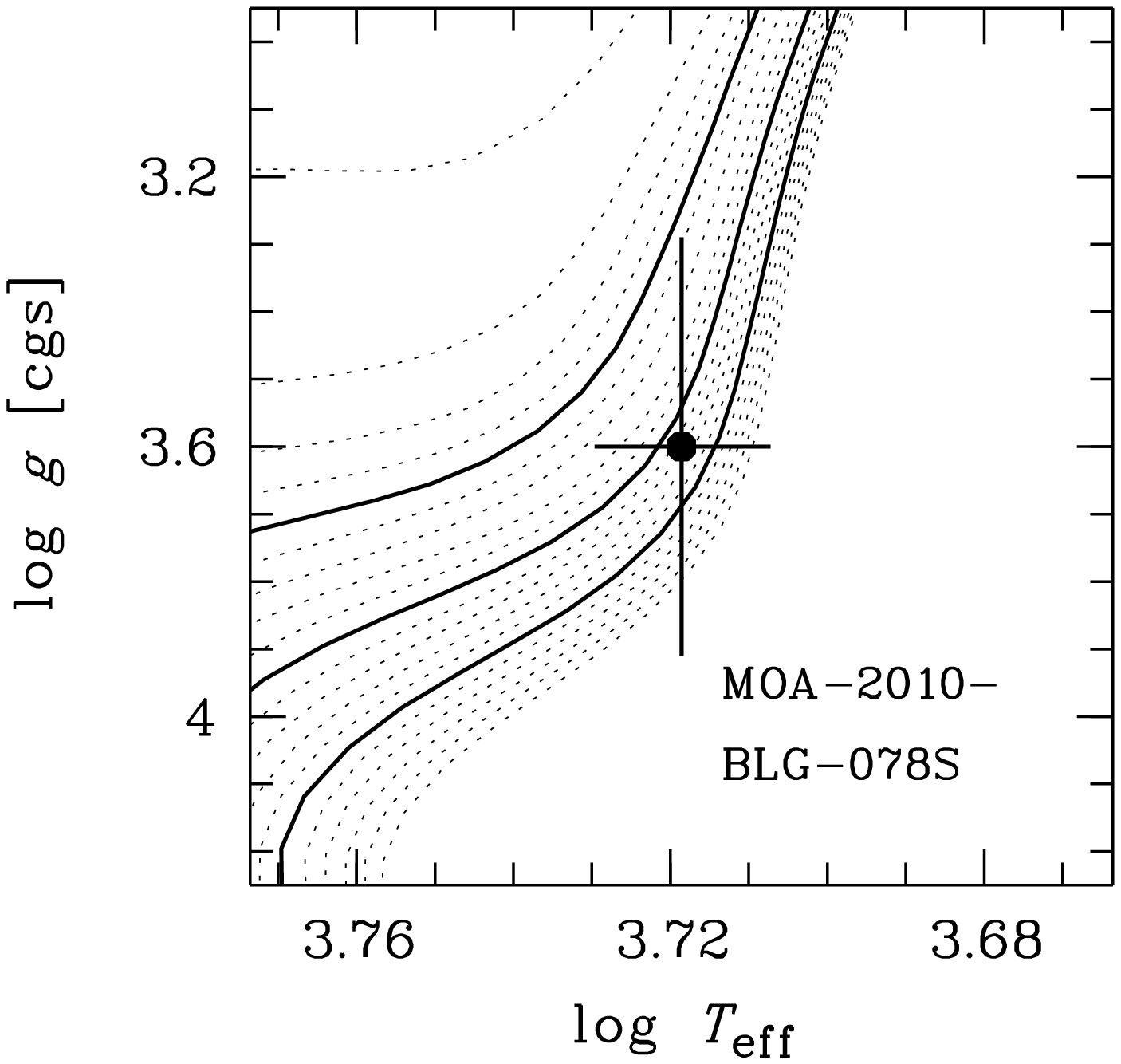}
\includegraphics[angle=-90,bb=140 85 550 450,clip]{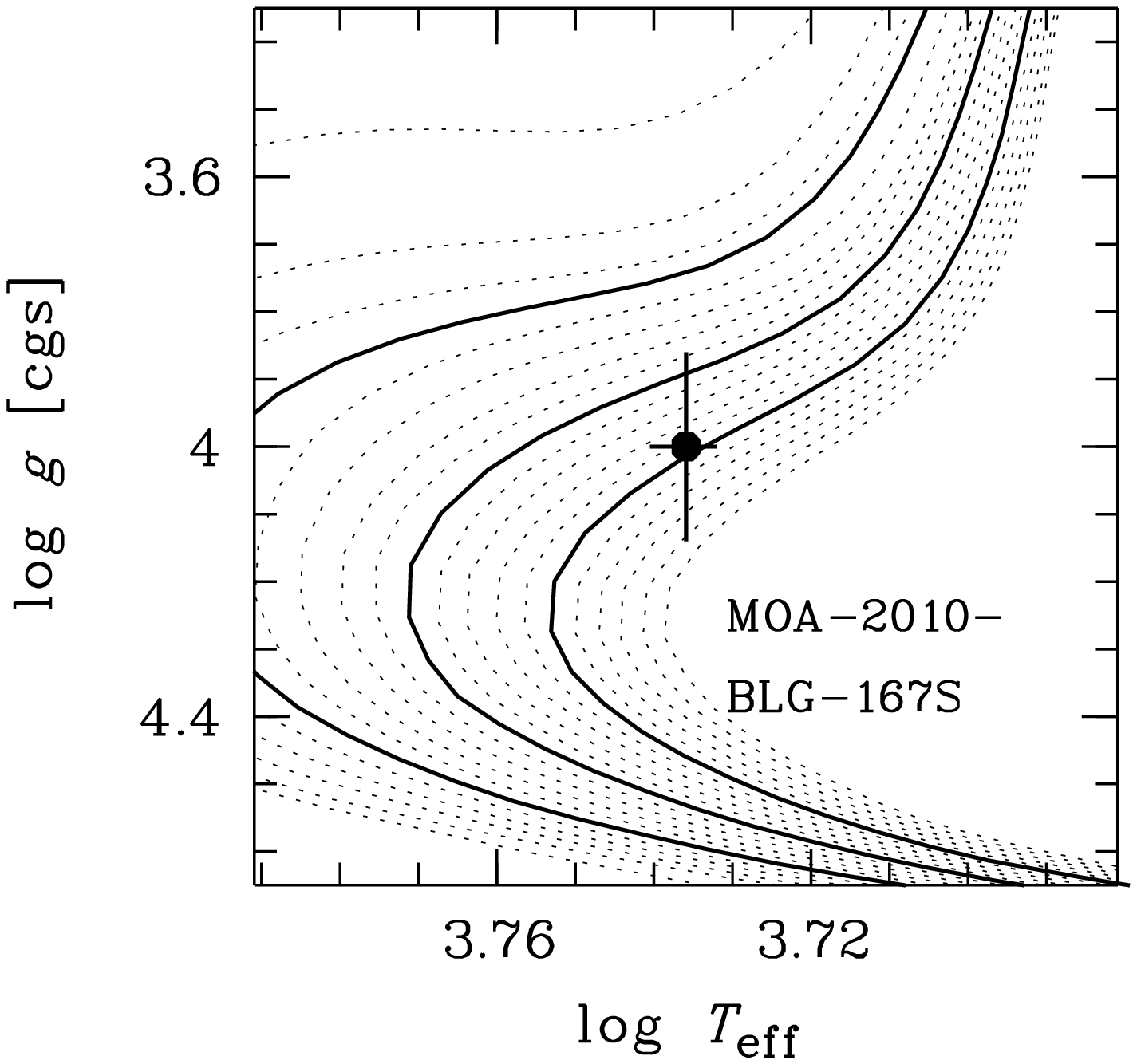}
\includegraphics[angle=-90,bb=140 85 550 450,clip]{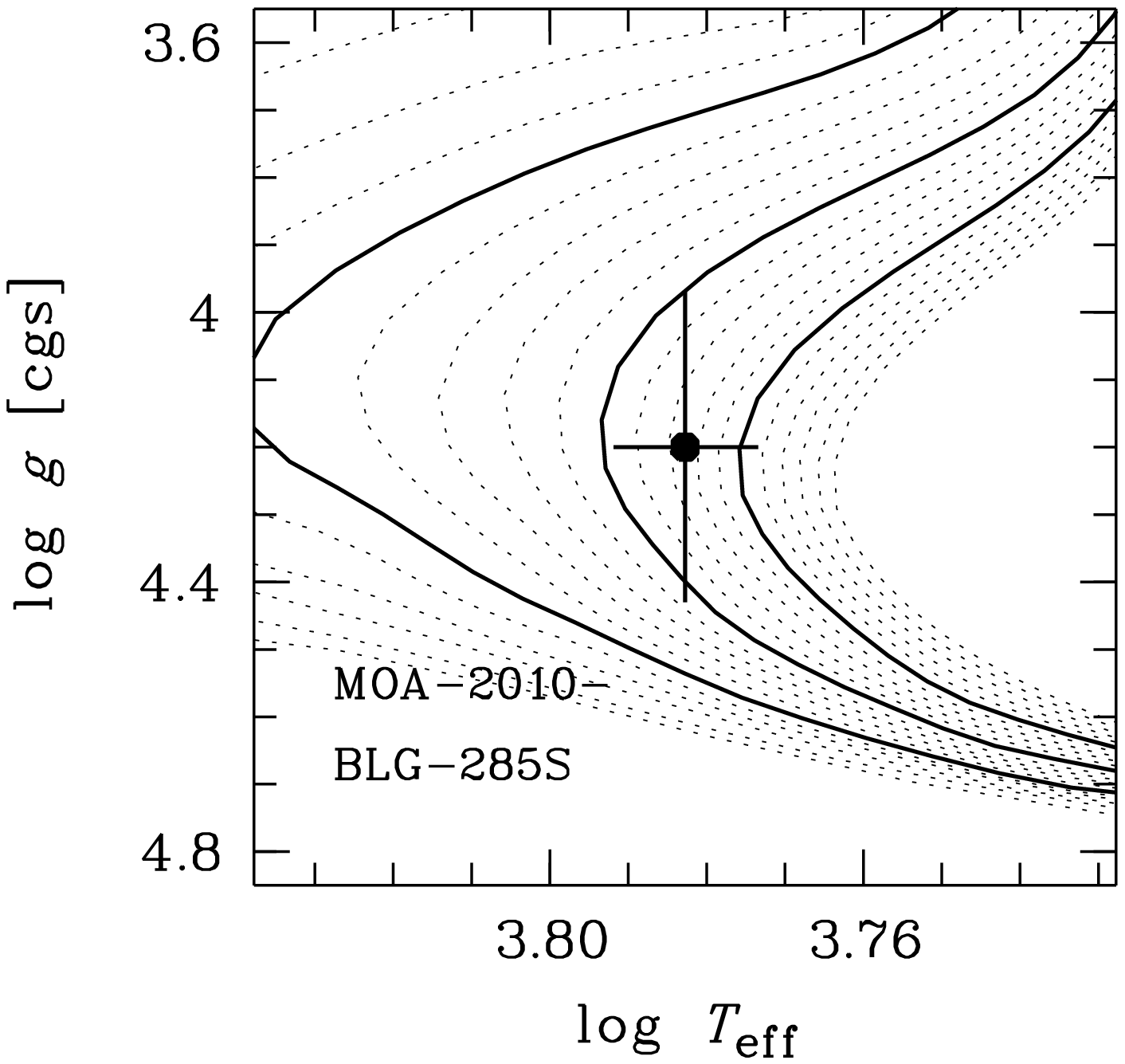}
\includegraphics[angle=-90,bb=140 85 550 485,clip]{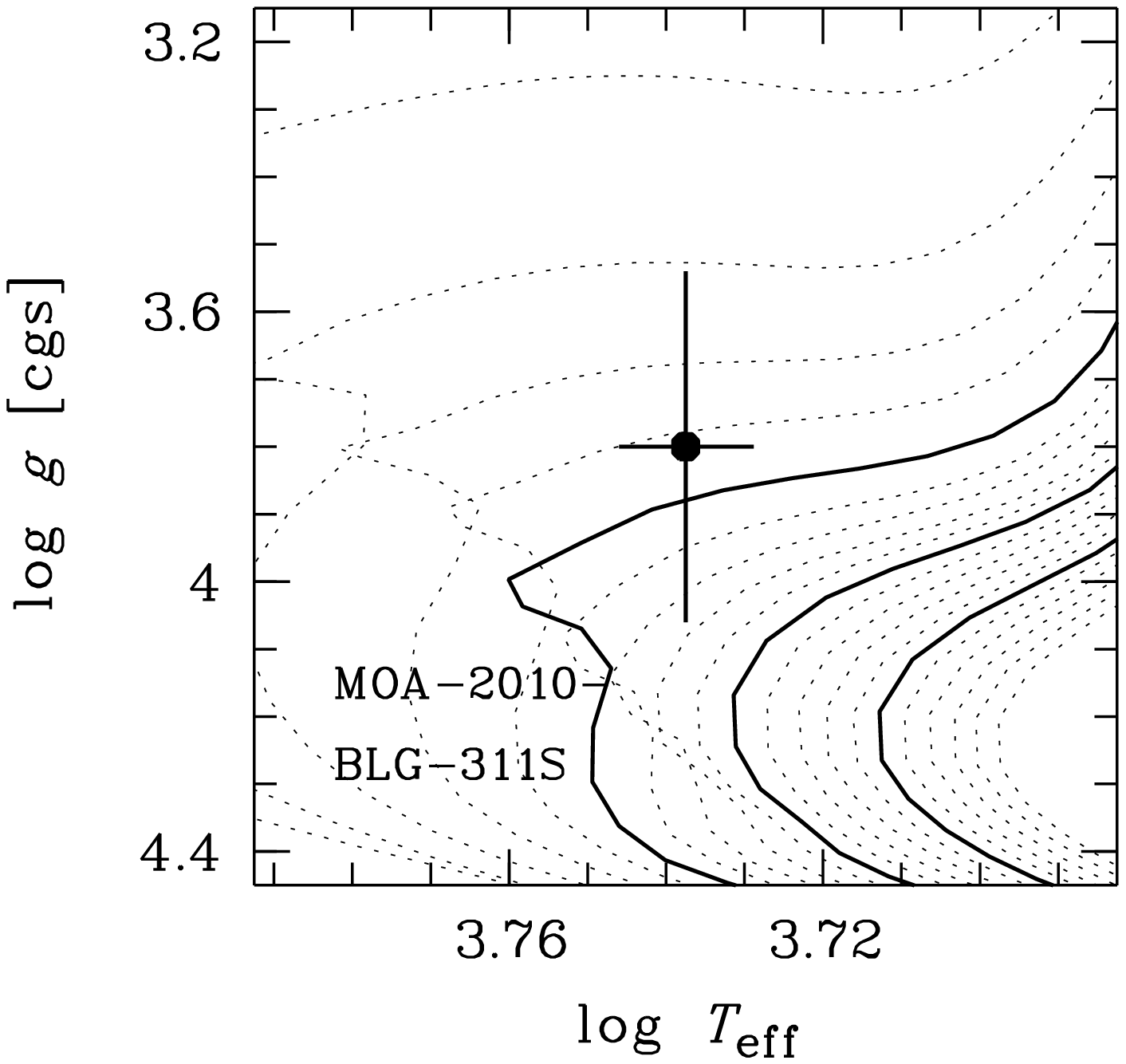}
}
\resizebox{\hsize}{!}{
\includegraphics[angle=-90,bb=140 30 550 450,clip]{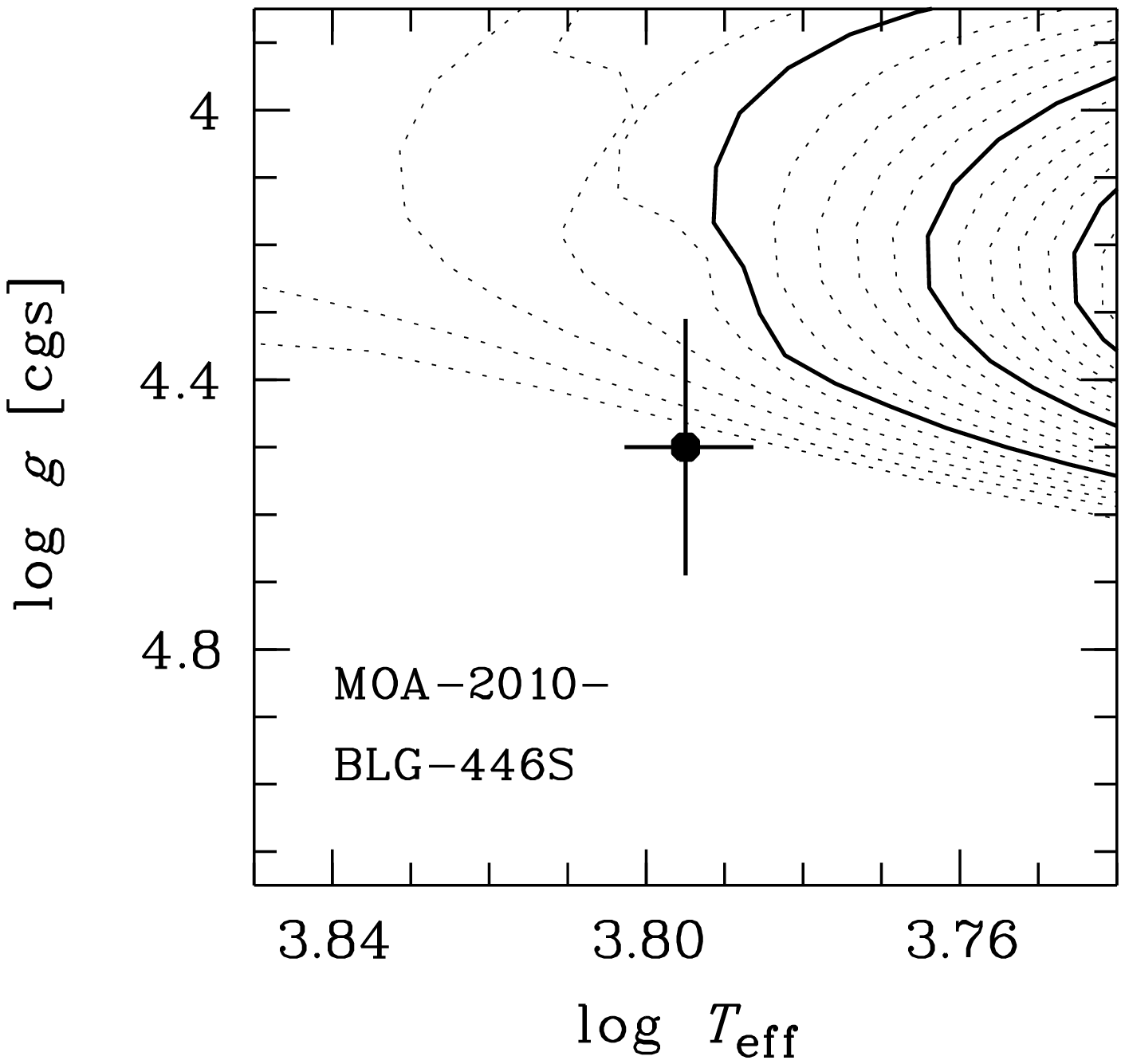}
\includegraphics[angle=-90,bb=140 85 550 450,clip]{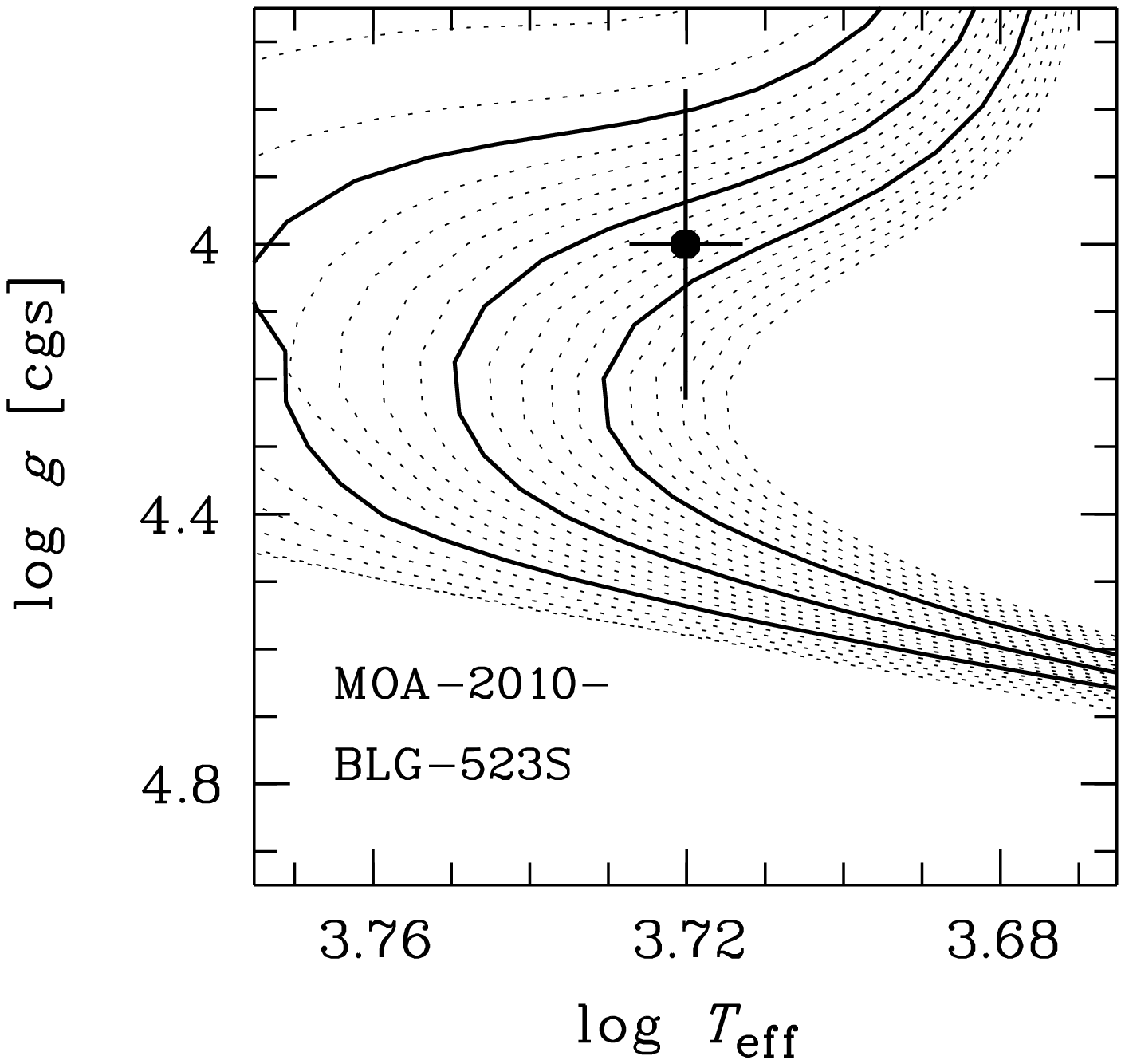}
\includegraphics[angle=-90,bb=140 85 550 450,clip]{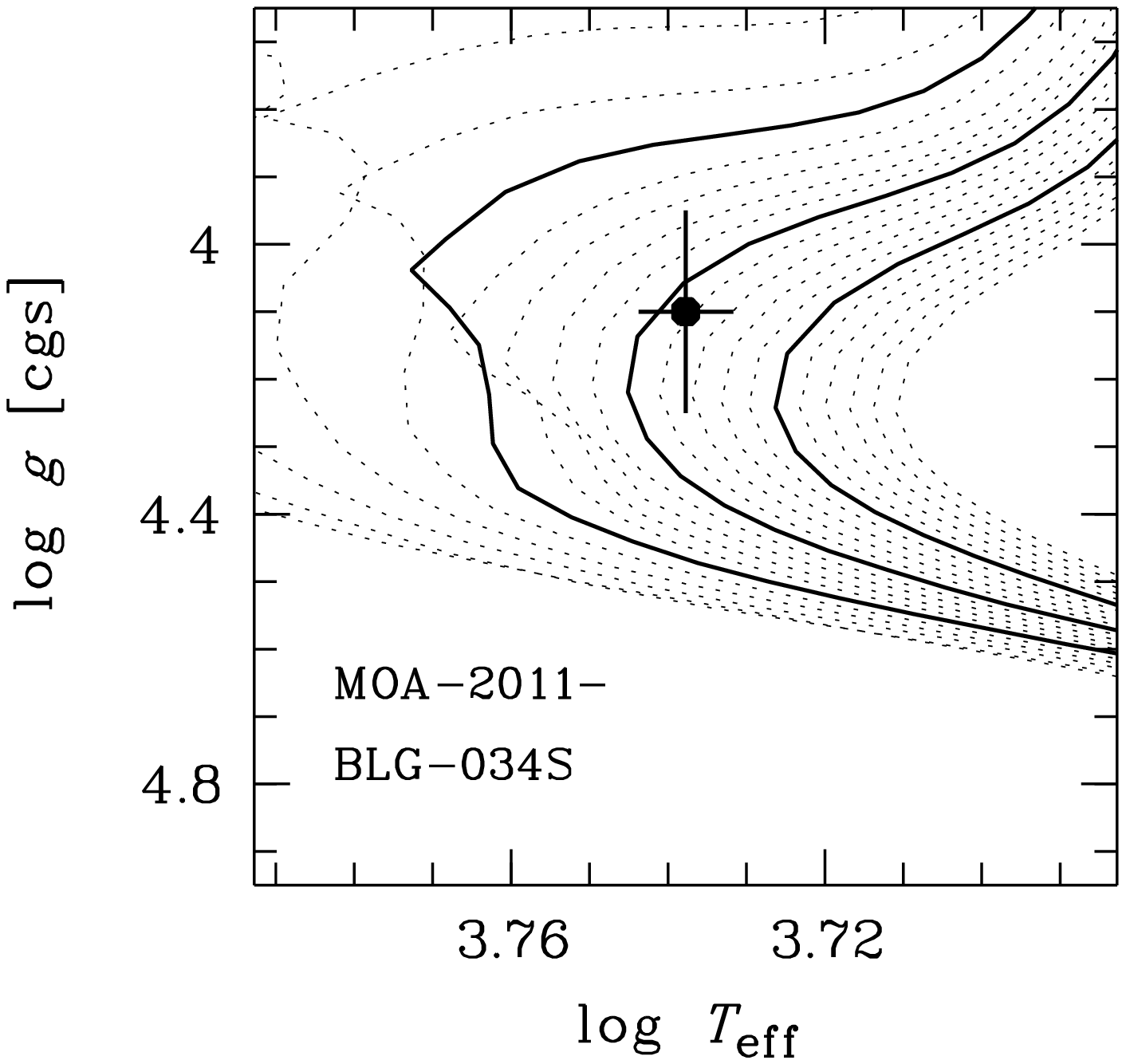}
\includegraphics[angle=-90,bb=140 85 550 485,clip]{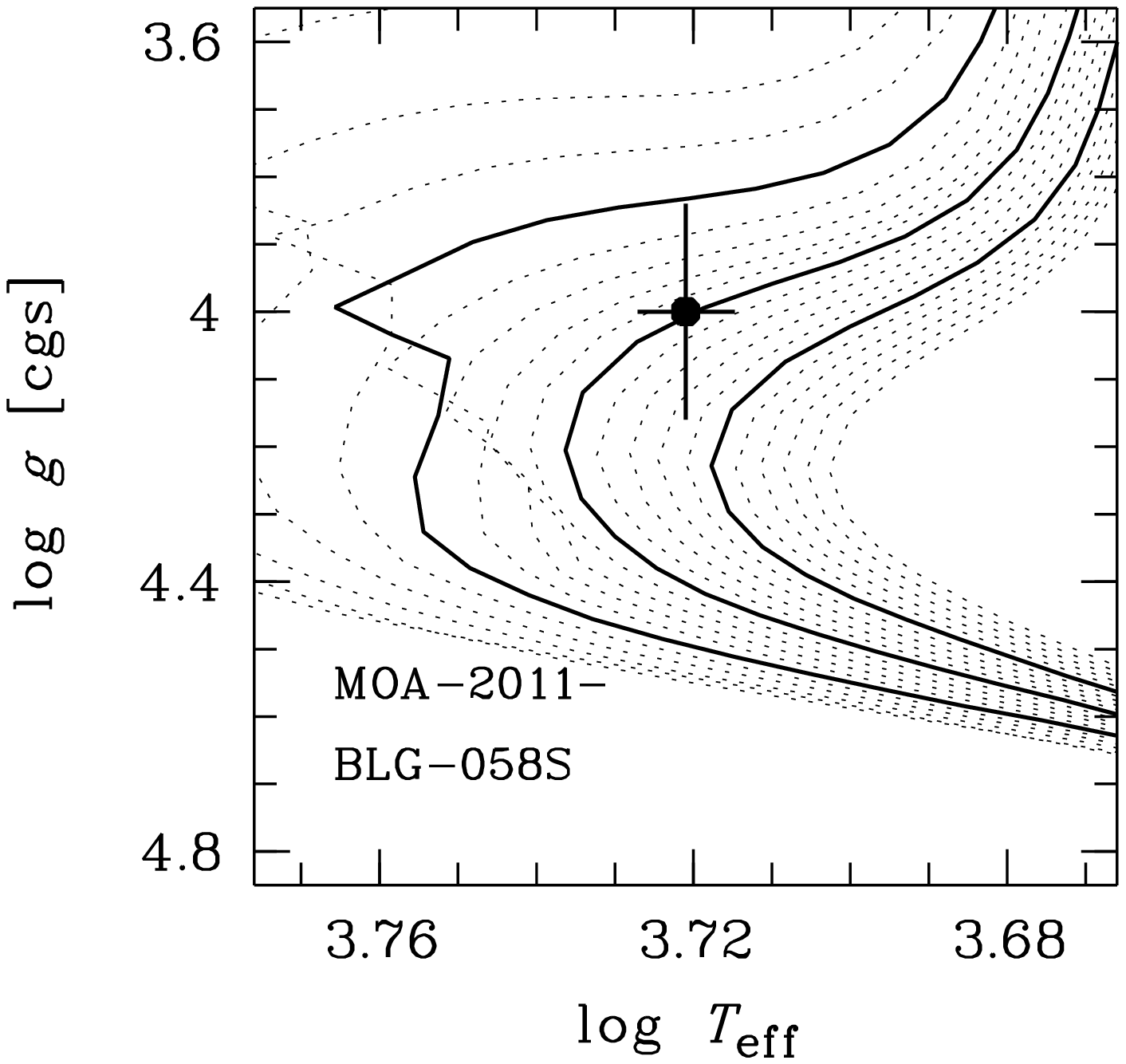}
}
\caption{
The 12 new microlensed stars plotted on the $\alpha$-enhanced
isochrones from \cite{demarque2004}. Each set of isochrones have been
calculated with the same metallicity and $\alpha$-enhancement as
derived for the stars. In each plot the solid lines represent
isochrones with ages of 5, 10, and 15\,Gyr (from left to right).
Dotted lines are isochrones in steps of 1\,Gyr, ranging from 0.1\,Gyr
to 20\,Gyr. Error bars represent the uncertainties
in $\teff$ and $\log g$ as given in Table~\ref{tab:parameters}.
Note that ages reported in Table~\ref{tab:irfm} need not coincide
exactly with the ages that may be read off directly from the figures.
The age determinations are based on probability distribution functions
and is described in Sect.~\ref{sec:analysis}.
\label{fig:ages}
}
\end{figure*}

   \begin{figure*}
   \resizebox{\hsize}{!}{
   \includegraphics[bb=30 320 545 510,clip]{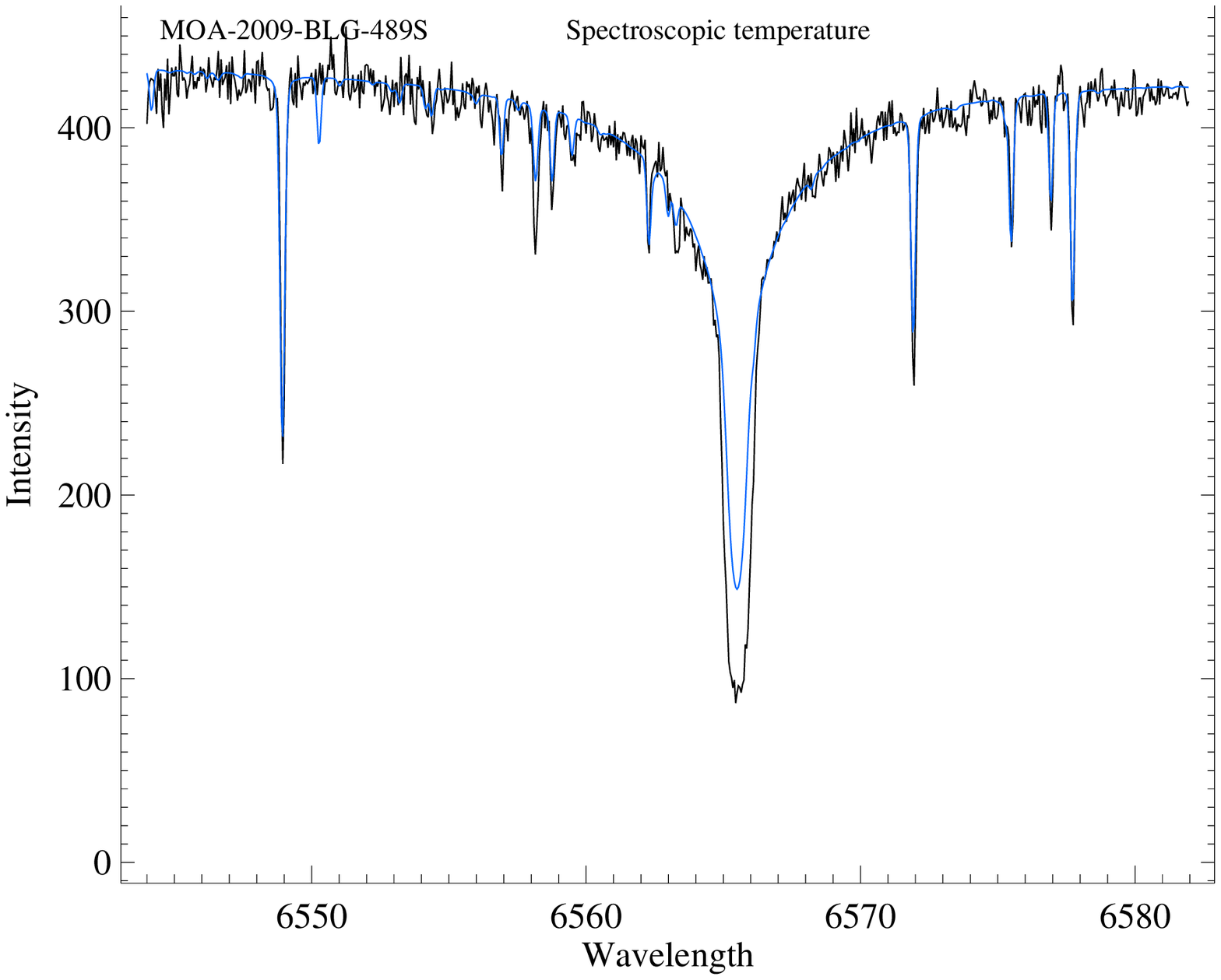}
   \includegraphics[bb=75 320 554 510,clip]{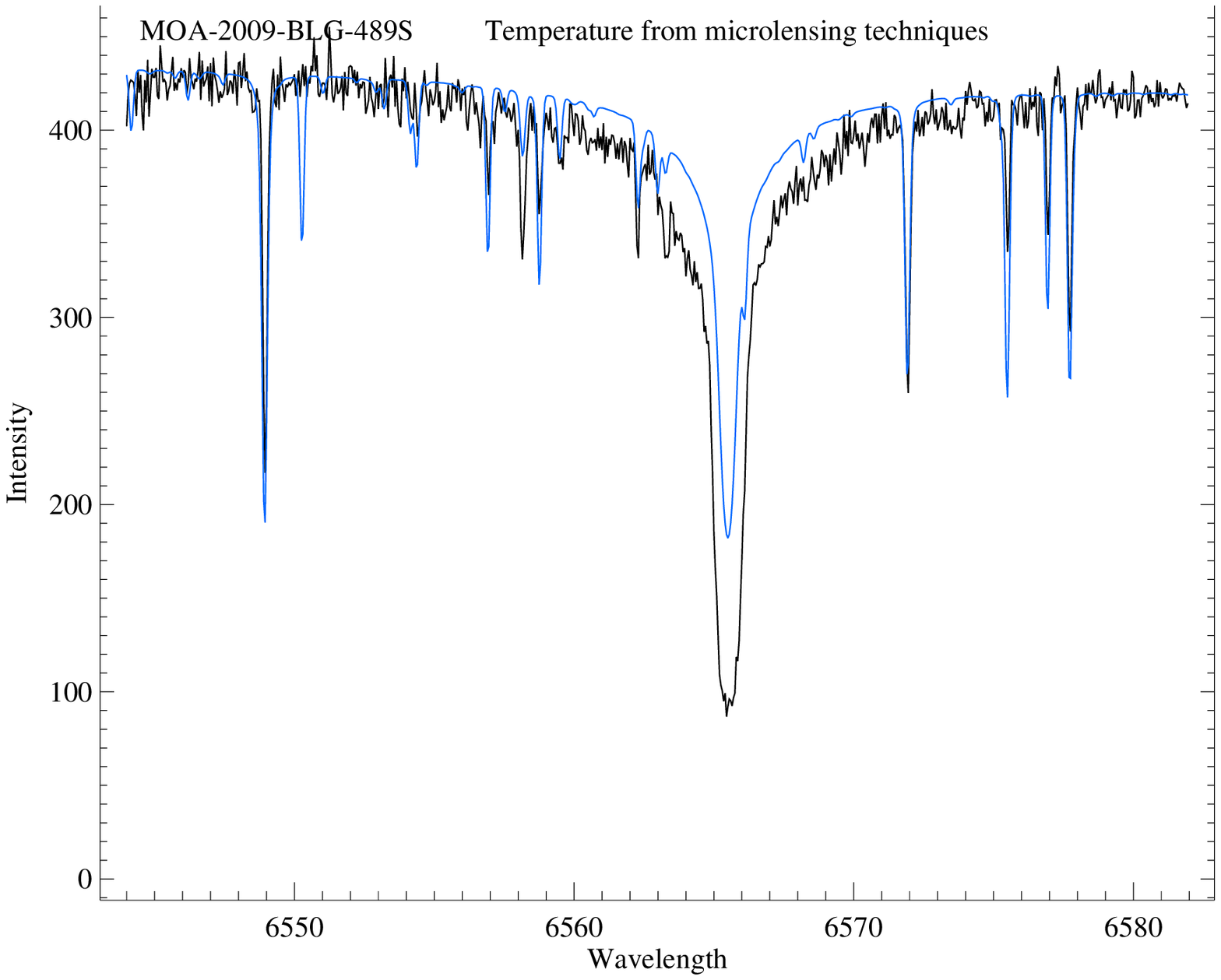}}
   \resizebox{\hsize}{!}{
   \includegraphics[bb=30 320 545 475,clip]{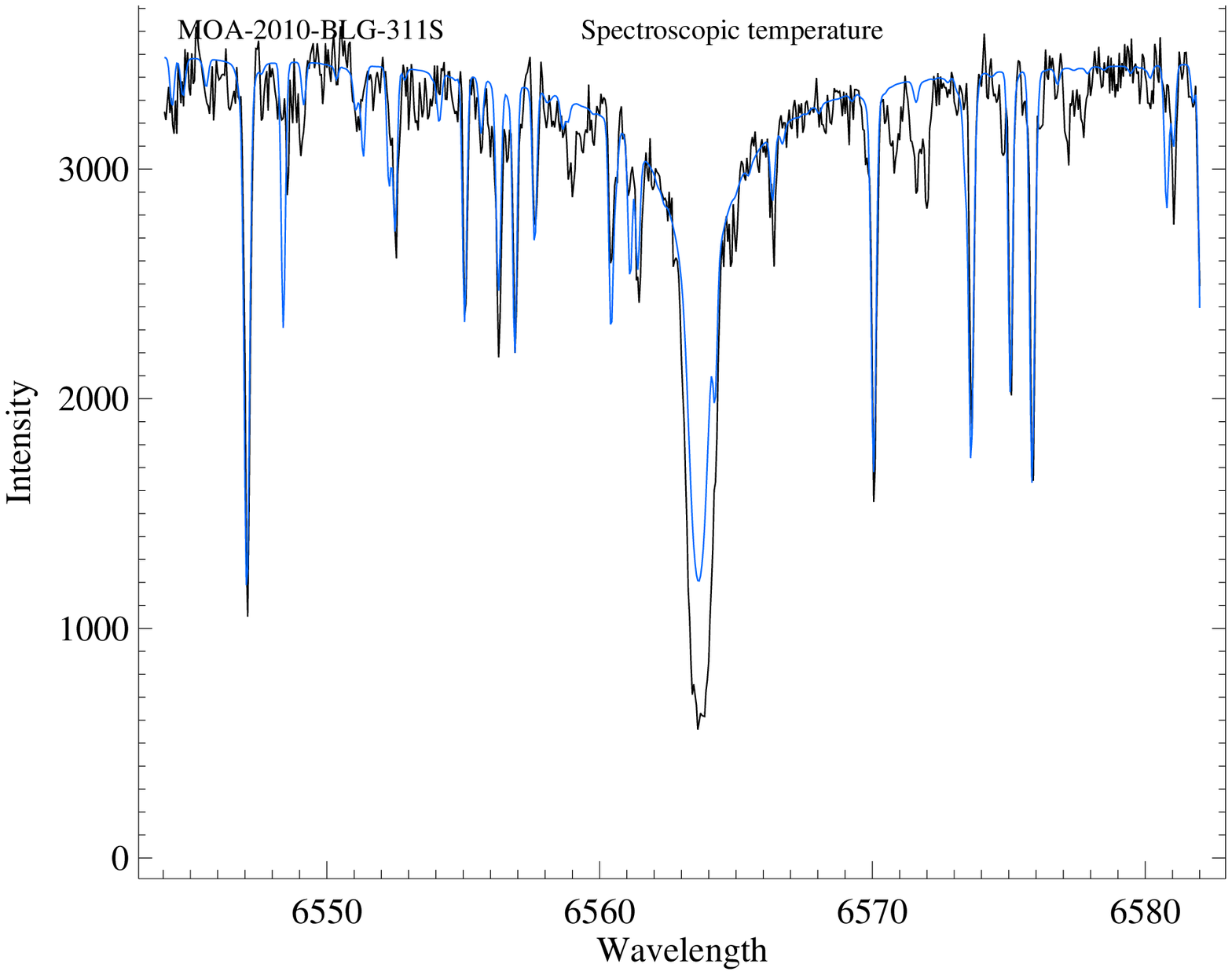}
   \includegraphics[bb=75 320 554 475,clip]{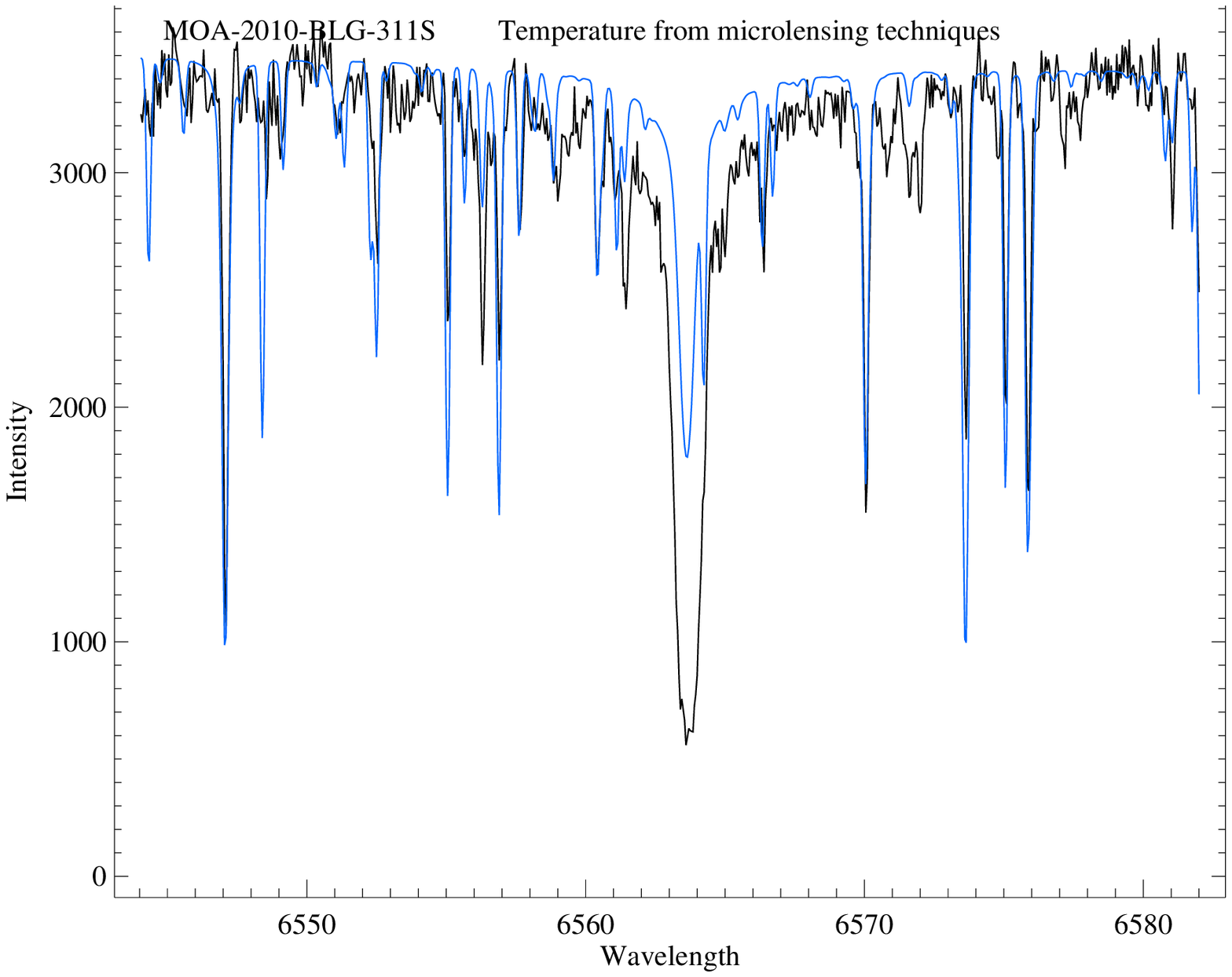}}
   \resizebox{\hsize}{!}{
   \includegraphics[bb=30 320 545 475,clip]{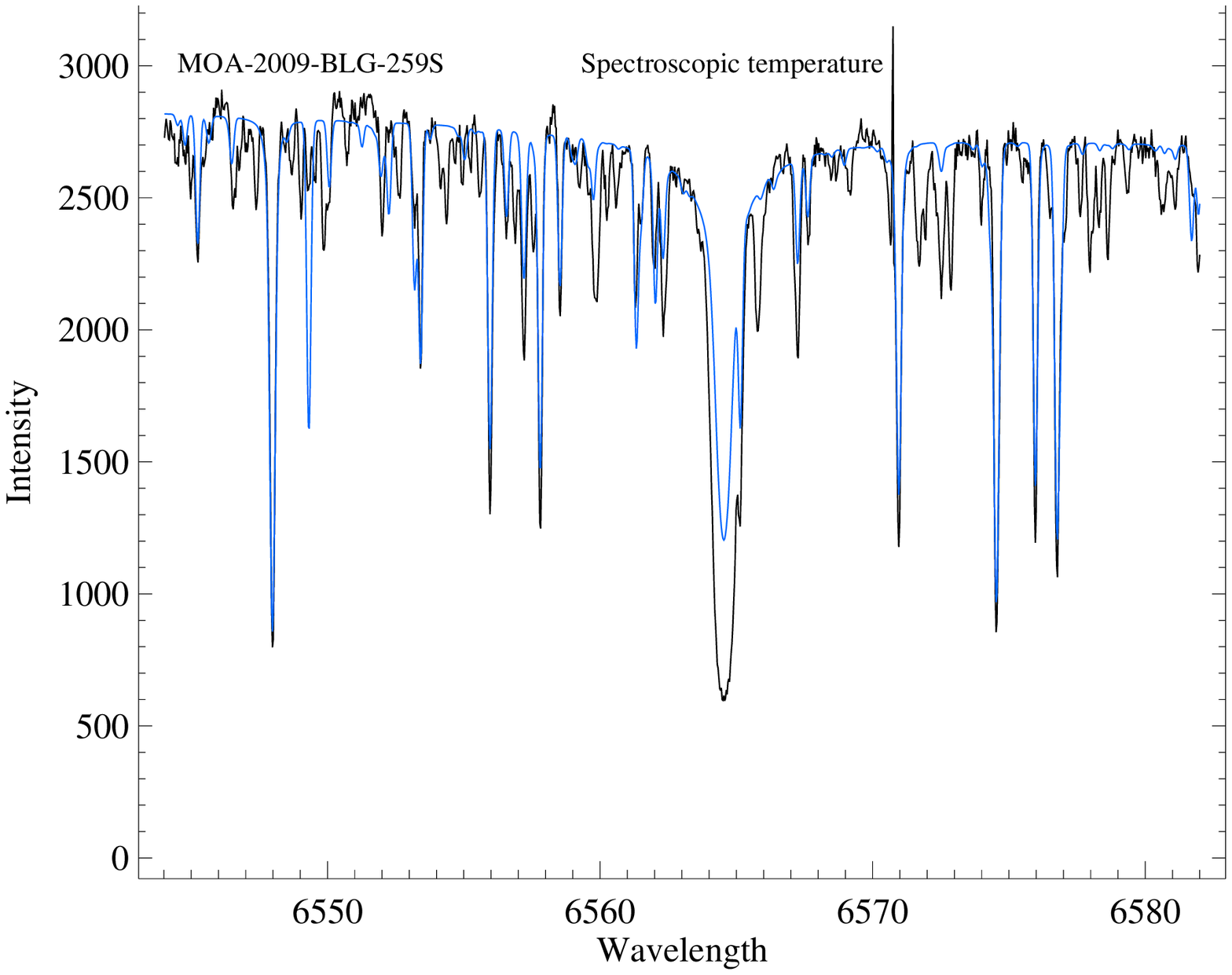}
   \includegraphics[bb=75 320 554 475,clip]{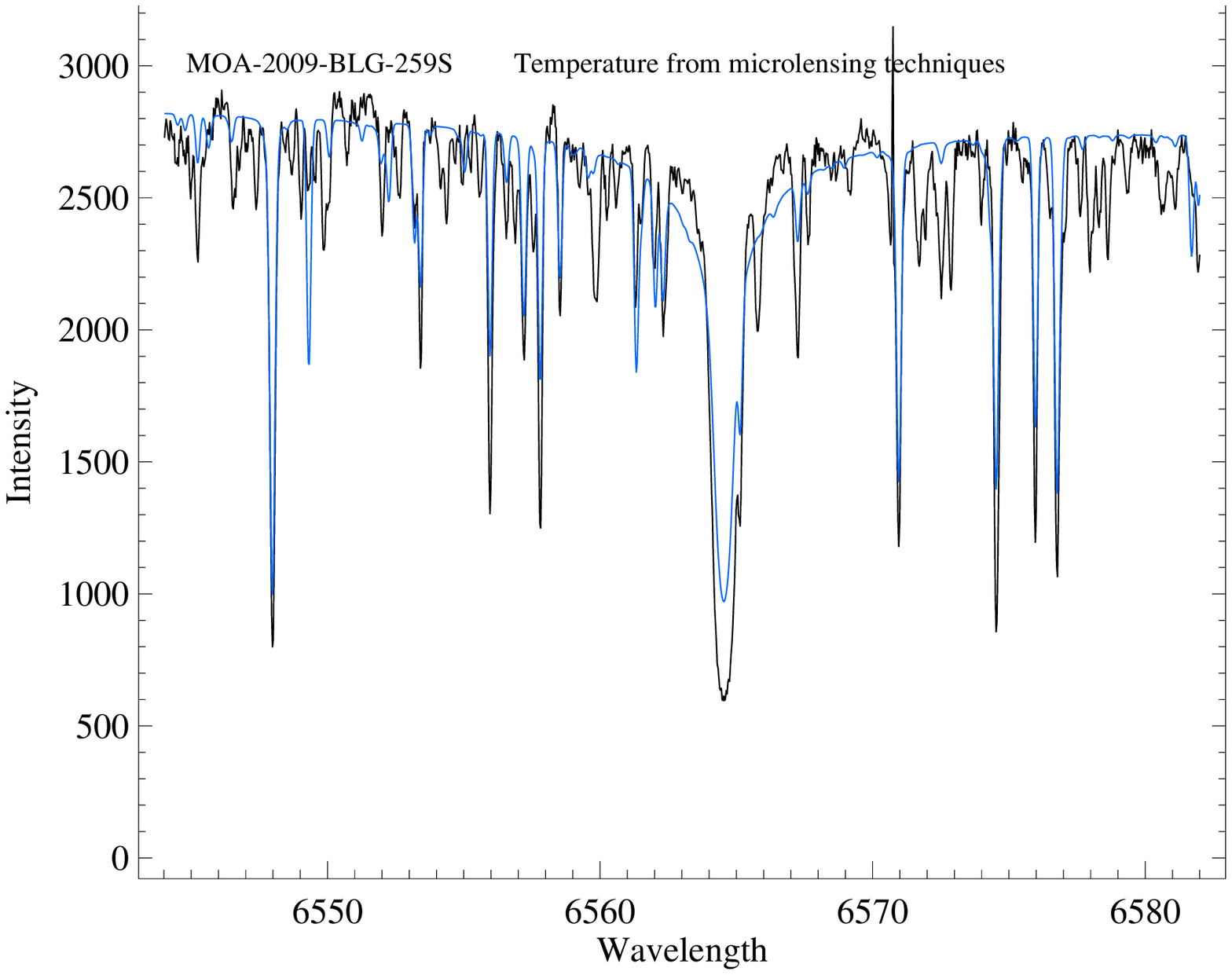}}
   \resizebox{\hsize}{!}{
   \includegraphics[bb=30 50 545 110,clip]{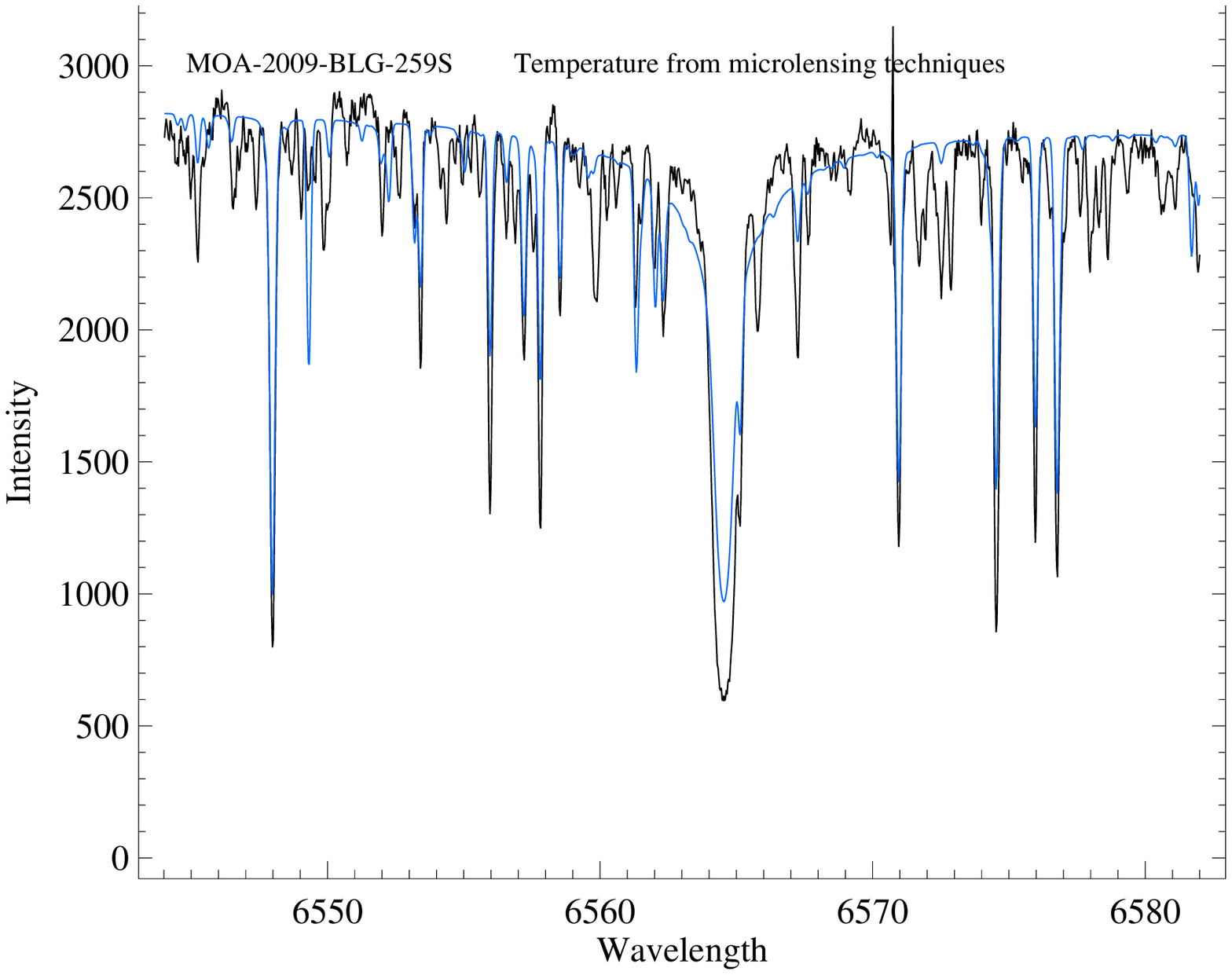}
   \includegraphics[bb=75 50 554 110,clip]{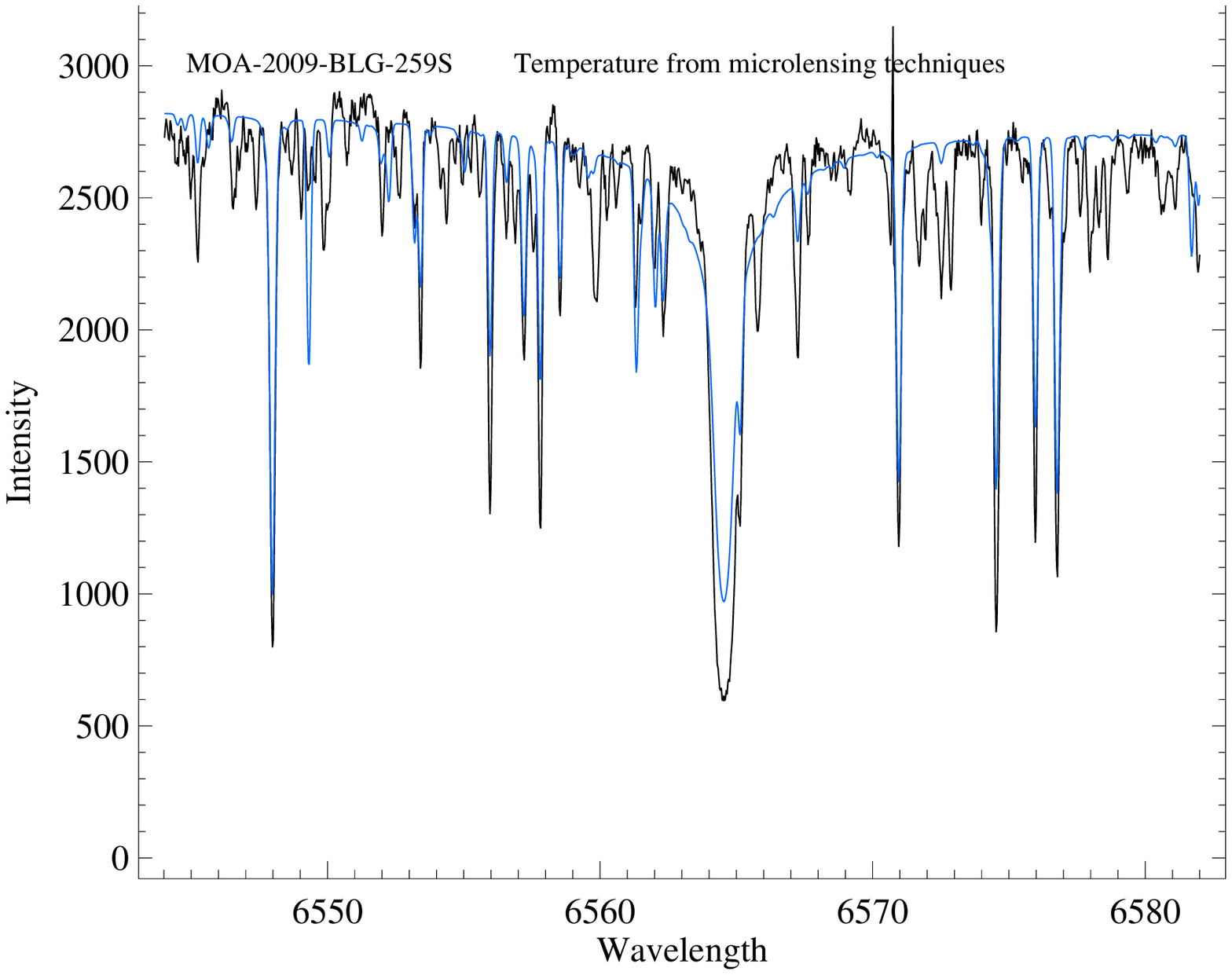}}
      \caption{Synthesis of the H$\alpha$ line at 6563\,{\AA} for the 
      three stars that show large deviations between the spectroscopic
      effective temperatures and the effective temperatures based on
      microlensing techniques. Observed spectra are shown with black lines,
      and the synthetic spectra with blue lines.
      Left-hand side shows the synthetic profiles
      based on spectroscopic temperatures and the right-hand side
      the profiles based on temperatures from microlensing techniques.
      The microlensing temperature for MOA-2010-BLG-311S is the
      one based on the original $(V-I)_0$ colour (i.e. $\teff = 4750$\,K).
      Note that no attempts to match individual lines has been made.
      The composition of the model atmospheres are simply scaled with
      [Fe/H] relative to the standard solar composition by
      \cite{grevesse2007}.
                          }
         \label{fig:halpha}
   \end{figure*}
\begin{table}
\centering
\caption{
Measured equivalent widths and calculated elemental abundances for the 12 new microlensed dwarfs.$^\dag$
\label{tab:ews}
}
\begin{tabular}{cccccccc}
\hline\hline
\noalign{\smallskip}
Element                           &  
$\lambda$                         &
$\chi_{\rm l}$                    &
\multicolumn{2}{c}{star 1}      &
 $\cdots$                         &
\multicolumn{2}{c}{star 10}       \\
\noalign{\smallskip}
                                  &
[{\AA}]                           &
[eV]                              &
$W_{\rm \lambda}$                 &
$\epsilon (X)$                    &
  $\cdots$                        &
$W_{\rm \lambda}$                 &
$\epsilon (X)$                    \\
\noalign{\smallskip}
\hline
\noalign{\smallskip}
\vdots &
\vdots &
\vdots &
\vdots &
\vdots &
$\cdots$ &
\vdots &
\vdots \\
\noalign{\smallskip}
\hline
\end{tabular}
\flushleft
{\tiny
$^{\dagger}$
For each line we give the $\log gf$ value, lower excitation energy ($\chi_{\rm l}$), equivalent width ($W_{\rm \lambda}$),
absolute abundance ($\log \epsilon (X)$).
The table is only available in electronic form at the CDS via anonymous ftp to 
{\tt cdsarc.u-strasbg.fr (130.79.128.5)} or via 
{\tt http://cdsweb.u-strasbg.fr/cgi-bin/qcat?J/A+A/XXX/AXX}.
}
\end{table}

\begin{table}
\centering
\caption{
LTE Li abundances and NLTE corrections from \cite{lind2009}. 
\label{tab:li}
}
\setlength{\tabcolsep}{1.5mm}
\begin{tabular}{rcccc}
\hline\hline
\noalign{\smallskip}
  \multicolumn{1}{c}{Object}                 &
  \multicolumn{1}{c}{$\log\epsilon ({\rm Li})_{\rm LTE}$}  &
  $\Delta_{\rm NLTE}$ &
  $\teff$ [K]    &
  [Fe/H]              \\
\noalign{\smallskip}
\hline
\noalign{\smallskip}
 OGLE-2006-BLG-265S$^\dagger$             &  0.96 & $+0.08$ & 5486 & $+0.47$ \\
 OGLE-2007-BLG-349S$^\dagger$             &  0.89 & $+0.11$ & 5229 & $+0.42$ \\
  MOA-2008-BLG-311S$^\dagger$             &  2.23 & $+0.01$ & 5944 & $+0.36$ \\
  MOA-2010-BLG-049S$\phantom{^\dagger}$   &  1.43 & $+0.07$ & 5738 & $-0.38$ \\        
  MOA-2010-BLG-285S$\phantom{^\dagger}$   &  2.21 & $-0.05$ & 6064 & $-1.23$ \\
  MOA-2010-BLG-523S$\phantom{^\dagger}$   &  1.64 & $+0.10$ & 5250 & $+0.09$ \\
\noalign{\smallskip}
\hline
\end{tabular}
\flushleft
{\tiny
$^\dagger$ Full set of stellar parameters and elemental abundances for
OGLE-2006-BLG-265S, OGLE-2007-BLG-349S, and MOA-2008-BLG-311S
are given in \cite{bensby2010}. 
}
\end{table}

\section{Analysis}
\label{sec:analysis}

\subsection{Stellar parameters and elemental abundances}

Stellar parameters and elemental abundances have been determined
through exactly the same methods as in our previous studies of 
microlensed bulge dwarfs \citep{bensby2009,bensby2010}. Briefly, we use 
standard 1-D plane-parallel model stellar atmospheres calculated with the 
Uppsala MARCS code \citep{gustafsson1975,edvardsson1993,asplund1997}. 
(For consistency with our previous analyses, we continue to
use the old MARCS models. As shown in \citealt{gustafsson2008},
the differences between the new and old MARCS models are very small
for our types of stars, i.e. F and G dwarf stars.) Elemental abundances
are calculated with the Uppsala EQWIDTH program using equivalent widths
that were measured by hand using the IRAF task SPLOT. Effective 
temperatures were determined from excitation balance of abundances from 
\ion{Fe}{i} lines, surface gravities from ionisation balance between 
abundances from \ion{Fe}{i} and \ion{Fe}{ii} lines, and the microturbulence
parameters by requiring that the abundances from \ion{Fe}{i} lines
are independent of line strength. 

An error analysis, as outlined in \cite{epstein2010},
has been performed for the microlensed dwarf stars.
This method takes into account the uncertainties in the four 
observables that were used to find the stellar parameters, i.e. the
uncertainty of the slope in the graph of \ion{Fe}{i} abundances versus lower
excitation potential; the uncertainty of the slope in the graph of
\ion{Fe}{i} abundances versus line strength; the uncertainty
in the difference between \ion{Fe}{i} and \ion{Fe}{ii} abundances; 
and the uncertainty in the difference
between input and output metallicities. The method also accounts for
abundance spreads (line-to-line scatter) as well as how the average
abundances for each element reacts to changes in the stellar parameters.
Compared to \cite{epstein2010}, who
used \ion{Ca}{i} lines to constrain the microturbulence
parameter, we use the same \ion{Fe}{i} lines that were used 
for the effective temperature.
Although the variables are not fully independent, they are only weakly
correlated. Thus the total error bar is dominated by the 
uncertainties in the slopes.

Stellar parameters and error estimates are given in
Table~\ref{tab:parameters}, elemental abundance ratios
in Table~\ref{tab:abundances2}, and measured equivalent widths and abundances
for individual lines in Table~\ref{tab:ews}.

We have also determined NLTE Li abundances in 6 of the 26 microlensed 
dwarf stars through line profile fitting of the $^7$\ion{Li}{i} resonance 
doublet line at 670.8 nm. The 1-D NLTE corrections are
taken from \cite{lind2009}. The Li line was not detected in the remaining
20 stars, and the spectra are not of sufficient quality to estimate
interesting upper limits. The method we use to determine Li abundances is
fully described in \cite{bensby2010li}
where we presented the Li abundance in MOA-2010-BLG-285S.
The Li abundances are listed in Table~\ref{tab:li}.

\subsection{Stellar ages and absolute magnitudes}
\label{sec:ages}

The determination of stellar ages has been greatly improved in relation 
to our previous work, by the use of probability distribution functions.
Furthermore, the Yonsei-Yale isochrones \citep{demarque2004} have been 
calibrated to reproduce the solar age and solar mass for the input 
solar stellar parameters ($\teff / \log\,g / [Fe/H] = 5777\,K/4.44/0.00$). 
This is described in detail in a forthcoming paper (Melendez et al., in prep.). 
In short, for a given set of ``observed" stellar parameters 
($\teff$, $\log\,g$, [Fe/H])
and theoretical isochrone points (effective temperature $T$, logarithm of surface gravity $G$, 
and metallicity $M$), the age is determined from an isochrone age probability distribution (APD): 

\begin{equation}
dP\mathrm{(age)} = \frac{1}{\Delta(\mathrm{age})}\sum_{\Delta\mathrm{(age)}} p\,(\teff,\log\,g,\feh,T,G,M)\ ,
\label{eq:dP}
\end{equation}
where $\Delta\mathrm{(age)}$ is an adopted step in age from the grid of isochrones and:

\begin{eqnarray}
p & \propto & \exp[-(\teff-T)^2/2(\Delta\teff)^2]\times   \nonumber \\
  &         & \exp[-(\log g-G)^2/2(\Delta\log g)^2]\times  \\
  &         & \exp[-(\feh-M)^2/2(\Delta\feh)^2]\ . \nonumber
\end{eqnarray}

The errors in the derived stellar parameters ($\Delta\teff$, etc.) are listed
in Table~\ref{tab:parameters}. 
The sum in Eq.~(\ref{eq:dP}) is made over a range of isochrone ages and in principle 
all values of $T,G,M$. In practice, the contribution to the sum from isochrone points 
farther away than $\teff\pm3\Delta\teff$, etc., is negligible. Therefore, 
the sum is limited to isochrone points within a radius of three times the errors
around the observed stellar parameters. 
The probability distributions are normalised so that $\sum dP=1$. 
The most probable age is obtained from the peak of the distribution 
while $1\,\sigma$ limits are derived from 
the shape of the probability distribution. 
Similar formalisms allow us to infer the stellar mass, absolute magnitudes,
and luminosities. 

For three very old stars
($>12$\,Gyr), both the peak and the upper error bar of the APD are not well defined. 
In those cases we adopted as the age the average of the APD peak and the median of 
the solutions, and the upper error bar was adopted as the scatter of the 
solutions. The probability distributions
for mass, absolute magnitudes, and luminosities are all well behaved.

Using the new improved
age determination method we have re-determined the ages for
all 26 stars. The ages, absolute magnitudes, and luminosities
are reported in Table~\ref{tab:irfm}, 
and Fig.~\ref{fig:ages} shows the twelve new 
microlensed dwarf stars together with the relevant Y2 isochrones.
As can be seen, a majority of the stars are either main sequence turn-off
stars or subgiant stars, while two or three are on the verge of
ascending the giant branch. 
The latter ones have just left the subgiant
branch, and should be un-affected by the internal
mixing processes that occur further up on the giant branch
\citep[see, e.g.,][for a review on mixing in stars]{pinsonneault1997}.

   \begin{figure}
   \resizebox{\hsize}{!}{\includegraphics[bb=18 155 592 480,clip]{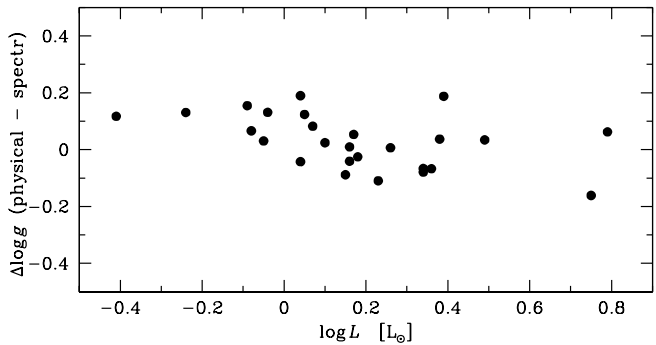}}
      \caption{Comparison between out spectroscopic surface gravities
      and the ``physical" $\log g$ that can be computed from the 
      luminosities and stellar masses inferred from isochrones.
      On average the physical $\log g$:s are 0.03\,dex higher (with a dispersion
      of 0.09\,dex).
                 }
         \label{fig:absmaglogg}
   \end{figure}

The masses and luminosities
presented in Table~\ref{tab:irfm} will yield a ``physical" $\log g$, computed
from the fundamental relationship that relates luminosity, stellar mass, and
effective temperature (see, e.g., Eq.~(4) in \citealt{bensby2003}).
Because a star with the best-fit age may not lie exactly at 
the spectroscopically measured $\teff$ and $\log g$, the the physical and
spectroscopic $\log g$:s agree within
$0.03\pm0.09$\,dex, rather than exactly, see Fig.~\ref{fig:absmaglogg}.

Probabilistic age determinations like ours are known to suffer 
from systematic biases mainly related to the sampling of isochrone 
data points \citep[e.g.,][their Sect.~4.5.4]{nordstrom2004}. 
Bayesian methods have been developed to tackle these problems 
\citep[e.g.,][]{pont2004,jorgensen2005,casagrande2011}
but we do not follow this approach and therefore 
our ages could be in principle further improved. Nevertheless, our current 
age determination is consistent with our previous work on thin/thick disk 
stars so, in a relative sense, our discussions regarding the ages of our 
sample stars are still reliable.

   \begin{figure}
   \resizebox{\hsize}{!}{\includegraphics[bb=18 155 592 718,clip]{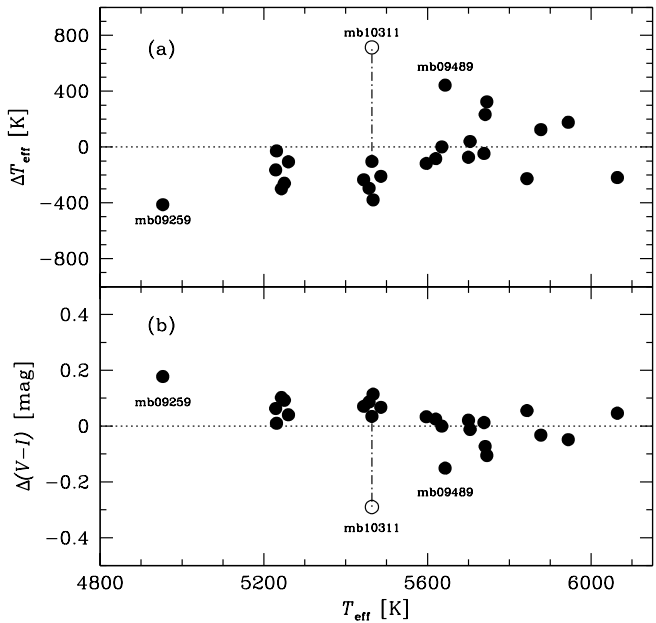}}
      \caption{
      (a) Difference between the spectroscopic temperatures
      and the temperatures from microlensing techniques versus effective
      temperature, and (b) the difference between $(V-I)_0$ colours 
      determined from microlensing techniques and those that can be 
      determined from
      the spectroscopic temperatures using the $IRFM$ calibration
      by \cite{casagrande2010}. The average differences are
      $\langle\Delta T_{\rm eff}\rangle = -80$\,K with a dispersion of
      217\,K, 
      and $\langle\Delta(V-I)_0\rangle =0.026$ with a dispersion
      of 0.073\,mag. 
      The three stars that show largest deviations between 
      microlensing and spectroscopic parameters are marked. For 
      MOA-2010-BLG-311S (mb10311) the original values are shown 
      with open circles, and the corrected value with filled 
      circles  (connected with the original values by a dash-dotted
      line).                 }
         \label{fig:irfm}
   \end{figure}

\subsection{Effective temperatures from microlensing techniques}
\label{sec:microteff}

As in \cite{bensby2010} we confront the effective temperatures
derived from spectroscopic principles with those
deduced from microlensing techniques \citep{yoo2004}.  The microlensing
method assumes that the reddening towards the microlensed source is 
the same as the reddening towards the red clump, and that the red clump 
in the Bulge has $(V-I)_{0} = 1.08$ \citep{bensby2010} and 
$M_{I} = -0.15$ (David Nataf, private communication). 
The de-reddened magnitude and colour are then derived from 
the offsets between the microlensing source and the red clump, in 
the instrumental colour-magnitude diagram (CMD).  As the lenses for the
MOA-2010-BLG-285S and -523S events likely are binaries,
and need more work, no absolute magnitudes could be estimated for these
stars. Also, the $(V-I)_{0}$ colour for MOA-2010-BLG-446S could not be
determined because there are too few clump stars close to this star
in the colour-magnitude 
diagram. From the colour-[Fe/H]-$\teff$ calibrations by \cite{casagrande2010} 
we can determine which temperatures the $(V-I)_{0}$ colours 
correspond to. The absolute de-reddened magnitudes and colours
and the effective temperatures are given in Table~\ref{tab:irfm}.

Even though the effective temperatures 
from microlensing techniques and spectroscopic principles are
in reasonable agreement, there are three events for which the differences
are disturbingly large; more than 400\,K for MOA-2009-BLG-259S and 
MOA-2009-BLG-489S, and more than 700\,K for MOA-2010-BLG-311S 
(Fig.~\ref{fig:irfm}a). Note that stellar parameters for MOA-2009-BLG-489S
were published in \cite{bensby2010}.
To check the temperatures for these stars, and get a third opinion, we calculate
synthetic spectra for the H$\alpha$ line wing profiles with the latest version 
(v2.1 oct 2010) of 
the Spectroscopy Made Easy (SME) tool \citep{valenti1996} 
using the latest 1-D LTE plane-parallel
MARCS model stellar atmospheres \citep{gustafsson2008}, and line data
from VALD \citep{vald_1,vald_3}. In Fig.~\ref{fig:halpha} we show
the observed line
profiles and synthetic line profiles for the Balmer H$\alpha$ line
for these three stars, using both the spectroscopic temperatures
(left-hand side plots) and the temperatures from microlensing
techniques (right-hand side plots). We see that the synthetic spectra
based on spectroscopic temperatures reproduce the wings of the
H$\alpha$ line almost perfectly, while the spectra based on
temperatures from microlensing techniques clearly do not match the
observed wing profiles. 

We could not find anything that was obviously wrong with the photometry
of MOA-2009-BLG259S and MOA-2009-BLG-489S, although differential reddening 
is noticeable in the
field of MOA-2009-BLG-489S (Nataf, private communication, see also 
Table~\ref{tab:irfm}). However, for
MOA-2010-BLG-311S there was only one highly-magnified $V$ point, so
the $(V-I)_0$ colour could not be robustly estimated by the standard
techniques used by \cite{yoo2004}.  In an alternative attempt to determine
the $(V-I)_0$ colour for MOA-2010-BLG-311S we first determined the
instrumental $(I-H)$ colour, making use of the $H$-band data that are
simultaneously taken with $I$ band on the SMARTS ANDICAM camera at the
1.3\,m CTIO telescope.  We
then converted from instrumental $(I-H)$ to $(V-I)$ by constructing a
$VIH$ colour-colour diagram for field stars, which is analogous to the
technique for optical colour transformations \citep{gould2010}.  
Now we find a colour of $(V-I)_{0}=0.75$ for MOA-2010-BLG-311S, 
which is in good agreement 
with the ``spectroscopic" value of 0.78. Note,
however, that this method cannot be blindly applied to $VIH$
transformations because (at the depth of the SMARTS $H$-band images)
the colour-colour diagram is constructed from giants and sub-giants,
while the transformation applies to dwarfs.  Additionally, there is a 
giant-dwarf bifurcation
of the $VIH$ colour-colour relations for $(V-K)_0> 3$
\citep{bessel1988}, which corresponds to $(V-I)_0> 1.3$ (see, in particular,
\citealt{dong2006}).  Nevertheless, since most microlensed dwarfs 
(including this one), have bluer colours than this limit, the method
can usually be applied, if necessary.

For the 24 microlensed dwarf stars that have microlensing $(V-I)_0$ colours, 
including the revised colour for MOA-2010-BLG311S,
we find that our spectroscopic temperatures are $80\pm217$\,K lower
than the ones based on the colour-[Fe/H]-$\teff$ relationships
(see Fig.~\ref{fig:irfm}a). Note that the colours for the previous 
microlensing events in \cite{bensby2010} were estimated under the assumption that
the red clump in the bulge has $(V-I)_{0}=1.05$. Hence, those colours have been
adjusted (by adding 0.03\,mag) and new effective temperatures were determined.
For clarity, all 26 events are listed in Table~\ref{tab:irfm}.
The \cite{casagrande2010} calibrations can also be used to see what $(V-I)_0$ 
colours the spectroscopic effective temperatures correspond to. For the new 
stars these are 
listed in Table~\ref{tab:irfm}, and the comparison 
between photometric and ``spectroscopic" $(V-I)_0$ colours are shown in 
Fig.~\ref{fig:irfm}. 
On average the spectroscopic colours are in good agreement with the
ones from microlensing techniques being only slightly lower 
(the difference is $0.02\pm0.07$\,mag, see Fig.~\ref{fig:irfm}b). 

This difference is due to that we now use the calibrations by 
\cite{casagrande2010}, while we in \cite{bensby2010} used the ones
by \cite{ramirez2005}. If we were to use the \cite{ramirez2005}
calibrations the difference would be $0.00\pm0.06$\,mag. 
This means that the $(V-I)_{0}$ value for the bulge red clump 
that was revised from 1.05 to 1.08 in \cite{bensby2010} could be 
revised back again to 1.06.

\section{Results}

   \begin{figure}
   \resizebox{\hsize}{!}{\includegraphics{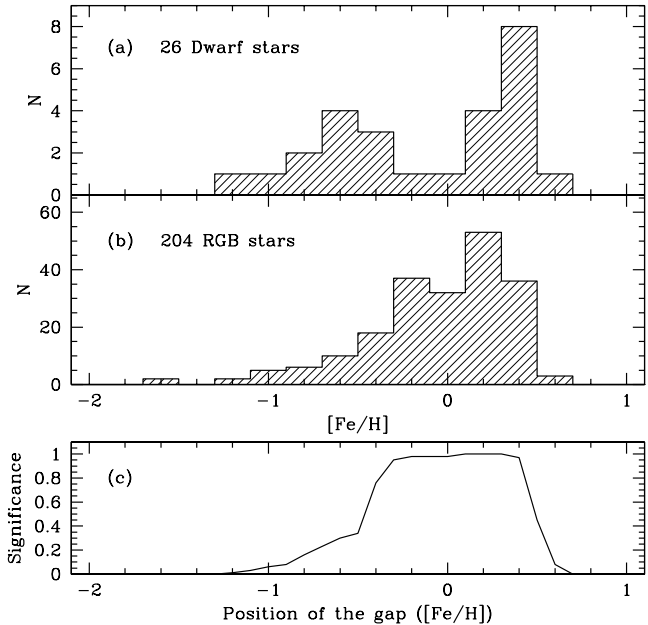}}
      \caption{
      {\bf a)} MDF for our 26 microlensed dwarf stars in the bulge. 
      {\bf b)} MDF for 204 RGB stars in the bulge, from \cite{zoccali2008}. 
      {\bf c)} The significance for the failure of reproduce a gap vs. 
      position of a gap in the dwarf 
      star MDF (see explanation in Sect.~\ref{sec:gap}).
              }
         \label{fig:bimod}
   \end{figure}

\subsection{The metallicity distribution}
\label{sec:gap}

In Fig.~\ref{fig:bimod}a we show the 
MDF for our sample of 26 dwarf stars in the bulge. Since the typical
error in [Fe/H] is about 0.1\,dex, we choose a bin size for the histogram
of 0.2\,dex. 
This ensures that any feature that we detect in the distribution is not due 
to error statistics, i.e., features that could originate from the 
measurement uncertainties are suppressed by the size of each bin.
 We note that this MDF appears
bimodal with peaks at $\rm [Fe/H]\approx -0.6$ and $\approx +0.3$\,dex, 
and a gap of only two
stars at $\rm [Fe/H]\approx -0.1$. In Fig.~\ref{fig:bimod}b, for comparison, we
show the MDF for 204 RGB stars in the bulge from \cite{zoccali2008}. Given 
that the MDF for the RGB stars only has one prominent peak
(although asymmetric),
there is a significant difference between the RGB and the dwarf
star MDFs. A Kolmogorov-Smirnov test between the two distributions yields
a $p$-value of 0.11. Even though this is not low enough to reject the 
null-hypothesis that the MDFs are drawn from the same distribution, 
under the commonly used limit of 0.05, it is low enough to warrant 
further investigation. 

The most prominent difference is the deficiency of stars in the dwarf
star MDF, situated almost where the RGB MDF peaks. The question we ask is: what
is the probability of obtaining such a deficiency, similar to the one in
the dwarf star MDF, if 26 stars are drawn randomly from the RGB MDF? To
answer this question, we construct a simple significance test. First, we
draw 26 stars randomly from the RGB MDF. Given that there are 204 stars in
the RGB sample, we assume the RGB MDF to be representative of the
complete underlying MDF.  Next, we check if the sample of 26 randomly drawn stars
has a deficiency of stars in the MDF at ${\rm [Fe/H]}= x$. We
define the deficiency of stars in the MDF as a gap with two stars or less
in a metallicity range of 0.3\,dex (see Fig. 5a). The gap in Fig.~\ref{fig:bimod} 
spans about 0.4\,dex. However, given that the typical error in
[Fe/H] for the dwarf stars is about 0.1\,dex, we use a range of 0.3\,dex. We
iterate this process $10^5$ times. We define the significance as the fraction of
iterations that fail to reproduce a gap in the randomly generated dwarf
star MDF. Finally, we sample over the position of the gap, x, from $\rm [Fe/H] =-2$ 
to $\rm [Fe/H]=+1$ in steps of 0.1\,dex. 

Figure 5c shows our result
from the test. As can be seen, the existence of a gap of stars in the
dwarf star MDF at $\rm [Fe/H]= -0.1$ is unlikely. This indicates that the 
gap in the MDF is most likely not caused by low number statistics.
Additionally, we
ran the test with a gap range of 0.2 and 0.4\,dex and found that it did not
affect the result significantly.
Given the possibility of systematic offsets in [Fe/H] for giants and
dwarfs, how would a shift in the giant MDF affect the significance level
as a function of [Fe/H]? As we sample over all [Fe/H] in the test above,
the only effect of shifting the giant MDF is a shift of the
significance curve in Fig.~\ref{fig:bimod} by the same amount.

\begin{figure}
\resizebox{\hsize}{!}{
\includegraphics[bb=18 170 592 718,clip]{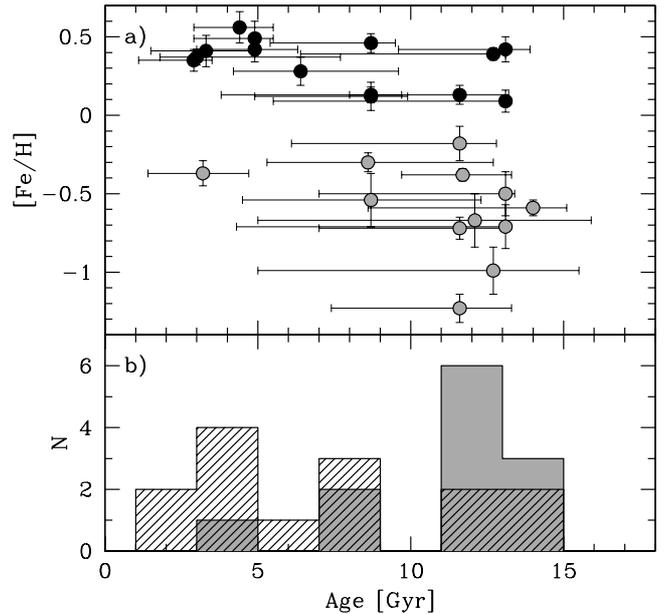}}
\caption{{\sl Top panel} show ages versus [Fe/H] for the now in total 26 
microlensed dwarf and subgiant stars in the Bulge. Stars with 
$\rm [Fe/H]>0$ are marked with black circles and stars with $\rm [Fe/H]<0$ 
with grey circles.
{\sl Bottom panel} shows age histograms for the metal-rich stars
(hatched histogram) and metal-poor stars (grey histogram).
\label{fig:agefe}}
\end{figure}

\subsection{Ages and metallicities}
\label{sec:agefe}

Figure~\ref{fig:agefe}a shows the age-metallicity diagram for the
26 microlensed dwarf and subgiant stars. At sub-solar metallicities
the stars are predominantly old with ages between 9 and 13\,Gyr
(see Fig.~\ref{fig:agefe}b), and no apparent trend of increasing or 
decreasing ages with [Fe/H] is seen. The average age and $1\sigma$ 
dispersion is $11.0\pm2.9$\,Gyr.
This is similar to what is found for thick disk dwarfs at the 
same metallicity in the  solar neighbourhood \citep[e.g.,][]{fuhrmann1998,bensby2007letter2}.

Furthermore, looking again at the metal-poor population, all stars, except 
MOA-2010-BLG-446S at $\rm [Fe/H]=-0.37$, seem to have the same age
within the errorbars. If we exclude this star the average age
becomes $11.7\pm1.7$\,Gyr, indicating that the age spread in
metal-poor population is indeed very small. Regarding MOA-2010-BLG-446S,
we do not rule out the possibility that it is an interloper from the
Galactic disk.

The stars at super-solar metallicities on the other hand
show a wide range of ages from the youngest disk, being only a 
few billion years old, to the oldest halo, as old as the Universe 
(see Fig.~\ref{fig:agefe}). The average age and $1\sigma$ dispersion 
is $7.6\pm3.9$\,Gyr for the  stars at super-solar [Fe/H].

If the ages are weighted by the inverse squares of their errors, the 
average ages and $1\sigma$ dispersions become $5.9\pm3.4$\,Gyr
for the metal-rich stars, and $11.7\pm 1.5$ for the
metal-poor stars ($9.5\pm4.2$\,Gyr when including MOA-2010-BLG-446S).

\subsection{Absolute magnitudes}
\label{sec:absmag}

In order to further investigate whether our microlensed dwarf and subgiant
stars are indeed in the
Bulge, we have compared the absolute $I$ magnitudes obtained from the 
microlensing photometry with the absolute magnitudes obtained from
the spectroscopic stellar parameters.

The spectroscopic absolute magnitudes were obtained with the same methods 
and the same set of isochrones employed to determine the ages 
(see Sect.~\ref{sec:analysis}). The absolute magnitudes from the 
microlensing technique, $M_{I,\mu lens}$, were obtained by comparing 
the instrumental $I$ magnitude of the microlensing source, $I_{instr,s}$, 
with the instrumental $I$ magnitude of the bulge red clump, 
$I_{instr,clump}$, assuming that the absolute magnitude of the clump is 
$M_{I,clump} = -0.15$, i.e.
\begin{equation}
M_{I,\mu lens} = I_{instr, s} - I_{instr, clump} -0.15
\end{equation}
This means that the $M_{I,\mu lens}$ will exactly equal
$M_{I,\mu source}$ if: 
\begin{enumerate}
\item The instrumental clump centroid is correctly measured.
[Clump centroids are probably accurate to about 0.1\,mag.
 This is the typical measured difference in the clump centroids
 of different data sets of the same field and then transformed
 from one to another.]
\item The instrumental source magnitude is
correctly measured. [In most cases this is extremely accurate 
(of order 0.02\,mag or less).]
\item The extinction towards the source is the same as the mean extinction 
towards the clump. [Most fields have modest differential extinction.  
There is very little systematic bias.  The random error due to 
differential extinction is probably less than 0.05\,mag.]
\item The absolute magnitude of the bulge
clump is $M_{I,clump}=-0.15$. [Note: $M_{I,clump}=-0.25$ is the value that 
usually has been used in bulge microlensing studies.  However, recent studies
indicate that it probably is incorrect, and 
should be $-0.15$ (D. Nataf, private communication).]
\item  The source is at the same distance as the clump.
[The sources are certainly {\it not} at the same distance as the clump.
They are thought to be in the bulge, and therefore at a range of
distances similar to the range of distances of clump stars.  However,
the sources must be behind the lenses. Thus, if the latter are in the
bulge (thought to be about 60\% of lensing events) then the former
must be preferentially towards the back of the bulge.  Hence, on 
these grounds alone one would expect them to be of order
0.1\,mag behind the clump. However, from Fig.~\ref{fig:allevents} it is clear that
the spectroscopic sample is drawn preferentially from positive
longitudes, which constitute the near side of the Galactic bar,
which has a slope of roughly $\rm 0.1\,mag\,deg^{-1}$.  We take this as
indicating a strong bias towards brighter sources (which then require
less magnification to enter our sample).  Of course, {\it within}
any given line of sight, this bias is working on a much narrower baseline,
i.e., the intrinsic dispersion in the bulge depth, which is order
0.15\,mag.  We therefore estimate that the two biases approximately
cancel out, and that there is an intrinsic dispersion of source distances
of 0.15\,mag.]
\end{enumerate}

   \begin{figure}
   \resizebox{\hsize}{!}{
   \includegraphics[bb=18 155 592 560,clip]{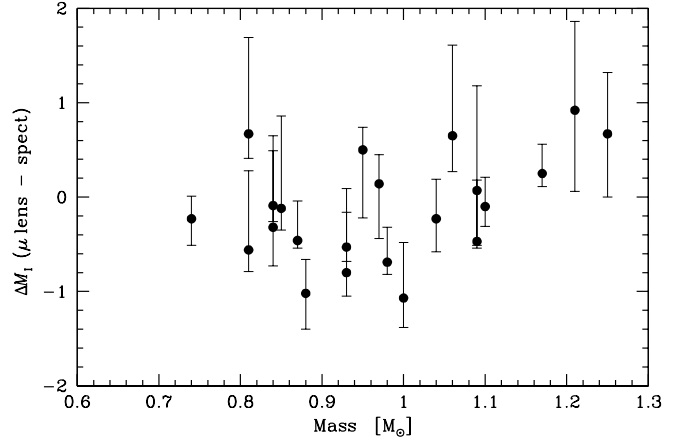}}
      \caption{The difference
      between the absolute $I$ magnitudes from microlensing
      techniques and from spectroscopy versus stellar mass (derived
      from spectroscopy). Error bars represent the uncertainties in the 
      spectroscopic values.
          }
         \label{fig:absmagdiff}
   \end{figure}

Figure~\ref{fig:absmagdiff} shows the difference between the absolute $I$
magnitudes from the two methods versus stellar mass (as determined from 
spectroscopy). The spectroscopic values are on average higher by 
$-0.13$\,mag (with a dispersion of 0.56\,mag).
The error bars represent the uncertainties in the
spectroscopic values. Even though the individual points appears to be 
basically consistent at the 1-$\sigma$ level, there is a tentative
trend in the sense that the spectroscopic values are higher for low
masses while microlensing values are higher for high masses.

To formally evaluate the significance of the difference between the
microlensing and spectroscopic estimates of $M_I$, we first estimate
an error of 0.2\,mag in the microlensing $M_I$ values, which is the
quadrature sum of the 5 enumerated error sources identified above.
We add this in quadrature to the (asymmetric) errors shown in
Fig.~\ref{fig:absmagdiff}, and evaluate $\chi^2$ as a function of assumed offset.
We find $\Delta M_{\rm I} = -0.10\pm 0.15$, with $\chi^2_{\rm min} = 12.48$
for 21 degrees of freedom.  The fact that the mean is within
$1\,\sigma$ of zero implies that the sample is consistent with being
entirely drawn from bulge stars.  Since $\chi^2$ is less than the
number of dof, there is no evidence for scatter beyond
that due to measurement errors (which includes our estimate of a 0.15 mag
dispersion in distances relative to the clump).  Of course, this does
not prove that all of our sources are in bulge: there could be disk
interlopers.  However, the fact that microlensing models place the
overwhelming majority of lensed sources in the bulge, combined with
the demonstrated consistency between the $M_I$ determined from spectra
and those determined from microlensing (based on the assumption that
they have the same mean distance as the bulge clump) strongly supports
a bulge location for the great majority of our sources.

\subsection{Colour-magnitude diagram}

   \begin{figure}
   \resizebox{\hsize}{!}{
   \includegraphics[bb=18 155 592 480,clip]{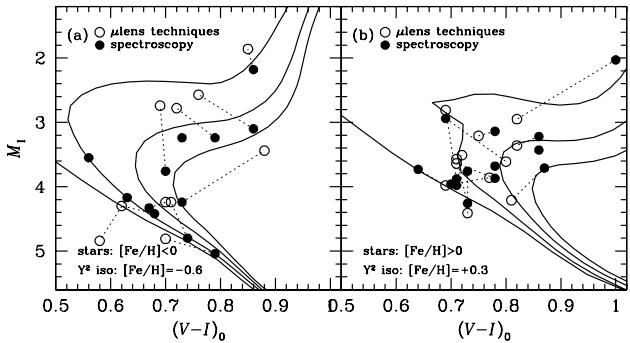}}
      \caption{(a) and (b) show the $(V-I)_0$ versus $M_{I}$ colour-magnitude
      diagram using values from microlensing techniques (open circles)
      and spectroscopy (filled circles). (a) shows the stars with
      $\rm [Fe/H]<0$ and (b) the stars with $\rm [Fe/H]>0$. Each CMD also
      show the Y$^2$ isochrones (1, 5, 10, and 15\,Gyr) from 
      \cite{demarque2004}. For the metal-poor sample the isochrones have
      $\rm [Fe/H]=-0.6$ and for the metal-rich sample  $\rm [Fe/H]=+0.3$.
      The spectroscopic and microlensing values have been connected
      with a dashed line for each star (4 stars do not have microlensing
      values).
          }
         \label{fig:absmag}
   \end{figure}

Figure~\ref{fig:absmag} shows our stars in the HR diagram. We consider 
the metal-poor and metal-rich stars separately, and we include de-reddened 
colours and absolute magnitudes estimated directly from the spectra 
(filled circles, see Sect.~\ref{sec:ages}) as well as from the
micro-lensing technique (open circles, see Sect.~\ref{sec:microteff}). 
The two values are connected for each star by a dotted line (note that 
4 stars do not have microlensing values, see Table~\ref{tab:irfm}).
The metal-poor stars (Fig~\ref{fig:absmag}a)
are spread from the main-sequence to the bottom of the red giant
branch. The age of the population of these stars is younger if
we consider the magnitudes and colours derived by micro-lensing
techniques (between 5 and 10 Gyr), while, if we use the spectroscopically
determined colours and magnitudes, the age is about 5  Gyr larger with
a very nicely defined sub-giant branch. Regradless of these differences
it is clear that we are dealing with a population that is old.

For the metal-rich stars (Fig~\ref{fig:absmag}b) the situation is somewhat different.
Here we have no stars that have evolved sufficiently to trace the
sub-giant branch. Additionally, the spectroscopy and micro-lensing
techniques yield rather similar results for almost all stars. The
stars are clustered in the turn-off region and there seems to be
both rather young stars (5 Gyr with very little room for a large
change) up to stars as old as 10 Gyr. There are quite a few stars
that potentially cluster around the 1 Gyr isochrone as well.

In summary, when we consider the high and low metallicity stars as two 
stellar populations, and study their location in an HR diagram, we see 
that the metal-poor population is fairly evolved and includes essentially 
older stars, whilst the metal-rich population is less evolved and tentatively 
includes some rather young stars.

\subsection{Abundance trends} \label{sec:abundances}

\begin{figure*}
\resizebox{\hsize}{!}{
\includegraphics[bb=18 220 592 470,clip]{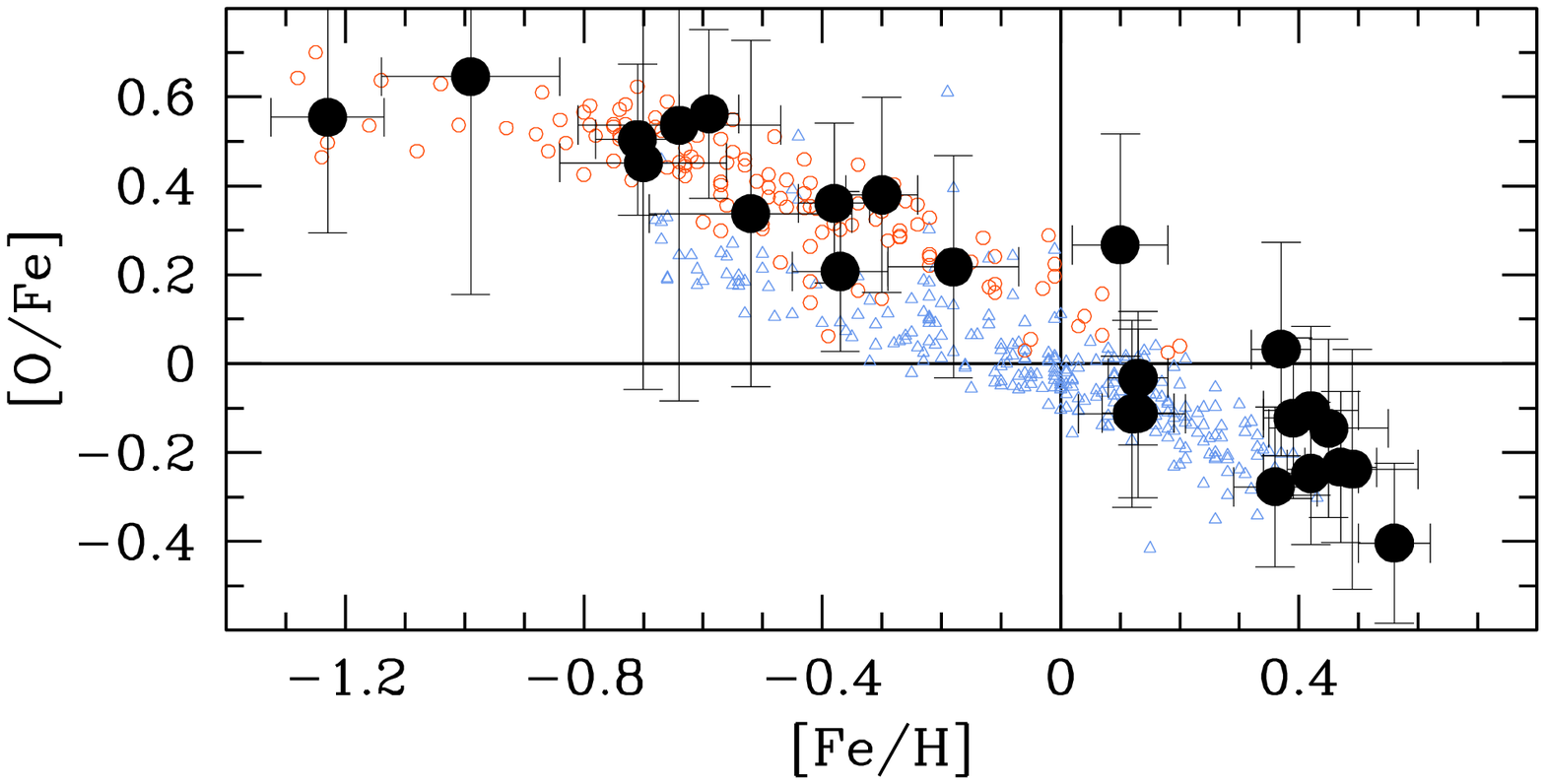}
\includegraphics[bb=18 220 592 470,clip]{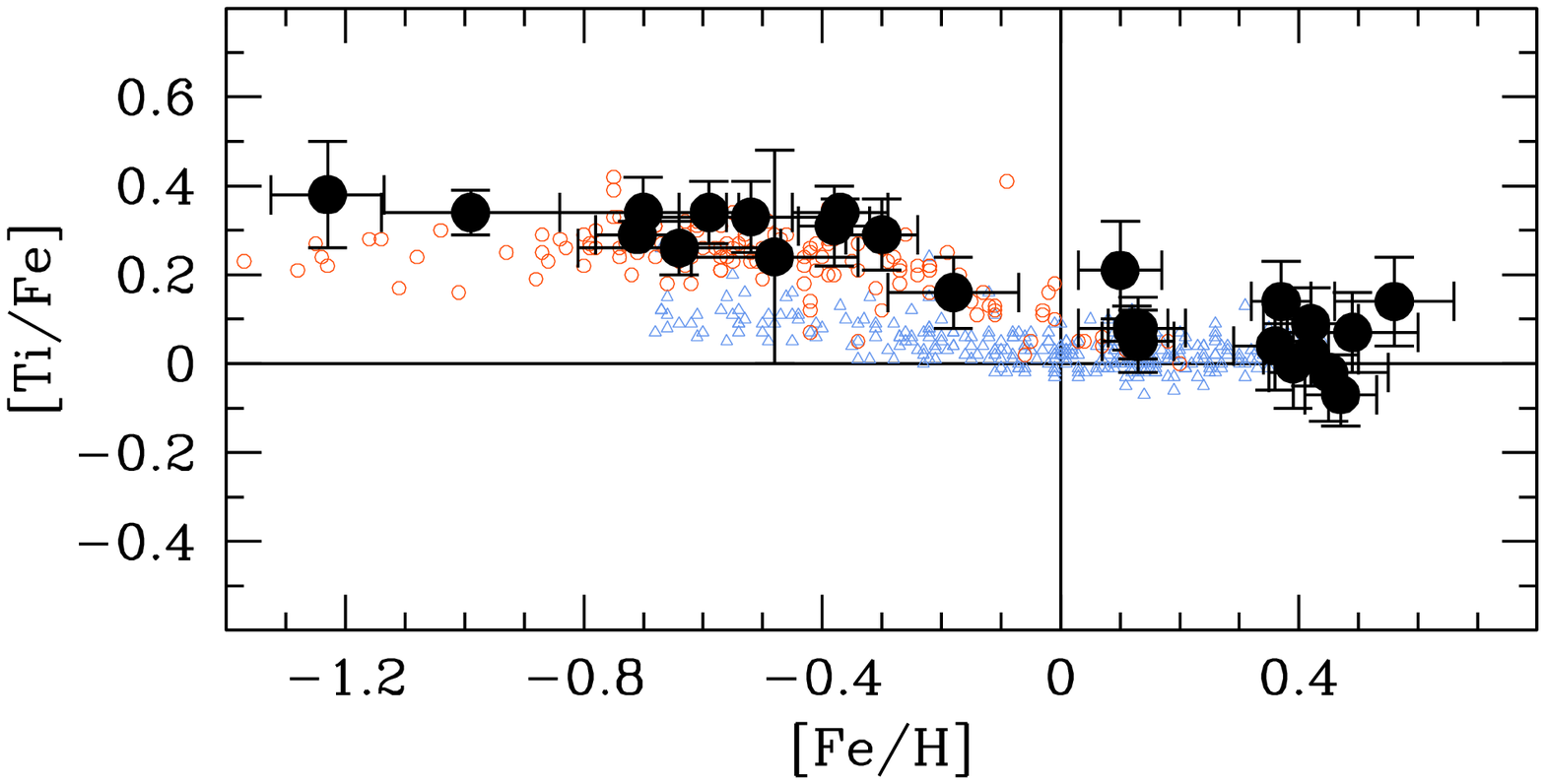}}
\resizebox{\hsize}{!}{
\includegraphics[bb=18 220 592 450,clip]{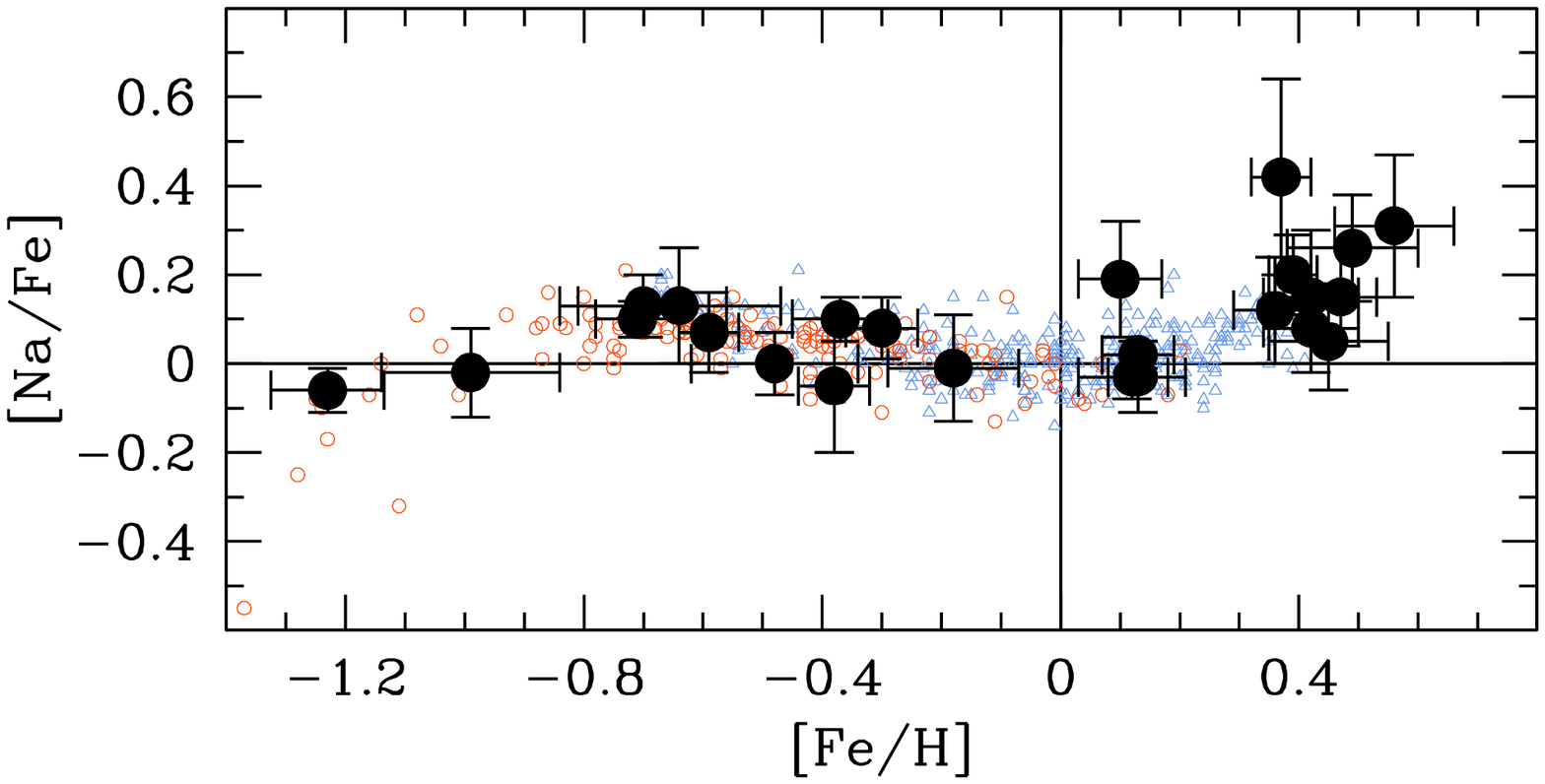}
\includegraphics[bb=18 220 592 450,clip]{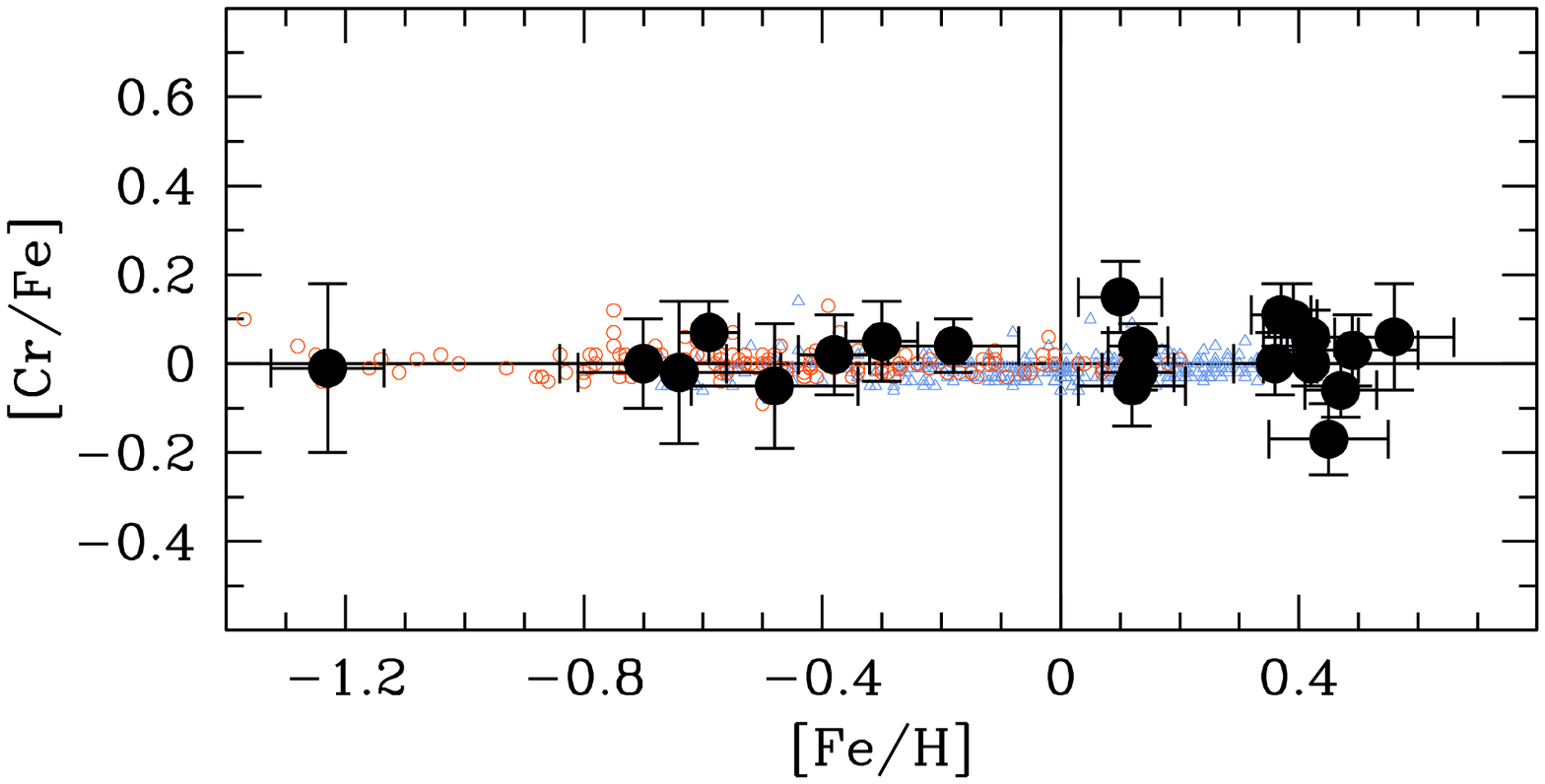}}
\resizebox{\hsize}{!}{
\includegraphics[bb=18 220 592 450,clip]{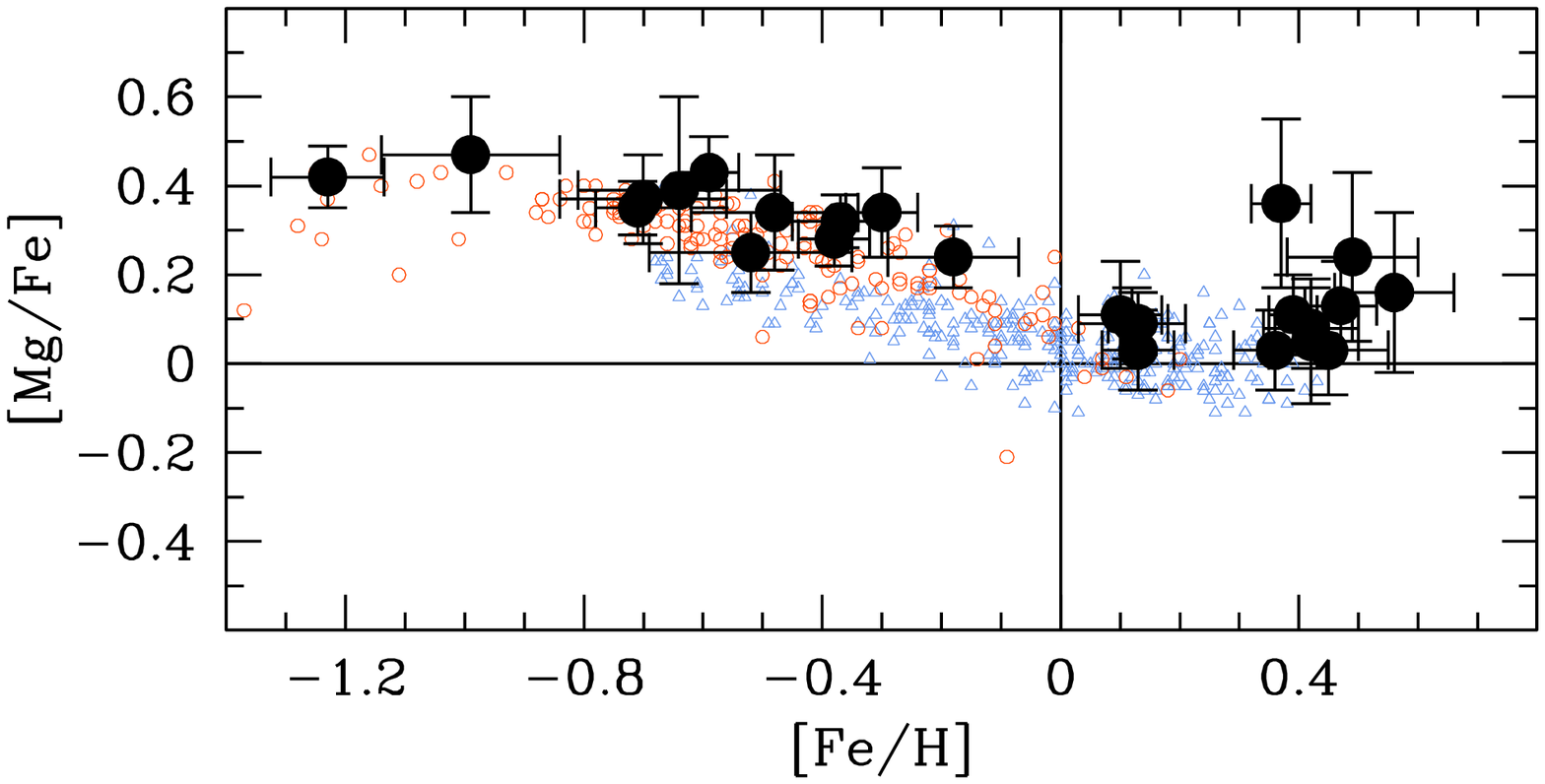}
\includegraphics[bb=18 220 592 450,clip]{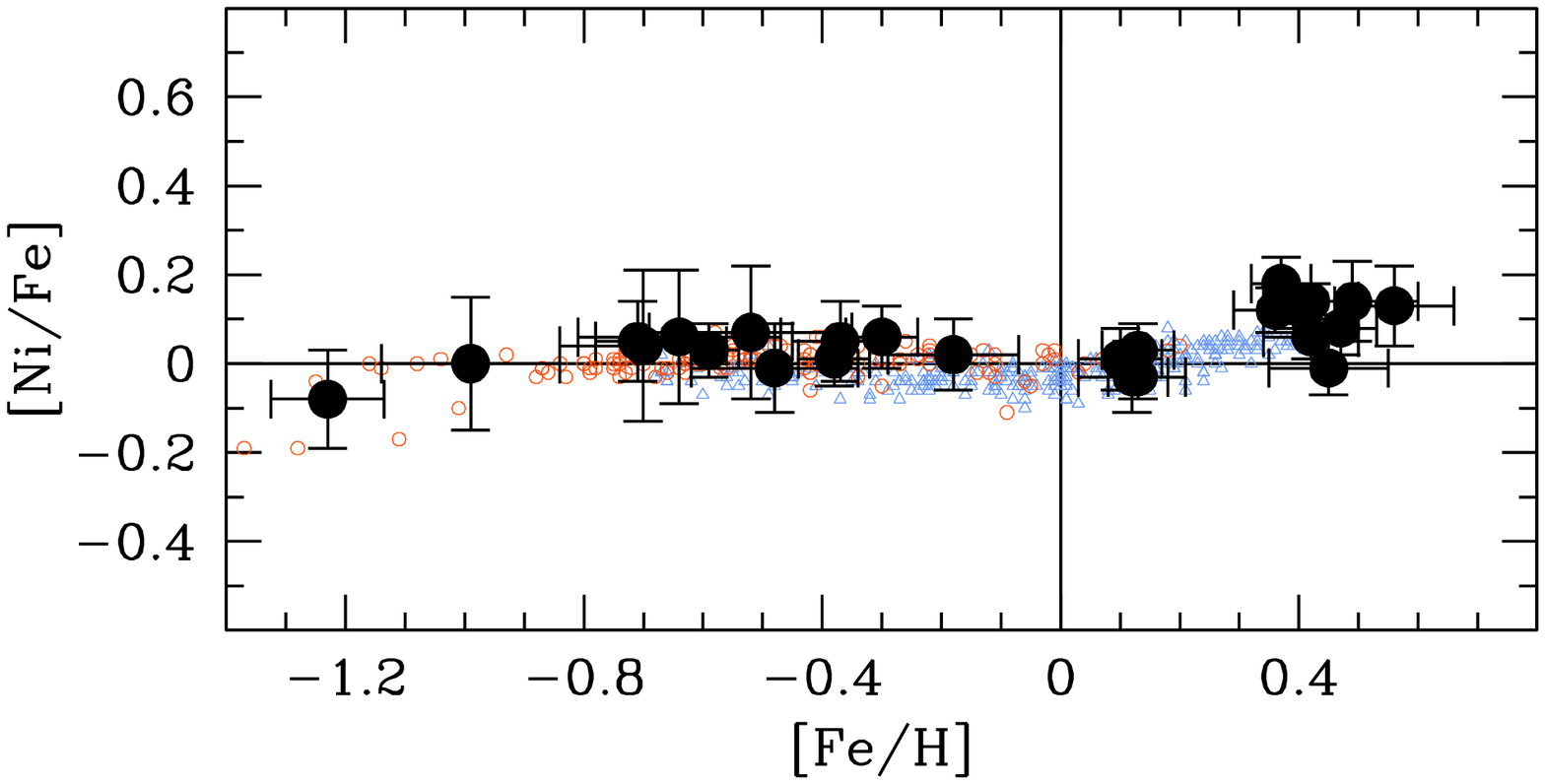}}
\resizebox{\hsize}{!}{
\includegraphics[bb=18 220 592 450,clip]{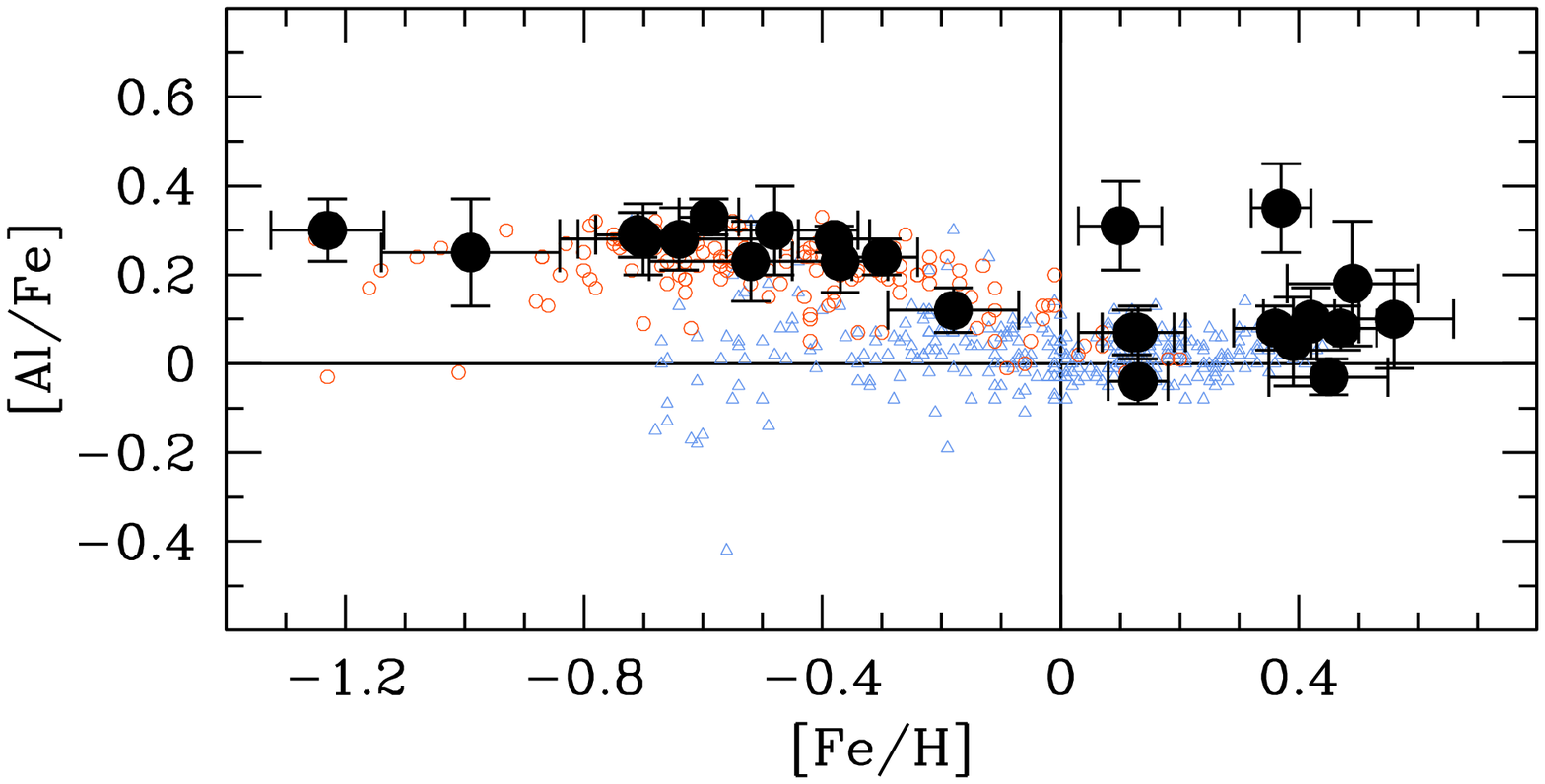}
\includegraphics[bb=18 220 592 450,clip]{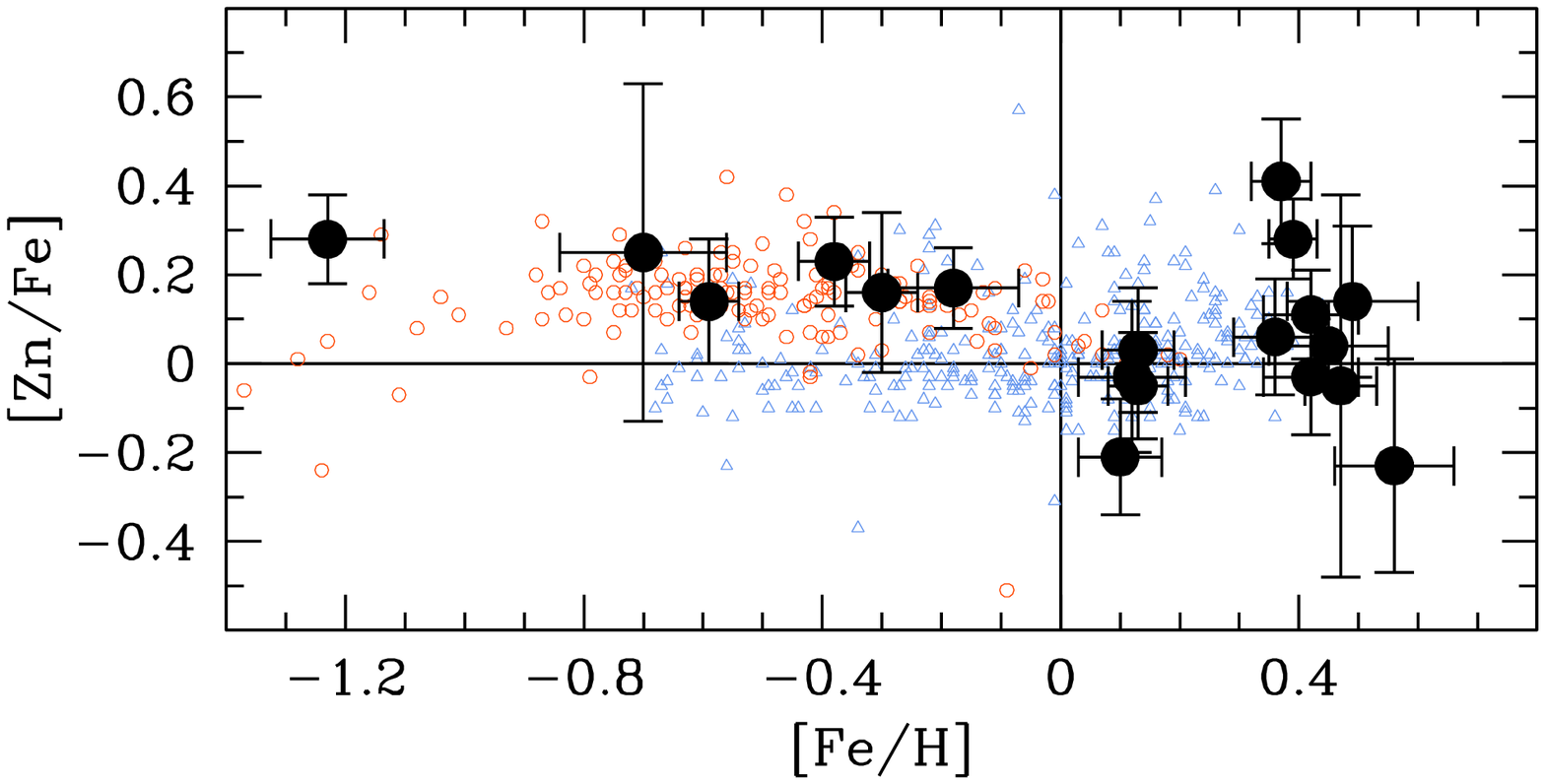}}
\resizebox{\hsize}{!}{
\includegraphics[bb=18 220 592 450,clip]{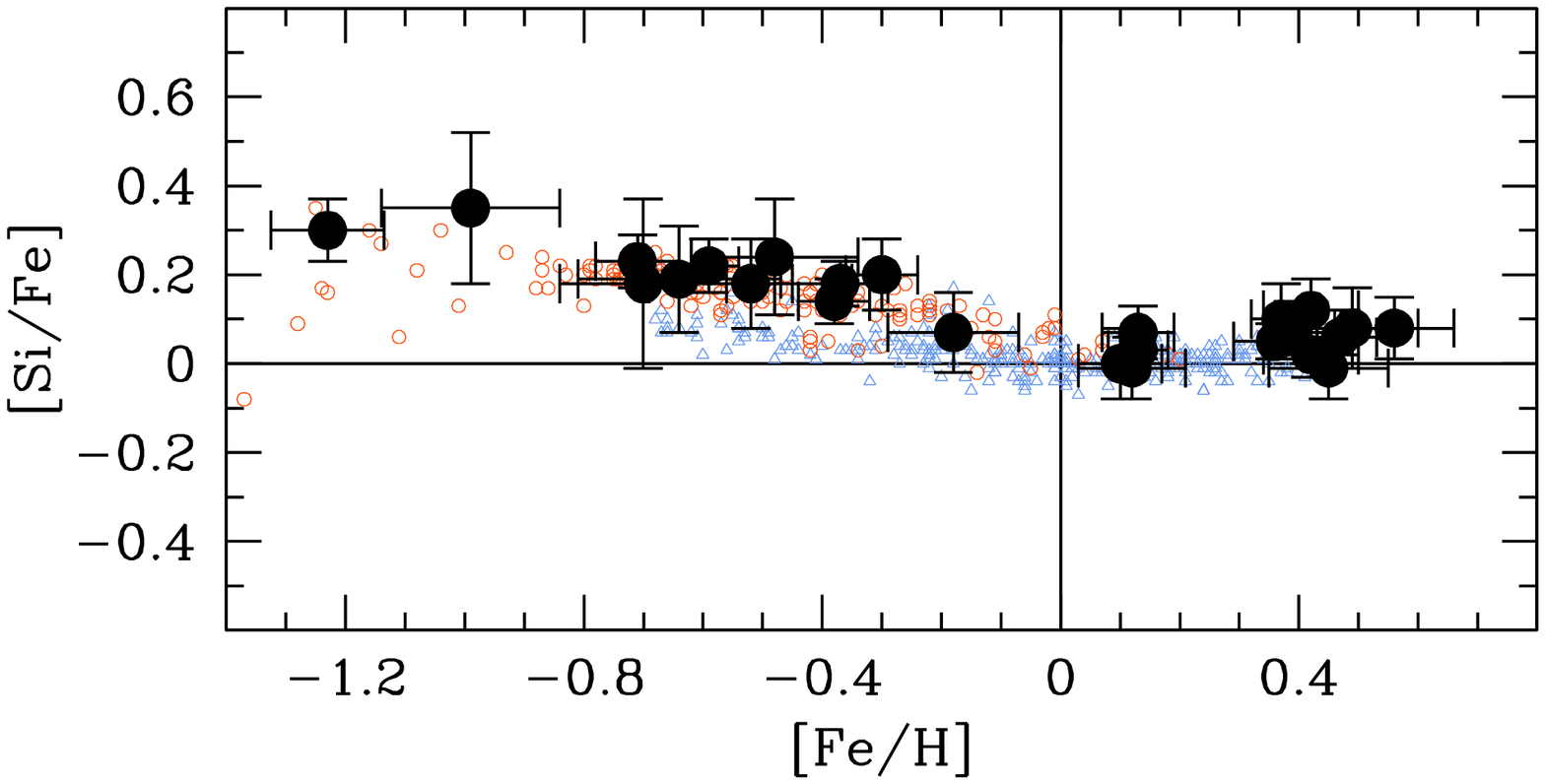}
\includegraphics[bb=18 220 592 450,clip]{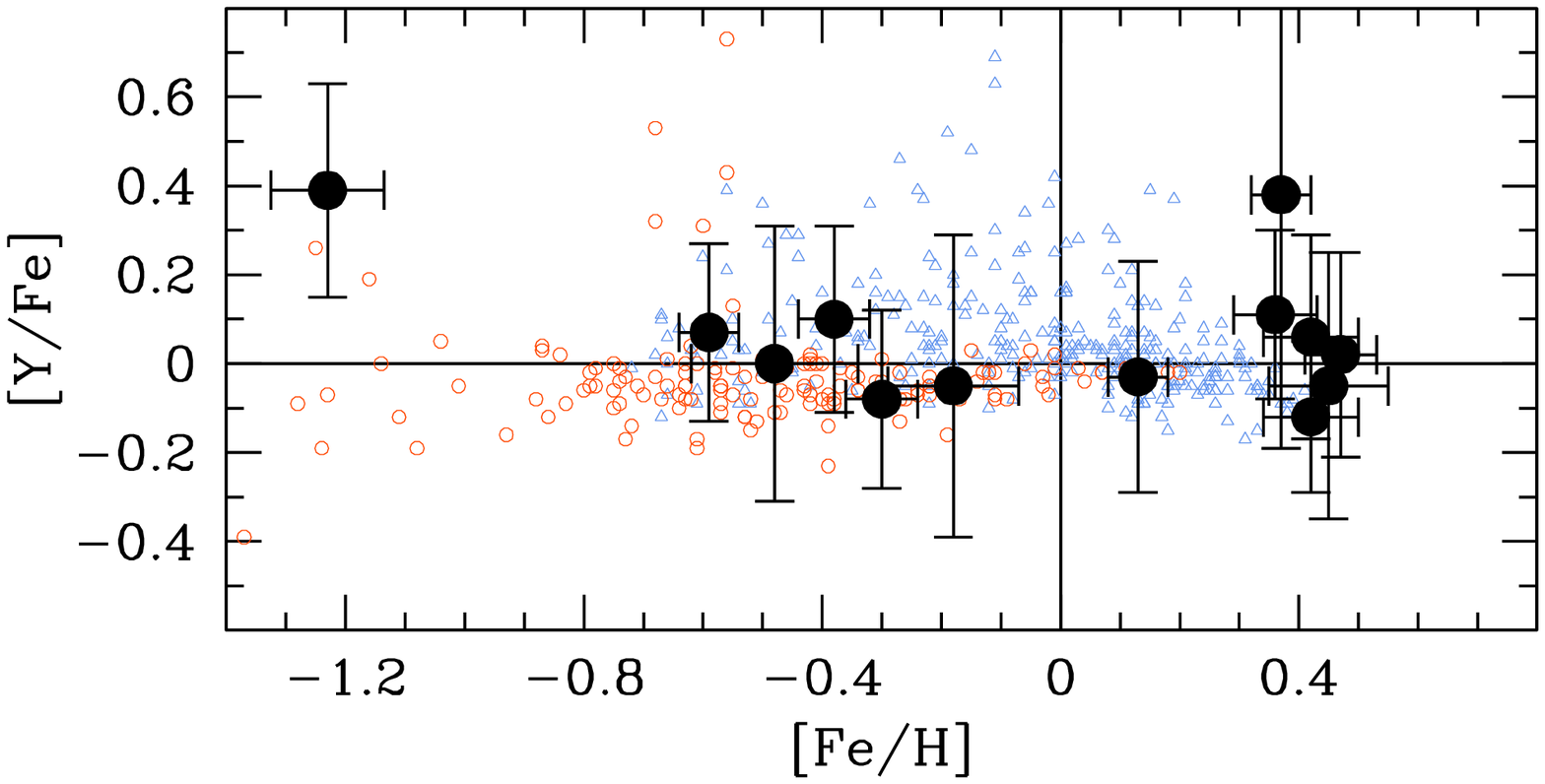}}
\resizebox{\hsize}{!}{
\includegraphics[bb=18 170 592 450,clip]{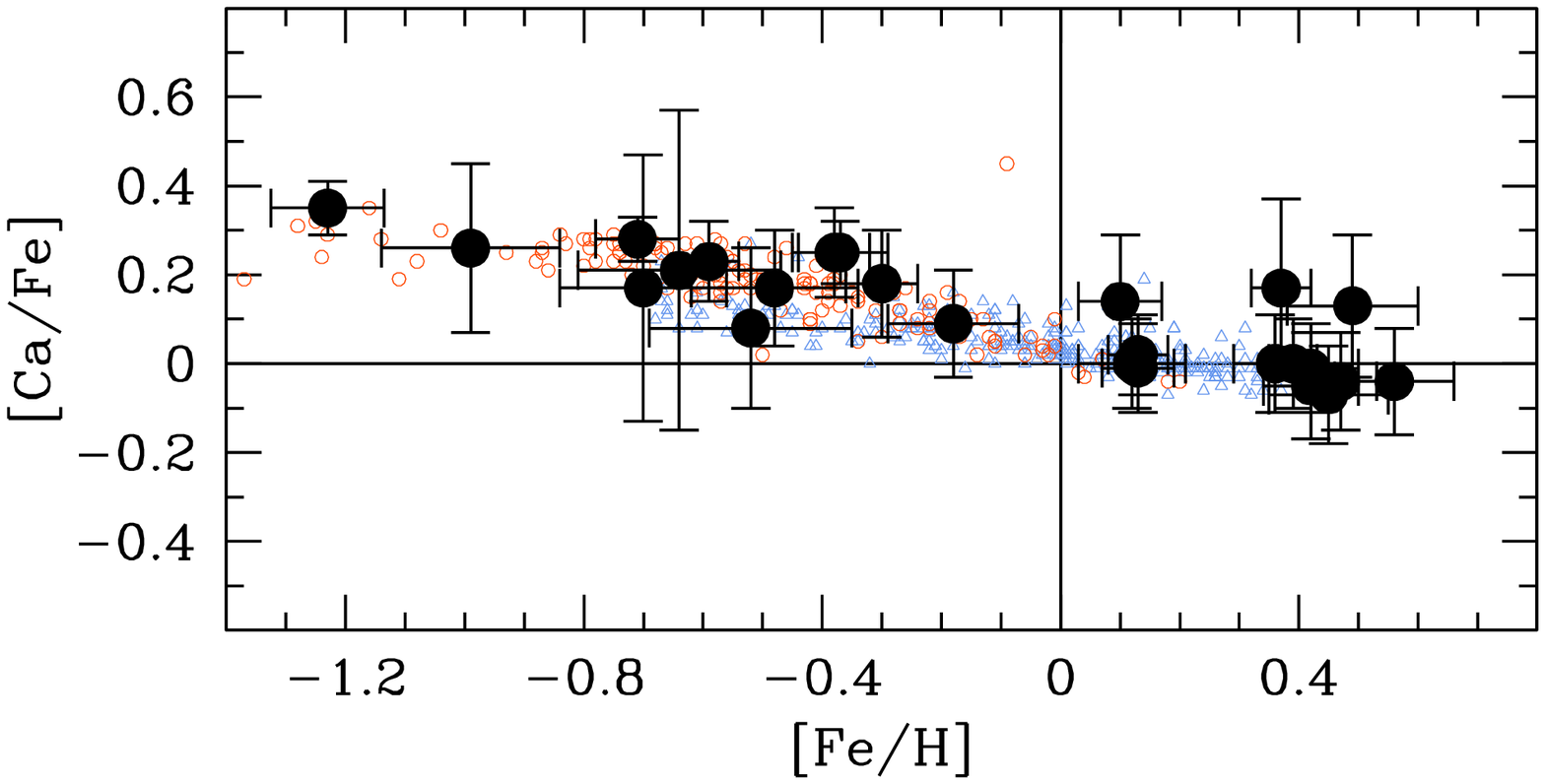}
\includegraphics[bb=18 170 592 450,clip]{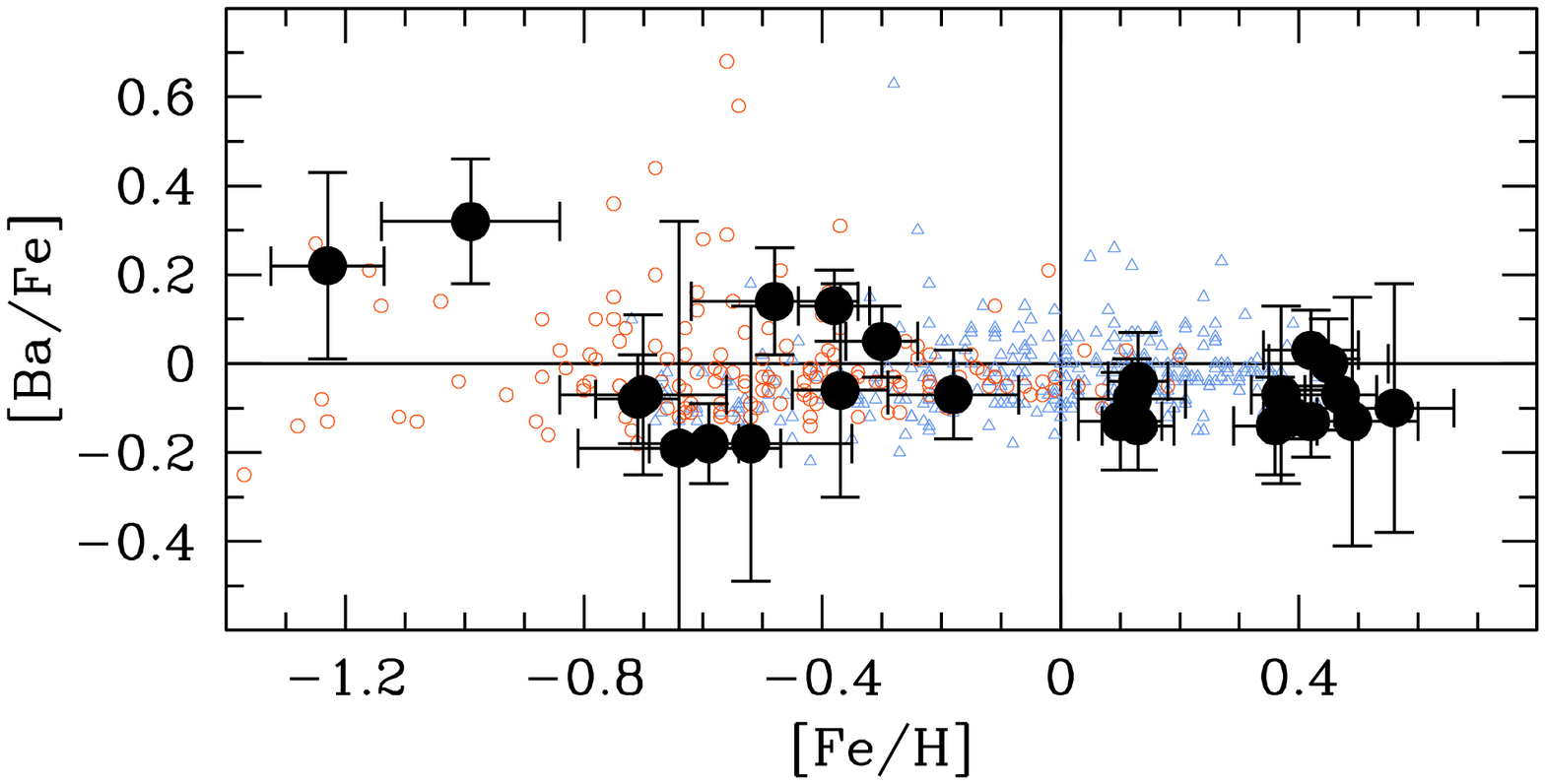}}
\caption{Filled bigger circles show the abundance results for 25 of the 26 
microlensed dwarf and subgiant stars in the Bulge (only stellar parameters
were determined for the star from \citealt{epstein2010}
that we analysed in \citealt{bensby2010}).
Thick disk and thin disk stars from \citep[][and in prep.]{bensby2003,bensby2005} 
are shown as small circles red and blue circles, respectively.
The error bars represent the total uncertainty in the abundance ratios
and have been calculated according to the prescription in \cite{epstein2010},
see Sect.~\ref{sec:analysis}.
\label{fig:abundances}}
\end{figure*}

\subsubsection{Similarities to the Galactic thick disk}

From the analysis of OGLE-2008-BLG-209S, which was the first microlensed
subgiant star with a metallicity below solar, we got 
the first hint from a chemically un-evolved star that the metal-poor 
parts of the bulge might be similar 
to those of the Galactic thick disk \citep{bensby2009}. In 
\cite{bensby2010} where the sample was 
expanded to 15 microlensed dwarf and subgiant stars, out of which 7 had 
$\rm [Fe/H]<0$, it was clear that the metal-poor bulge, in terms of both 
elemental abundance ratios as well as stellar ages, is similar to the 
Galactic thick disk. With this study we add another five stars with sub-solar
metallicities, and another seven with super-solar metallicities. 
The updated abundance plots are shown in Fig.~\ref{fig:abundances} and it 
is clear that the new stars further strengthens the similarity between 
the metal-poor bulge and the Galactic thick disk as seen in the
solar neighbourhood. The agreement is almost 
perfect for all the  $\alpha$-elements and only for the two $s$-process 
elements Ba and Y are there some discrepancies, and then for only one or 
two stars at the lowest [Fe/H].
Given the error bars and the fact that it is only 1-2 stars, it
is difficult to judge if the differences for Ba and Y are real.
We also note the large dispersion in [Zn/Fe] at high metallicities,
which most likely is caused by the fact that the Zn abundances for 
a majority of the stars are based on only one spectral line.
 
The similarity between the bulge and the local thick
disk abundance trends was actually first recognised from studies of K giants 
\citep{melendez2008} and later by \cite{alvesbrito2010} and \cite{gonzalez2011}. 
Recently it has been shown that this 
similarity also holds when comparing bulge giant stars with K giant 
stars from the thick disk in the inner parts of the Galaxy 
\citep{bensby2010letter}.

\subsubsection{Rapid enrichment and time scale for SNIa in the bulge}

The high level of $\alpha$-element abundances relative to Fe are signatures 
of rapid star formation where massive stars, which have short lifetimes, are 
the main contributors of chemical enrichment of the interstellar 
medium when they explode as core collapse SNe produce and expel
large amounts of $\alpha$-elements \citep[see, e.g.,][]{ballero2007}. 
This will result in high [$\alpha$/Fe] ratios at low [Fe/H], and we see
this for the microlensed dwarf stars at $\rm [Fe/H]\leq-0.4$. 
Once low- and intermediate-mass stars starts to play a significant r\^ole 
in the chemical enrichment, through the explosions of SNIa, which produce 
relatively little of the $\alpha$-elements, there will be a downturn 
(knee) in the [$\alpha$/Fe]--[Fe/H] abundance plot. This is what we see for 
the microlensed bulge dwarfs at $\rm [Fe/H]\approx-0.4$. This means that 
the star formation in the bulge continued for at least as long as 
it takes for the SNIa to contribute significantly to the build-up of
elements. Estimates of the SNIa time scale depends on the type of environment and
ranges from a few hundred million years to 1-2 billion years 
\citep[e.g.,][]{matteucci2001b}. In Fig.~\ref{fig:agefe} we see that 
the bulge dwarfs with sub-solar 
[Fe/H] predominantly have ages around 10\,Gyr, and the lack of an 
age-metallicity relation for the metal-poor dwarfs (Sect.~\ref{sec:agefe}) 
could indicate that SNIa started 
to contribute to the chemical enrichment soon after star formation started 
in the bulge. 
Even though the internal measurement errors in the old stars of the sample is 
still quite large to be constraining timescales of 100\,Myr versus 1\,Gyr,
this could mean that the timescale for SNIa in the bulge was short.
If not, the stars with [Fe/H] just above the knee would have
been significantly younger than those on the plateau, but they are not. 
A short time-scale of a few hundred million years for SNIa is found for 
elliptical galaxies while the time-scale in the disks of spiral galaxies
are a few billion years \citep[e.g.,][]{matteucci2001b}.
Given the similarities we see between the metal-poor bulge and the
thick disk it is worthwhile to mention that the
thick disk in the solar neighbourhood shows a 
possible age-metallicity relation \citep{bensby_amr,haywood2006,bensby2007letter2}, 
with the stars that are more metal-rich than the knee having
ages that are 1 to 2 Gyr younger than those more metal-poor than the
position of the knee. 
Now, we have only 1-2 microlensed dwarf stars with 
$\rm -0.4\lesssim [Fe/H] \lesssim 0$, 
and it might well be, once we have a larger sample, that those 
stars turn out to be, on average, younger than the more metal-poor ones.  
Also note that there is a possibility that the lack of an age-metallicity 
relation could be due to a lot of mixing (either caused by the bar, or caused by the 
merging of sub-clumps). Such mechanisms
have been proposed to be responsible for both the observed 
age-velocity relation and the absence of age-metallicity relation in the solar 
neighbourhood \citep[e.g.,][]{minchev2010}.

\subsubsection{The metal-rich bulge}

Regarding the metal-rich bulge, most stars seem to follow the trends 
as outlined by the thin disk as seen in the solar neighbourhood. 
However, there are two stars that have elevated abundance ratios that stand out: 
MOA-2010-BLG-523S at $\rm [Fe/H]=+0.09$
and MOA-2010-BLG-259S at $\rm [Fe/H]=+0.37$. From Table~\ref{tab:parameters}
we see that these are two of the more evolved stars in the sample.
In order to investigate whether the unusually high Na abundances in 
MOA-2010-BLG-523S and MOA-2009-BLG-259S for their metallicities
could be due to the assumption of LTE, we have performed 1D non-LTE
calculations following \cite{lind2011} for the four employed \ion{Na}{i} lines. 
We find that while the non-LTE effects are significant ($-0.1$\,dex) they 
are essentially the same in all our targets as well as in the Sun, leaving 
our derived LTE-based [Na/Fe] ratios unaffected. Non-LTE can thus not 
explain the atypical Na abundances of these two stars. Another possibility
could be that the stellar parameters for these stars are erroneous.
In order to bring down the [Na/Fe] ratio from $+0.19$ to zero for 
MOA-2010-BLG-523S, and to bring down  [Na/Fe] ratio from $+0.4$ to
$+0.2$ for MOA-2009-BLG-259S, we need to increase the effective temperatures 
with more than 500\,K or increase the surface gravities by $\sim 1$\,dex. 
As the current temperatures of 5250\,K and 4953\,K well reproduce the wing 
profiles of their H$\alpha$ lines, it is unlikely that erroneous effective 
temperatures are responsible for the high abundance ratios that we see
for some elements in this star. Also, both of these stars have more
than 10 \ion{Fe}{ii} lines measured and the \ion{Fe}{i}-\ion{Fe}{ii}
ionisation balances are well constrained, making a shift of 1\,dex in surface 
gravity unlikely. For now we have to accept the fact that these two stars
show elevated levels for some elements that does not agree with the other
microlensed bulge dwarf stars at similar [Fe/H].
 
Notable is that the uprising trends of [$X$/Fe] with [Fe/H] at very high 
metallicities that are seen for the disk stars for many elements, e.g., 
[Na/Fe], [Al/Fe], and in particular the iron-peak element [Ni/Fe],
is also present in the microlensed bulge dwarfs
(see Fig.~\ref{fig:abundances}).

We furthermore note that some of the characteristics of MOA-2009-BLG-259S and 
MOA-2010-BLG-523S (high Na and Al) seem typical of second generation 
globular cluster stars \citep[e.g.,][]{carretta2009b}. 
Stars resembling globular cluster second generation stars have 
been found in the halo \citep{martell2010} and their eventual presence in 
the bulge might gauge the role that globular clusters had, through losing 
stars and/or disintegrating in the formation of the bulge. However, 
\cite{martell2010} have not analysed O, Na, Mg, and Al, so it is still
to be proved whether their sample indeed originated from globular clusters.
Also, a connection with the globular cluster O-Na anti-correlation for
MOA-2009-BLG-259S and MOA-2010-BLG-523S is
unlikely given that their O, Na, Al, and Mg abundances are all high (and 
several other elements too for one of the two stars).

\begin{figure}
\resizebox{\hsize}{!}{
\includegraphics[bb=18 144 592 718,clip]{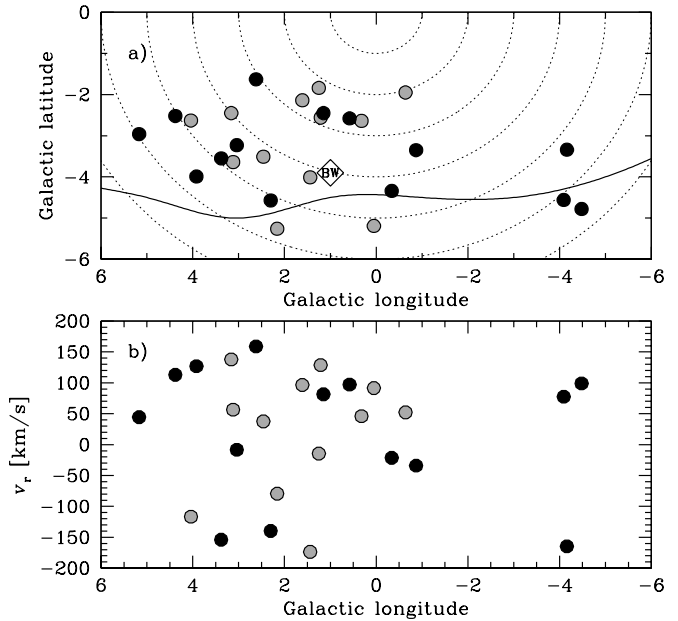}}
\caption{
        Positions, radial velocities, and metallicities for
        the now in total 26 microlensed dwarf and subgiant stars. 
        Grey circles mark stars that
        have $\rm [Fe/H]<0$ and black circles those that have $\rm [Fe/H]>0$.
        The curved line in (a) shows
        the outline of the southern Bulge based on observations with
        the COBE satellite \citep{weiland1994}, and the concentric
        dotted lines show the angular distance from the Galactic centre
        in $1^{\circ}$ intervals. 
        \label{fig:allevents}}
\end{figure}

\subsection{Positions and radial velocities}

Figure~\ref{fig:allevents}a shows the positions on the sky for 
the full sample of microlensed dwarf stars in the bulge. 
They are all located between
galactic latitudes $-2^{\circ}$ to $-5^{\circ}$,  similar to Baade's
window at $(l,\,b)=(1^{\circ},-4^{\circ})$. 
The reason why most stars are located at positive
galactic longitudes and negative latitudes is related to  the
extinction / geometry of the bulge.
As shown in, e.g., \cite{kerins2009}, the optical depth, based on stars, is higher 
in these regions and therefore the chances for microlensing
to occur is higher here. Also, MOA and OGLE  have so far mainly surveyed
the bulge at negative latitudes as these are regions of lower extinction.

We find that there is essentially no difference
in the velocity distributions between the metal-poor and metal-rich
samples (see Fig.~\ref{fig:allevents}b). 
The average velocities are almost identical, $21.9\pm98.7$ and
$19.7\pm109.0\,\kms$ for the two samples.
The high velocity dispersion is typical for what has been seen in
large surveys of the bulge, for instance the BRAVA
survey, which observed and determined radial velocities for
several thousand giants \citep{howard2008}. They found velocity dispersions
ranging between $80-120\,\kms$ at these galactic longitudes.

   \begin{figure}
   \resizebox{\hsize}{!}{
   \includegraphics[bb=18 150 470 718, clip]{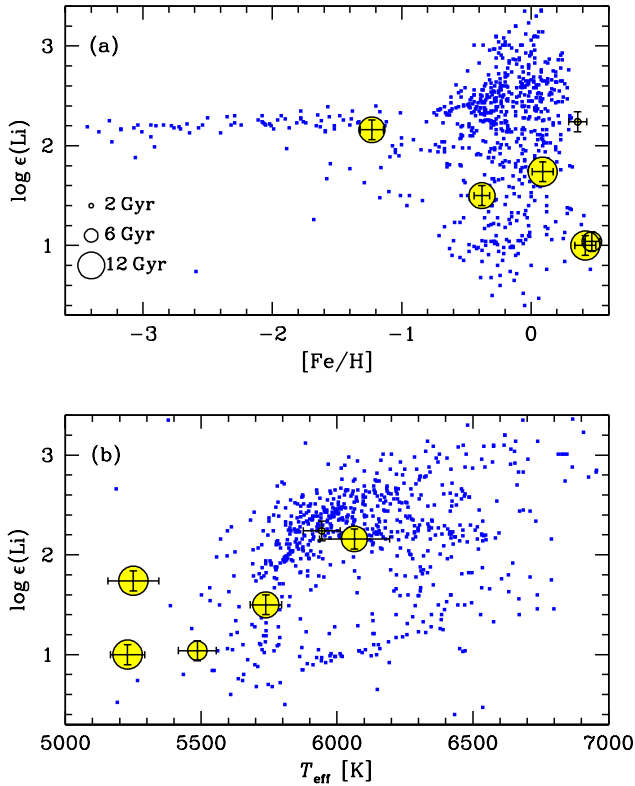}}
      \caption{(a) Li abundance versus [Fe/H], and (b) versus $\teff$. 
      The large circles indicate the microlensed bulge dwarfs 
      listed in Table~\ref{tab:li}. The ages of the stars have been
      coded by the sizes of the circles as indicated in (a). 
      Comparison data (small blue dots) 
      come from \cite{melendez2010li}, \cite{lambert2004}, and 
      \cite{ghezzi2010fe,ghezzi2010li}.
                   }
         \label{fig:life}
   \end{figure}

\subsection{Lithium abundances in the bulge}

As we reported in \cite{bensby2010li}, MOA-2010-BLG-285S is the first 
metal-poor dwarf star for which Li has been clearly detected in the 
Galactic bulge. It has a Li abundance which is fully consistent with 
what is seen in other metal-poor dwarf stars in the Galactic disk and 
halo at this effective temperature and metallicity \citep[e.g.,][]{melendez2010li}. 
Figure~\ref{fig:life}a shows that the star lies on the metal-rich end 
of the Li Spite plateau \citep{spite1982}. Combined with its old age, 
MOA-2010-BLG-285S is an excellent confirmation that the bulge did not 
undergo a large amount of Li production or astration in its earliest 
phases.

The Li abundances in the other five more metal-rich dwarf stars,
in which it could be measured,
show a wide range, most of them significantly lower
than the Li Spite plateau.  MOA-2008-BLG-311S, which has
a higher temperature ($\teff=5944$\,K) than the other 
metal-rich stars, is the one with the highest Li abundance.
Most likely, the Li in these metal-rich stars
represent abundances that were obtained
by depletion during their main-sequence lifetimes
from higher initial values. Observations of
intermediate-age solar-metallicity stars both in clusters such
as M67 \citep[e.g.,][]{jones1999} and in the field \citep[e.g.,][]{baumann2010} 
show that they have a large range of Li values, which can be explained by
a range of ages, masses, and initial rotation values.

The only other detections of Li in the Galactic bulge are
from observations of RGB and AGB stars
\citep[e.g.,][]{gonzalez2009} in which the 
atmospheric Li abundance have been altered. Dwarf and subgiant
stars with effective temperatures greater than about 5900\,K 
are the only reliable tracers of Li \citep[e.g.,][]{weymann1965,boesgaard1986}.

\section{Why are the dwarf and giant MDFs different?}
\label{sec:zoccali}

The MDF is a key observable used to constrain chemical evolution models
\citep[e.g.,][]{chiappini1997,kobayashi2006}. Therefore
it is important to discuss in detail the differences in the MDFs of 
dwarf and giant stars in the bulge.
If the gap in the dwarf MDF is real and if the bulge MDF 
truly is bimodal, why does the giant MDF not show this bimodality?
Below we discuss several possibilities in detail.

\subsection{The microlensed dwarf stars are not in the bulge?}

In \cite{bensby2010} we explored in detail whether 
the microlensed  dwarf and subgiant stars are located in the Bulge 
region, or in the disk on either this side or the far side of the Bulge.
We concluded then that the combination of kinematics,
colour-magnitude diagrams and microlensing statistics makes it
highly likely that the microlensed dwarfs that we are studying 
are members of a stellar population that belong in the Bulge.
In this study we have further investigated the bulge membership of
the microlensed bulge dwarfs. In Sect.~\ref{sec:absmag} we determined
the absolute magnitudes of the microlensed dwarfs and compared
them to those that can be determined from microlensing techniques.
We find that the two are in agreement at better than our $1\sigma$
measurement error of 0.15\,mag.  This is consistent with the
expectation (based on Galactic microlensing models) that the overwhelming
majority of microlensed sources towards these lines of sight are in
the bulge.

\subsection{Baade's window not representative?}

Could it be that the region towards Baade's window is not representative
for the Bulge, not even other close-by regions, and that this is the reason
for the discrepant MDFs from the RGB and microlensed dwarf star samples?
\cite{zoccali2008} find different MDFs in their fields ($b=-4^{\circ}$, 
$-6^{\circ}$, and $-12^{\circ}$) and claim that there is a vertical 
metallicity gradient in the bulge.
The \cite{zoccali2008} RGB comparison sample is located in Baade's window,
which indeed is a very small region of the Galactic bulge. The 26
microlensed dwarf stars have a wider spread in $l$ and $b$. However,
their average distance from the Galactic centre coincides very well with 
Baade's window (see Fig.~\ref{fig:allevents}).
Also \cite{brown2010} used WFC3 photometry
and two-colour plots to derive MDFs for 4 fields in the bulge 
including some close to the Galactic plane
($b=-2.15^{\circ}$, $-2.65^{\circ}$, $-3.81^{\circ}$, and $-4.72^{\circ}$). 
They found that K-S tests ruled out the possibility that they were drawn 
from the same distribution. They were working in reddening-free
measurements, and while not as good as spectroscopy, obviously,
it is a possible confirmation of MDF variations across the bulge.
Interesting, some of their MDFs look quite metal-rich (though they
had issues concerning their zero-point) and more double-peaked.

\subsection{Strong gradients in the bulge?}

The RGB comparison sample in Baade's window from \cite{zoccali2008}
is located $4^{\circ}$ from the Galactic plane as well as from
the Galactic centre, while our microlensed dwarf stars range 
2$^{\circ}$-5$^{\circ}$ from the Galactic plane and 
2$^{\circ}$-6$^{\circ}$ from the Galactic centre (see Fig.~\ref{fig:allevents}).
Would it be possible that there are gradients present in the 
metal-rich bulge population that are not picked up by the RGB sample
in Baade's window? Possibilities include:
  	 \begin{itemize}
	 \item There are two co-existing bulge populations - one metal-poor and one 
	 metal-rich. However, there is a very steep spatial abundance gradient in what 
	 we pick up as the metal-rich dwarf population, and at the distance of 
	 Baade's window the [Fe/H] is much lower than close to the plane or close to
	 the Galactic centre. Hence, the metal-rich stars can not be seen in the
	 RGB sample from Baade's window.
	 \item There is a number density gradient in the metal-rich population,
	 meaning that at the distance of Baade's window the number of very metal-rich
	 stars has dropped and is very low. Hence, the metal-rich population is missing
	 in the RGB sample.
	 \end{itemize}
\cite{babusiaux2010} use the \cite{zoccali2008} data and they find
a metal-rich and a metal-poor population. They mention that
the metal-rich group seems to drop out as they move away from
Baade's window, perhaps supporting the second possibility
wherein the number density of metal-rich stars drops quickly.
From the same stellar sample \cite{gonzalez2011} show that these 
two populations separate also in their $\rm [\alpha/Fe]$ ratios.
However, the fields they are considering are located at $b=-4^{\circ}$, 
$-6^{\circ}$, and $-12^{\circ}$, is a significantly larger range
than that covered by the microlensed dwarf stars ($b=-2^{\circ}$, 
to $-6^{\circ}$, see Fig.~\ref{fig:allevents}) within which the effect should
be less dramatic.
We also note the claim by \cite{rich2007}, which is based on IR spectroscopy, 
that there is no metallicity gradient in the inner bulge ($|b|<4^{\circ}$).

   \begin{figure*}
   \resizebox{\hsize}{!}{
   \includegraphics[bb=18 144 592 718,clip]{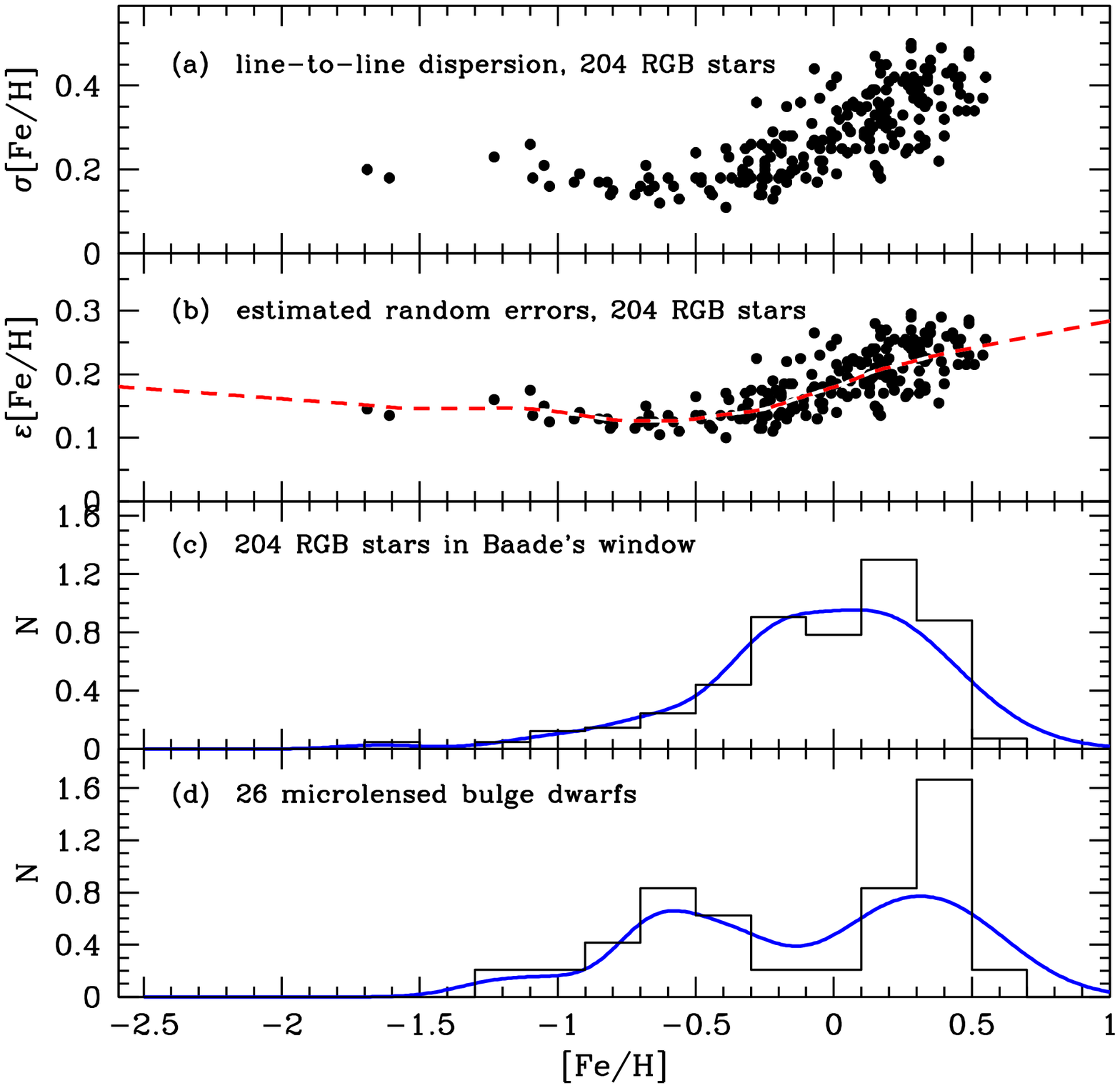}
   \includegraphics[bb=18 144 592 718,clip]{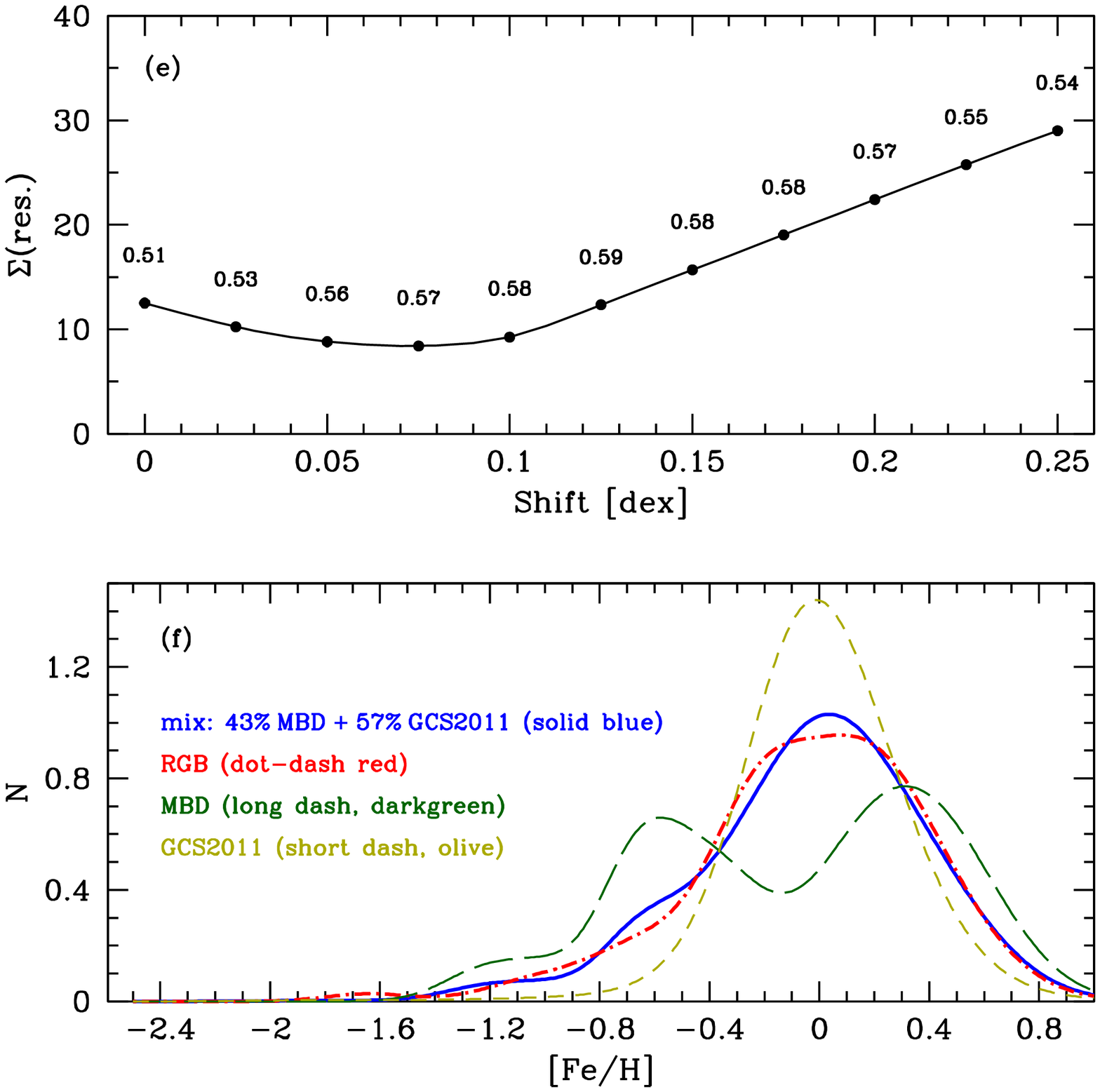}}
      \caption{
      (a) Line-to-line dispersion of \ion{Fe}{i} abundances for the 204
      RGB stars from \cite{zoccali2008};
      (b) The estimated random errors of the 204 RGB stars from \cite{zoccali2008}.
      Dashed red line represents the running median (extrapolated outside
      the data limits);
      (c) The MDF for the 204 RGB stars. The blue line is a smoothed version
      where each star is represented by a Gaussian having with a width
      equal to the estimated random error as shown in (b);
      (d) The MDF for the microlensed dwarf stars. The blue line is a smoothed
      version where each star is represented by a Gaussian with a width
      as given by the dashed line in (b);
      (e) shows how the sum of the minimum residual between 
      a joint disk GCS and microlensed dwarf MDF and the RGB MDF
      vary with the shift of the GCS disk sample. For each value,
      the fraction of the GCS disk MDF is shown. The minimum
      sum of residuals is reached when the disk GCS MDF is shifted
      $+0.075$\,dex, and when the fraction of GCS disk MDF is 57\,\%.
      The MDFs and the smeared MDFs for the RGB sample in Baade's window 
      (f) The smoothed MDFs for the RGB sample (red dash-dotted line),
      the microlensed dwarf sample (green long-dashed line), the GCS2011
      disk sample (olive coloured short-dashed line). The solid blue line
      shows the mix of the MDFs from the GCS2011 disk sample and the
      microlensed dwarf sample that best match the RGB MDF.
      All MDFs have been normalised to unit area (hence $N$ on the ordinates
      are arbitrary numbers).
              }
         \label{fig:contamination}
   \end{figure*}

\subsection{Uncertainties in the analysis?}
\label{sec:problems}

The analysis of giants is more difficult than that of 
dwarf stars. This is because blending is severe for giants, caused 
by their lower effective temperatures, which will result in spectra 
with stronger lines from neutral elements and a lot of molecular lines. 
The blending will become even more severe for metal-rich stars,
which could be the reason for spuriously high abundance ratios
in some studies \citep[see, e.g.,][]{lecureur2007}. 
Regarding problems in the analysis of solar-metallicity giants,
\cite{santos2009} reported that their standard analysis
of giant stars in open clusters results in a much narrower metallicity 
distribution (with a total spread between different clusters 
of only about 0.1 dex) than that obtained using dwarf stars,
which show that there is a cluster-to-cluster spread in
metallicity about 2.5 times higher than that found using giants.
The problem seems to be related to blends by atomic and molecular 
lines, as the use of the line list by \cite{hekker2007}, which has 
been checked for blends, results in a total metallicity spread 
derived from open cluster giants similar to that derived from dwarfs 
(Fig. 1 in \citealt{santos2009}). Thus, standard analysis of giant 
stars may result in a compression of the MDF for metallicities around 
solar or higher, perhaps erasing the metallicity gap we have found 
in the MDF of the bulge.

The RGB star spectra of \cite{zoccali2008} have a resolution of 
$R\sim20 000$, whilst our microlensed dwarf stars are observed with 
$R\sim40 000$ to 60\,000. That blending might be a problem in metal-rich
giants can be seen in Fig.~\ref{fig:contamination}a which shows how 
the line-to-line dispersion of the Fe abundances for the \cite{zoccali2008} 
RGB sample in Baade's window vary with [Fe/H], 
reaching dispersions of 0.5\,dex, and in some cases even higher. 

To better account for the large dispersions seen in the
RGB sample we have in Figure~\ref{fig:contamination}c constructed an
alternative MDF for the RGB sample. Instead of plotting a standard histogram,
each star is represented by a normalised Gaussian, defined by the specific
mean value and error in [Fe/H] for each star.
It would have been be good if we had error estimates similar to the ones 
that were calculated in Sect.~\ref{sec:analysis} for the microlensed 
dwarfs, but the only information
we have from \cite{zoccali2008} is the line-to-line dispersion for each of 
the 204 stars in Baade's  window and that in average 60-70 \ion{Fe}{i} 
lines have been used to determine the stellar parameters and [Fe/H]. 
However, Bensby et al.~(in prep.) have performed
an error analysis following \cite{epstein2010} for 703 F and G stars 
(dwarfs, subgiants, and a handful of low-luminosity giants.). 
It is found that the calculated errors in [Fe/H] correlate with the 
line-to-line dispersion of the \ion{Fe}{i} abundances. 
As the analysis of a majority of the 703 stars have used many more 
\ion{Fe}{i} lines than the 60-70 that \cite{zoccali2008} have used, we 
select a subsample of $\sim 100$ stars from the 703 stars that have 100 
or less \ion{Fe}{i} lines measured and used in the analysis. We then perform 
a simple linear regression between the line-to-line dispersion and the error 
in [Fe/H] for those stars. This regression line
has a constant of 0.044 and a slope of 0.50, allowing us to
transform the line-to-line dispersions into 
estimated errors in [Fe/H] for the 204 RGB stars. This plot is shown in 
Fig.~\ref{fig:contamination}b. 
It should be cautioned that the spectra used by \cite{zoccali2008}
have lower resolutions and lower $S/N$ than the spectra used by 
Bensby et al.~(in prep.), and that the stellar
parameter determination methodology is different (gravity is not a
spectroscopically tuned parameter in \citealt{zoccali2008}).
Additionaly, and most important,
the \cite{zoccali2008} sample consists of giants while the Bensby et al.~(in prep.)
have analysed dwarf and subgiant stars. However, it is likely that the stellar
parameters are less well determined when the line-to-line scatter is higher.

The total MDF curve shown in Fig.~\ref{fig:contamination}c is then the sum of 
the Gaussians for all 204 RGB stars, normalised to unit area.
It is interesting to see how the large uncertainties
smears the MDF, and especially, now showing a less steep drop-off at 
the metal-rich end.
The question now is what happens if the same type of uncertainties
are added to the sample of microlensed dwarf stars?
Can it amount for an erasure of the gap such that the dwarf MDF resembles
the RGB MDF?

Assigning each star in the microlensed dwarf sample an uncertainty
given by the dashed line in Fig.~\ref{fig:contamination}b, we then construct
a normalised gaussian for each of the 26 microlensed dwarf stars. These are
merged to form a new MDF for the microlensed dwarf sample, and is shown
in Fig.~\ref{fig:contamination}d. Although the double-peak still is present,
it is clear that it has been smeared out.  However, the
double-peak remains. Hence, we can not blame uncertainties
in the giant analysis to be the sole source for the discrepant MDFs, unless the
error consists in a compression of the [Fe/H] values (see discussion earlier).

\subsection{Disk contamination in RGB sample?}
\label{sec:contamination}

A certain degree of contamination in the RGB sample in Baade's window
is expected. Based on the Besan{\c c}on model of the Galaxy \citep{robin2003},
\cite{zoccali2008} estimated a contamination of $10-15$\,\% in their 
RGB sample in Baade's window, divided more or less equally between 
foreground thin disk stars, located $2-5$\,kpc from the Sun, and thick 
disk stars located within the bulge. 
The estimation of the thick disk contamination is based on the assumption 
in the Besan{\c c}on model that the thick disk follows an exponential 
radial distribution that peaks in the Galactic centre. As also stated in 
\cite{zoccali2008}, these contaminations are poorly determined as we 
know very little about the stellar populations in the inner Galactic disk, 
and essentially nothing about possible thin and thick disk stellar 
populations within the bulge. We therefore find it worthwhile to investigate
how much contamination is actually needed for the microlensed dwarf MDF and 
the Baade's window RGB MDF to agree.

The bulge has a radius of about 1.5\,kpc \citep{rattenbury2007}, meaning 
that the foreground disk stars, if they are located in front of the Bulge, 
should be located, at most, approximately 6\,kpc from the Sun. Baade's 
window is located at $b=-4^{\circ}$, which at
this distance from the Sun translates into a vertical distance
of $\sim 400$\,pc.  Assuming that the inner disk
contains both a thin and a thick disk (see first results
in \citealt{bensby2010letter}), and that they have
similar scale heights as in the solar neighbourhood, this is a distance
from the plane where the thin disk is dominating. Hence, it is
likely that if the giant sample is contaminated with 
foreground stars, it should presumably be by thin disk stars. 

Assigning each star an uncertainty in [Fe/H] given by the dashed
curve in Fig.~\ref{fig:contamination}b,
we create two normalised MDFs: one for the bulge (green dashed curve in 
Fig.~\ref{fig:contamination}f.) and one for the disk (olive coloured dashed 
curve in Fig.~\ref{fig:contamination}f.). 
The normalised bulge MDF is based on the 26 microlensed dwarf stars,
and the normalised disk MDF on the disk stars from the re-calibration
of the Geneva-Copenhagen Survey by \cite[][hereafter GCS2011]{casagrande2011}. 
In the solar neighbourhood the disk MDF peaks slightly below solar 
\citep[e.g.,][]{casagrande2011}. As the Galactic disk is likely  
to exhibit a shallow metallicity gradient 
\citep[e.g.,][and references therein]{maciel2010,cescutti2007}
the disk closer to the bulge would therefore have a slightly higher average 
metallicity. We will therefore leave the shift of the GCS2011 MDF as a 
free parameter. 

We then construct a joint MDF from the smeared out dwarf star MDF and 
the smeared out GCS2011 disk MDF, with varying fractions of the two 
(from 0 to 1 in steps of 0.1).
Then we compare this mixed MDF with the Baade window RGB MDF (as given in 
Fig.~\ref{fig:contamination}a) and determine for which mixture the sum
of the residuals is minimised. This procedure is done for several different
metallicity shifts of the GCS2011 disk MDF. The results are shown
in Fig.~\ref{fig:contamination}e which
shows how the minimum sum of the residuals changes with the
degree of contamination for eleven different
shifts of the GCS2011 MDF. The lowest values  are achieved for a shift 
of approximately $+0.075$\,dex of the GCS2011 MDF and a mix of 57\,\% GCS2011
and 43\,\% dwarf MDF. As can be seen
in Fig.~\ref{fig:contamination}f, the mixed bulge-disk MDF and the RGB MDF are 
similar and agree quite well. Note that the exact shift of the GCS2011 disk
MDF seems to be of little importance. All shifts suggest a disk 
contamination fraction of 50 to 60\,\%.

Is such a high foreground contamination of the RGB sample realistic?
If 50 to 60\% of the bulge RGB samples are actually foreground
disk stars, this might be visible in the kinematics of the
RGB sample. However, the velocity dispersion 
($\sigma_{v_{\rm r}}$) is similar in the RGB sample and in the microlensed dwarf
sample.  Also stars with disk kinematics would introduce a
metallicity-dependent bias in the RGB sample, since the
foreground contamination only is significant near solar metallicity 
(between [Fe/H] of $-0.3$ and $+0.1$).
However, \cite{babusiaux2010} found no sign
of disk kinematics within that metallicity range.
Furthermore, the BRAVA survey RGB stars should also presumably
be contaminated by foreground disk stars.  However, there is good
agreement between the $v_{\rm r}$ distribution of the microlensed
dwarfs and that of the BRAVA giants.

\subsection{The magnification puzzle}

\begin{figure}
\resizebox{\hsize}{!}{
\includegraphics[bb=20 170 590 530,clip]{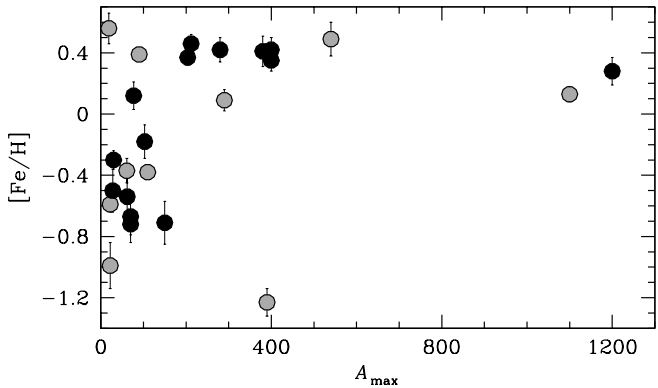}}
\caption{
[Fe/H] versus $A_{\rm max}$. Grey circles mark
the new stars and black circles the stars in \cite{bensby2010}.
\label{fig:puzzle}
}
\end{figure}

In \cite{cohen2010puzzle} a peculiar relation between maximum 
magnification ($A_{\rm max}$) and [Fe/H] was discovered. Using
16 microlensing events available at that time it was demonstrated 
that a very strong correlation between metallicity and 
maximum magnification existed. Several possible explanations were given, 
but all had to be rejected, leaving the puzzle unsolved.
Now, with additional data points in hand, we again look
at this issue. As can be seen in Fig.~\ref{fig:puzzle} it is evident
that most of the events with the highest magnifications have super-solar
metallicities, and that lower magnification events dominate at sub-solar
metallicities. However, the new data points (grey circles)
appears to weaken the relation as we now have high-magnification
events at low metallicities and low-magnification events at high
metallicities, combinations that were not present in \cite{cohen2010puzzle}.
A Spearman rank test between [Fe/H] and $A_{\rm max}$ now
gives a correlation of 0.42 and a significance that it
deviates from zero of 0.033.
A Spearman correlation of zero indicates that there is no 
tendency for [Fe/H] to either increase or decrease when $A_{\rm max}$ 
increases. This means that we can not rule out the null hypothesis 
that the two properties are uncorrelated.
So, even though there is still a statistical significance 
that the two parameters are correlated, it is weaker than in 
\cite{cohen2010puzzle} who found a significance of 0.005,
i.e. strongly correlated.
More microlensing events are needed in order to resolve whether the
$A_{\rm max}$ puzzle is real or not.

In any case, assuming that it is real -- how would that affect the microlensed
dwarf star MDF? Since all stars observed so far, without exception,
has turned out to be either metal-poor {\it or} metal-rich, the only
effect would probably be that the relative sizes of the two peaks change. The most 
likely scenario is that we are missing more low-magnification
events than high-magnification events. If so, that would mean that the
metal-poor peak is under-estimated. The gap around solar metallicities
would probably persist. To investigate what the effects might be on
the results regarding a possible foreground contamination of the RGB sample,
we repeat the analysis shown in Fig.~\ref{fig:contamination}f
with two times as many stars in the metal-poor peak of the
microlensed dwarf stars MDF. This results in that the best fit
is now found for a mix of 65\,\% disk stars from GCS2011 (now shifted
0.125\,dex) and 35\,\% microlensed dwarf stars. Thus, if the
$A_{\rm max}$ -- [Fe/H] puzzle is real, and should be accounted
for, the possible foreground contamination of the RGB sample could 
be even higher.

\subsection{Conclusion}

In summary, we find no definitive answer to why the bulge MDF look
different depending on if it is based on RGB stars or microlensed dwarf 
stars. Abundance analysis of dwarf stars is certainly coupled to smaller
errors and uncertainties than that of giant stars. Also, the resolution and 
$S/N$ of the microlensed dwarf star spectra are higher than for the 
RGB stars in Baade's window. Furthermore, statistics show that 
the microlensed dwarf stars are highly likely to be located in the Bulge 
region. However, our estimate of the foreground disk contamination is at 
least four times as high as what the Besan{\c c}on model used by 
\cite{zoccali2008} predicts, and is hard to digest.
Given that the knowledge of the
inner disk is very poor, and also the fact that the reddening is uncertain 
and often very patchy, it is clear that 
it is very difficult to get a handle on the disk contamination in the 
RGB sample.

\section{The origin of the bulge}

\begin{table}
\centering
\caption{
Properties for the metal-rich and metal-poor bulge sub-samples$^{\dag}$.
\label{tab:properties}
}
\setlength{\tabcolsep}{5.5mm}
\begin{tabular}{ccc}
\hline\hline
\noalign{\smallskip}
                                  &  Metal-poor bulge    &   Metal-rich bulge \\
                                  & $\rm [Fe/H]<0$       &   $\rm [Fe/H]>0$   \\
\noalign{\smallskip}
\hline
\noalign{\smallskip}   
$N$                               & 12              &  14             \\                              
$\rm  [Fe/H] $      & $-0.60\pm0.30$  &  $0.32\pm0.15$  \\
$\rm Age$        & $11.7\pm 1.7$\,Gyr$^{\ddag}$   &  $7.6\pm3.9$\,Gyr    \\
$ v_{\rm r}$        & $22\pm99\,\kms$       &  $20\pm109\,\kms$     \\
$\rm [\alpha/Fe]$  & $0.29\pm0.06$   &  $0.08\pm0.05$              \\ 
\noalign{\smallskip}
\hline
\end{tabular}
\flushleft
{\tiny
{\bf Notes.} 
$^{\dag}$ The values given in the table are the mean values 
and their associated dispersions. Also, note that the $\alpha$-element 
abundance is here defined as the mean of 
the Mg, Si, and Ti abundances.\\
$^{\ddag}$ Excluding MOA-2010-BLG-446S. If this star is included the average
age change to $11.2\pm 2.9$\,Gyr
}
\end{table}

From our studies of  26 microlensed dwarf and subgiant stars
we find that the bulge MDF is bimodal, and Table~\ref{tab:properties}
summarises the properties of the two sub-populations. 
The metal-poor sample has an average metallicity of $\rm [Fe/H]\approx-0.60\pm0.30$,
which is in perfect agreement with the mean metallicity that is found for
thick disk stars at the solar circle ($\rm [Fe/H]=-0.6$, \citealt{carollo2010}).
The metal-rich sample shows a very high-metallicity, and narrow, MDF
peaking at $\rm [Fe/H]=+0.32\pm0.16$. The metallicity dispersion in the metal-poor
sample is twice the dispersion in the metal-rich one.
For the metal-poor bulge we find that almost all stars are old with ages in 
the range 9-13\,Gyr, while the stars of the metal-rich 
population have ages that span from the oldest stars of the Galactic halo 
($>12$\,Gyr) to the youngest stars of the Galactic disk.
These age and metallicity differences indicate that the bulge consists of 
two distinct stellar populations; the metal-poor bulge and the
metal-rich bulge, respectively.   
Furthermore we find that the abundance trends of the 
metal-poor bulge is similar to those of the Galactic thick disk as seen in 
the solar neighbourhood, i.e., enriched $\rm [\alpha/Fe]$ ratios signalling
a formation on short time-scales. The presence of a ``knee" in the $\rm [\alpha/Fe]$
abundance plot, coupled with the lack of an age-metallicity relation, could
favour a short time-scale for SN\,Ia in the bulge. 
The similarity between the metal-poor bulge and nearby 
thick disk abundance trends are also seen in studies of giant stars 
\citep[e.g.,][]{melendez2008,alvesbrito2010,ryde2010} and now this similarity has
also been seen between bulge giants and inner Galactic disk giants, 
3 to 5\,kpc from the Sun \citep{bensby2010letter,bensby2011letter}.
The abundance trends of the metal-rich 
population partly resemble those of the thin disk in the solar 
neighbourhood. We see no difference in the radial
velocity distributions of the two populations, both show the same
high velocity dispersions that is expected for a bulge population
\citep{howard2009}.

The observational constraints above points to that the Milky Way
has a bulge with a complex composition of stellar populations.
The similarities between the metal-poor bulge and the thick disk 
suggest that they might be tightly connected. Possible scenarios could 
be that the metal-poor bulge formed from an old Galactic disk 
(secular evolution, evidence for secular origins of bulges 
have been seen in many external galaxies, e.g., \citealt{genzel2008}), 
that the metal-poor bulge and the old disk formed simultaneously 
(but separately), or that the metal-poor bulge and the thick disk are 
indeed the same population. 
Simulations have also shown that it is possible for thick disks
and bulges to form together by rapid internal evolution in unstable, 
gas-rich, and clumpy disks \citep{bournaud2009}, which further could 
explain the similarities between the Galactic bulge and the thick disk.
Furthermore, from a differential analysis of K giants at Galactocentric
radii 4, 8, and 12\, kpc, \cite{bensby2011letter} found that it is likely 
that the scale-length of the thick disk is shorter than that of the thin disk.
This has consequences for the relative fractions of thin and 
thick disk stars in the Galactic plane, implying that in the bulge region
the two disks should be equally present at low galactic latitudes. 
Interestingly, the 26 microlensed bulge stars divide about equally
into being either metal-poor or metal-rich (see Fig.~\ref{fig:bimod}).
Could this mean that what we are seeing is a bulge solely made up
of thin and thick disk stars, and that the metal-rich peak in the microlensed
bulge dwarf MDF is a manifestation of the inner thin disk? The thin disk is 
known to show a strong metallicity gradient \citep[see, e.g.,][]{cescutti2007}, 
and if interpolated (linearly) to the bulge region it would reach 
$\rm [Fe/H]>+0.5$. A linear relationship all the way to the bulge is unlikely, 
but the strong gradient indicates that the average metallicity could end up at, 
or around, the metalliicty that we see for the 
metal-rich peak of the microlensed dwarf MDF ($\rm [Fe/H]\approx +0.3$). 
The lack of a radial gradient in
the thick disk could be explained through radial migration that 
has had time to wash out abundance gradients in this old population.
For the much younger thin disk, radial  migration has not had enough time 
to act, and a radial metallicity gradient is still present. These findings
could support the results from the BRAVA survey
which claim that there is no evidence of a classical bulge \citep{shen2010},
but that the Milky Way is a pure-disk galaxy. The formation of the bulge
from a two-component stellar disk has recently been modelled by \cite{bekki2011}.
We note that also \cite{babusiaux2010}
find evidence for two bulge populations. Their results are
based on the metallicities from \cite{zoccali2008}, now complemented with
kinematics, and they find that the metallicity gradient along the bulge
minor axis observed by \cite{zoccali2008} can be related to a varying mix 
a metal-rich population with average metallicity of $\rm [Fe/H] = +0.13$
that has bar-like kinematics and a metal-poor population with average metallicity
of $\rm [Fe/H]=-0.3$ that has thick disk-like kinematics.  
Regarding the metal-rich bulge we find a wide range of stellar ages.
As also discussed in \cite{babusiaux2010}, this is consistent
with a stellar population with old stars from the inner disk that have 
been re-distributed by the central bar, and whose young stars have formed
in star bursts triggered by the central bar.

\section{Implications for the bulge IMF}

The MDFs based on the microlensed bulge dwarfs and on red giants in BaadeÕs
window are clearly different. This has important consequences for chemical
evolution models, as the MDF strictly constrains which kind of models are
plausible. In particular, the peak of the MDF depends on the slope of the
initial mass function (IMF). The chemical evolution model of the bulge by
\cite{ballero2007} is based on the photometric MDF of giant stars by
\cite{zoccali2003} and on the spectroscopic MDF of giant
stars by \cite{fulbright2006}. \cite{ballero2007} suggested for the
bulge an IMF much flatter than in the solar neighbourhood, i.e., an IMF skewed
towards high mass stars, to explain the MDF of bulge giants by 
\cite{zoccali2003} and \cite{fulbright2006}. 

A more recent model by \cite{cescutti2011}, using now the
spectroscopic MDF of giant stars by \cite{zoccali2008}, also favours a flat
IMF to reproduce the peak of the MDF of bulge giants. 
However, those conclusions may have to be revised in view of our new MDF based
on microlensed dwarf stars. 
Our MDF has two well-defined peaks, one that we
associate with a metal-poor old bulge, and another one at 
super-solar metallicities associated with a younger population. 
Hence, the IMF no longer has to be flat to explain a single
peaked solar metallicity MDF that was made in 0.5 Gyr,
as in the \cite{cescutti2011} models. To reproduce the bimodal nature 
of the bulge MDF, probably a normal IMF can be used for the metal-poor bulge,
while contributions from Type Ia SN can probably explain the younger metal-rich peak.

Notice that our MDF is still poorly sampled (only about two dwarfs per
metallicity bin), thus, in order to provide even firmer constraints on
chemical evolution models, it is important to at least double the sample of
microlensed bulge stars observed at high resolution and high $S/N$. This will
provide tight constraints on our galactic bulge, and on a broader context will
also have important consequences on the modelling of external galaxies 
\citep[e.g.,][]{ballero2007b}

\section{Summary}

With the twelve events presented in this study the sample size
of microlensed dwarf and subgiant stars in the Galactic bulge
has grown to 26. All stars have been observed with high-resolution
spectrographs on 8-10 m class telescopes, which has allowed us
to determine detailed elemental abundances for many elements.
The following results and conclusions can be drawn with the 
current sample.
\begin{enumerate}
\item The MDF based on the microlensed dwarf and subgiant stars
is bimodal with one peak at $\rm [Fe/H]\approx -0.6$, one peak
at $\rm  [Fe/H]\approx +0.3$, and with essentially no stars
around solar [Fe/H]. This is  in stark contrast to the MDF
provided by giant stars in Baade's window which peaks at this
exact location.
\item A statistical test shows that it is unlikely to have 
a gap in the MDF based on the microlensed dwarf stars if 
the microlensed sample and the giant sample from \cite{zoccali2008} 
have been drawn from the same underlying population. 
We find no definitive answer to why the two distributions
look different. Possible explanations include:
large uncertainties in the analysis of the giant stars 
(especially the metal-rich ones); a higher level than predicted 
contamination of foreground disk giant stars in the giant sample 
(maybe as high as 60\,\%); low number of microlensed dwarf
stars, the difference will go away once more microlensing
events have been observed; a mix of all above.
\item The metal-poor bulge is very similar to the Galactic thick 
disk in terms of mean metallicity, elemental abundance trends, 
as well as stellar ages. The metal-rich bulge is complicated, resembling
the thin disk in terms of abundance ratios, but  the thin disk does 
not show a wide range in ages.
These findings could be evidence of the co-existence of two distinct
stellar populations within the bulge, each which might have 
its own formation scenario.
\item Combining the similarities between the metal-poor bulge and the 
thick disk with additional evidence from \cite{bensby2011letter}
and \cite{shen2010}, we speculate that the bulge is purely made from 
the stellar disk, where the metal-rich bulge would be 
the manifestation of the inner thin disk.
The origin of the metal-rich bulge population might also be 
coupled to the Galactic bar, including
old stars from the inner disk scattered by the bar and younger
stars whose formation was triggered by the bar.
\item 
Li abundances were presented for six of the microlensed stars. 
One star is located on the Spite plateau ($\rm [Fe/H]=-1.23$ and 
$\log\epsilon ({\rm Li}) = 2.16$),
indicating that the Spite plateau might be universal. The other five stars
are all metal-rich and show evidence of Li depletion due to their
lower effective temperatures.
\item 
The new events presented in this study have weakened the 
recently discovered, and un-explained, $A_{\rm max}$-[Fe/H] puzzle. 
We now have stars at high [Fe/H] that have low $A_{\rm max}$ 
and stars at low [Fe/H] that have low $A_{\rm max}$. 
More events are needed to reveal the true nature of this
puzzling relation.
\item  The recent claims by \cite{cescutti2011} of a very flat 
IMF in the bulge is not supported by the MDF and abundance trends as 
probed by our microlensed dwarf stars. 
\item Lastly, using the new colour-temperature calibration by
\cite{casagrande2010} and the now in total 26 microlensing events, we find
that the colour of the bulge red clump is $(V-I)_{0} = 1.06$.
\end{enumerate}

\begin{acknowledgement}

 T.B. was funded by grant No. 621-2009-3911 from The Swedish 
 Research Council.
 S.F. was partly funded by the Swedish Royal Academy of Sciences
 and partly by grant No. 2008-4095 from The Swedish Research Council.
 Work by A.G. was supported by NSF Grant AST\,0757888. 
 A.G.-Y. is supported by the Israeli Science Foundation, an
 EU Seventh Framework Programme Marie Curie IRG fellowship and the
 Benoziyo Center for Astrophysics, a research grant from the Peter 
 and Patricia Gruber Awards, and the William Z. and Eda Bess Novick 
 New Scientists Fund at the Weizmann Institute.
 S.L. research was partially supported by the DFG cluster of excellence
`Origin and Structure of the Universe'. 
J.M. thanks support from FAPESP (2010/50930-6), 
USP (Novos Docentes) and CNPq (Bolsa de produtividade).
J.G.C. was supported in part by  NSF grant AST-0908139.
J.C.Y. is supported by an NSF Graduate Research Fellowship.
Work by C.H. was supported by the grant from National Research Foundation 
of Korea (2009-0081561).
T.S. is supported by JSPS20740104 and JSPS23340044.
The MOA project is funded by JSPS20340052, JSPS22403003, and
JSPS23340064.
 We would like to thank Bengt Gustafsson, Bengt 
 Edvardsson, and Kjell Eriksson for usage of the MARCS model atmosphere 
 program and their suite of stellar abundance programs. Finally, we thank 
 anonymous referee for many and valuable comments.
  
\end{acknowledgement}

\bibliographystyle{aa}
\bibliography{referenser}

\end{document}